\documentclass[journal,10pt,transmag]{IEEEtran}

\usepackage{algorithm}
\usepackage[section]{placeins}

\usepackage{diagbox}

\usepackage{amsthm}

\makeatletter
\def\ps@IEEEtitlepagestyle{
  \def\@oddfoot{\mycopyrightnotice}
  \def\@evenfoot{}
}
\def\mycopyrightnotice{
  {\footnotesize
  \begin{minipage}{\textwidth}
  \centering
  Copyright~\copyright~2017 IEEE. Personal use of this material is permitted. However, permission to use this  \\ 
  material for any other purposes must be obtained from the IEEE by sending a request to pubs-permissions@ieee.org.
  \end{minipage}
  }
}

\newtheorem{proposition}{Proposition}

\newcommand{\norm}[1]{\left\Vert #1 \right\Vert}
\newcommand{\abs}[1]{\left\vert #1 \right\vert}

\usepackage{xcolor}   
\newcommand{\rev}[1]{{\color{black} #1}}

\newcommand{\secondrev}[1]{{\color{black} #1}}

\usepackage{graphicx}
\usepackage[cmex10]{amsmath}
\usepackage{amssymb,amsthm}
\usepackage{array}
\usepackage{url}
\usepackage{cite}
\usepackage{multirow}
\usepackage{color}
\usepackage{algorithm}
\usepackage[noend]{algpseudocode}

\DeclareMathOperator*{\argmin}{arg\,min}
\DeclareMathOperator*{\argmax}{arg\,max}

\usepackage{tikz,pgfplots}
\def\etanoisy{{\overline{\eta}_{\mathrm{noisy}}}}
\def\x{{\mathbf x}}

\def\z{{\mathbf z}}
\def\b{{\mathbf b}}
\def\vtil{{ \widetilde{\mathbf v} }}
\def\z{{\mathbf z}}
\def\P{{\mathbf P}}

\def\C{{\mathbb C}}

\def\P{{\mathbf P}}
\def\v{{\mathbf v}}
\def\w{{\mathbf w}}

\def\bPsi{{\boldsymbol{\Psi}}}

\def\bPhi{{\boldsymbol{\Phi}}}

\def\EE{{ \mathbb E }}

\pgfplotsset{compat=1.12}

\ifCLASSINFOpdf
\else
\fi
\begin{document}
\title{Learning-Based Compressive MRI} 

\author{
    \IEEEauthorblockN{    Baran G\"ozc\"u\textsuperscript{1}, Rabeeh Karimi Mahabadi\textsuperscript{1}, Yen-Huan Li\textsuperscript{1}, Efe Il{\i}cak\textsuperscript{2}, \\ Tolga \c{C}ukur\textsuperscript{2,3}, Jonathan Scarlett\textsuperscript{4}, and Volkan Cevher\textsuperscript{1}
    }
    \IEEEauthorblockA{\textsuperscript{1}Laboratory for Information and Inference Systems (LIONS), EPFL, Switzerland}
    
    \IEEEauthorblockA{\textsuperscript{2}National Magnetic Resonance Research Center (UMRAM), Bilkent University, Ankara, Turkey}
    
    \IEEEauthorblockA{\textsuperscript{3}Department of Electrical and Electronics Engineering, Bilkent University, Ankara, Turkey}
    
    \IEEEauthorblockA{\textsuperscript{4}Department of Computer Science \& Department of Mathematics, National University of Singapore, Singapore}
} 

\markboth{ IEEE Transactions on Medical Imaging }
{Shell \MakeLowercase{\textit{et al.}}: Bare Demo of IEEEtran.cls for Journals}

\maketitle

\begin{abstract}
    In the area of magnetic resonance imaging (MRI), an extensive range of non-linear reconstruction algorithms have been proposed that can be used with general Fourier subsampling patterns.  However, the design of these subsampling patterns has typically been considered in isolation from the reconstruction rule and the anatomy under consideration. In this paper, we propose a learning-based framework for optimizing MRI subsampling patterns for a specific reconstruction rule and anatomy, considering both the noiseless and noisy settings.  Our learning algorithm has access to a representative set of training signals, and searches for a sampling pattern that performs well on average for the signals in this set.  \secondrev{    We present a novel parameter-free greedy mask selection method, and show it to be effective for a variety of reconstruction rules and performance metrics.} Moreover we also support our numerical findings by providing a rigorous justification of our framework via statistical learning theory.

\end{abstract}


\IEEEdisplaynontitleabstractindextext
\IEEEpeerreviewmaketitle


\section{Introduction}

Magnetic resonance imaging (MRI) serves as a crucial diagnostic modality for scanning soft tissue in body parts such as the brain, knee, and spinal cord.  While early MRI technology could require over an hour of scan time to produce diagnostic-quality images, subsequent advances have led to drastic reductions in the scan time without sacrificing the imaging quality.

The application of MRI has served as a key motivation for compressive sensing (CS), a modern data acquisition technique for sparse signals.  The theory and practice of CS for MRI have generally taken very different paths, with the former focusing on sparsity and uniform random sampling of the Fourier space, but the latter dictating the use of {\em variable-density} subsampling. Common to both viewpoints, however, is the element of {\em non-linear decoding} via optimization formulations.

In this paper, we propose a {\em learning-based framework} for compressive MRI that is both theoretically grounded and practical.  The premise is to use training signals to {\em optimize the subsampling specifically for the setup at hand}.  

\secondrev{
    In more detail, we propose a novel greedy algorithm for mask optimization that can be applied to arbitrary reconstruction rules and performance measures. This mask selection algorithm is parameter-free, excluding unavoidable parameters of the reconstruction methods themselves.  We use statistical learning theory to justify the core idea of optimizing the empirical performance on training data for the sampling design problem. In addition, we provide numerical evidence that our framework can find good sampling patterns for different performance metrics such as peak signal-to-noise ratio (PSNR) and structural similarity (SSIM) index \cite{wang2004image}, and for a broad range of decoders, from basis pursuit and total variation to neural networks and BM3D.} Since our framework can be applied to arbitrary decoders, we also anticipate that it can benefit {\em future} decoding rules.

\paragraph*{Organization of the paper}

In Section \ref{sec:background}, we introduce the compressive MRI problem and outline the most relevant existing works, as well as summarizing our contributions.  In Section \ref{sec:algo}, we introduce our learning-based framework, along with its theoretical justification.  In Section \ref{sec:NUMERICAL}, we demonstrate the effectiveness of our approach on a variety of data sets, including comparisons to existing approaches. Conclusions are drawn in Section \ref{sec:CONCLUSION}.

\section{Background} \label{sec:background}

\subsection{Signal acquisition and reconstruction} \label{sec:signal}

In the {\em compressive sensing} (CS) problem \cite{Donoho2006}, one seeks to recover a sparse vector via a small number of linear measurements.  In the special case of compressive MRI, 
these measurements take the specific form of subsampled Fourier measurements, described as follows:
\begin{equation}
    \b = \P_{\Omega} \bPsi \x + \rev{\w},
\end{equation}
where $\bPsi \in \C^{p \times p}$ is the Fourier transform operator applied to the vectorized image,\footnote{The original image may be 2D or 3D, but we express it in its vectorized form for convenience.} $\P_{\Omega}: \C^{p} \rightarrow \C^{n}$ is a subsampling operator that selects the rows of $\bPsi$ indexed by the set $\Omega$, with $|\Omega| = n$, \rev{and $\w \in \C^n$ is possible additive noise.}  We refer to $\Omega$ as the {\em sampling pattern} or {\em mask}.

Given the measurements $\b$ (along with knowledge of $\Omega$), a {\em reconstruction algorithm} (also referred to as the {\em decoder}) forms an estimate $\hat{\x}$ of $\x$.  This algorithm is treated as a general function, and is written as follows:
\begin{equation}
    \hat{\x} = g(\Omega,\b). \label{eq:phi}
\end{equation}
A wide variety of decoding techniques have been proposed for compressive MRI; here we present a few of the most widely-used and best-performing techniques, which we will pursue in the numerical experiments in Section \ref{sec:NUMERICAL}.

In the general CS problem, decoders based on convex optimization have received considerable attention, both due to their theoretical guarantees and practical performance.  \rev{In the noiseless setting (i.e., $\w = 0$),} a particularly notable choice is {\em basis pursuit} (BP) \cite{Donoho2006}:
\begin{equation} \label{eq: BP}
    \hat{\x} = \argmin_{{\z} \,:\, \b = \P_{\Omega} \bPsi  {\z}} \| \bPhi {\z}\|_1
\end{equation}
 where  $\bPhi $ is a sparsifying operator such as the wavelet or shearlet transform.  A similar type of convex optimization formulation that avoids the need for the sparsifying operator is {\em total variation} (TV) minimization:
 \begin{equation} \label{eq: TV}
     \hat{\x} = \argmin_{{\z} \,:\, \b = \P_{\Omega} \bPsi  {\z}} \| {\z}\|_{\mathrm{TV}},
 \end{equation}
 where $\| {\z}\|_{\mathrm{TV}}$ is the total variation norm.
 
For the specific application of MRI, heuristic reconstruction algorithms have recently arisen that can outperform methods such as BP and TV, despite their lack of theoretical guarantees.  A state-of-the-art reconstruction algorithm was recently proposed in \cite{eksioglu2016decoupled} based on the {\em block matching and 3D filtering (BM3D)} denoising technique \cite{dabov2009bm3d} that applies principal component analysis (PCA) to patches of the image.  At a high level, the algorithm of \cite{eksioglu2016decoupled} alternates between denoising using BM3D, and reconstruction using regularized least squares formulations.  We refer the reader to  \cite{eksioglu2016decoupled} for further details, and to Section \ref{sec:NUMERICAL} for our numerical results.
 
Following their enormous success in machine learning applications, {\em deep neural networks} have also been proposed for MRI reconstruction.  We consider the approach of \cite{schlemper2017deep}, which uses a cascade of convolution neural networks (CNNs) interleaved with data consistency (DC) units.  The CNNs serve to perform de-aliasing, and the DC units serve to enforce data consistency in the reconstruction.  The deep network is trained by inputing the subsampled signals and treating the full training signal as the desired reconstruction.  We refer the reader to  \cite{schlemper2017deep} for further details, and to Section \ref{sec:NUMERICAL} for our numerical results.

\rev{Among the extensive existing literature, other relevant works include \cite{murphy2012fast} and \cite{uecker2014espirit}, where compressive sensing is unified with parallel MRI. In \cite{jin2016general}, a matrix completion framework is proposed for the parallel MRI setting. \secondrev{ In \cite{ravishankar2011mr}, \cite{ravishankar2013learning}, \cite{ravishankar2015efficient} and \cite{ravishankar2016data}, dictionary learning and faster transform learning methods are shown to provide considerably better quality of reconstructions compared to nonadaptive methods. In \cite{lingala2011accelerated}, \cite{yoon2014motion}, \cite{otazo2015low} and \cite{ravishankar2017low}, low rank models are used for improved results in dynamic MRI setting, for which dictionary-based approaches are also presented in \cite {lingala2013blind} and \cite{wang2014compressed}. Another notable work in the dynamic setting is  \cite{jung2009k},  which, in a compressive sensing framework, generalizes the previous work that exploits spatiotemporal correlations for improved frame rate \cite{tsao2003k}. Patient-adaptive methods for dynamic MRI also exist in the literature \cite{aggarwal2008patient,sharif2010patient}. A Bayesian approach is taken in \cite{bilgic2011multi} and \cite{duarte2012framework} for compressive MRI applications, whereas in \cite{huang2014bayesian}, Bayesian and dictonary learning approaches are combined. 

In \cite{lee2017deep}, a deep convolutional network is used to learn the aliasing artefacts, providing a more accurate reconstruction in the case of uniform sampling. In \cite{han2017deep}, this approach is applied to the radial acquisition setting. In \cite{sun2016deep}, a deep network is used to train the transformations and parameters present in an regularized objective function.  } Moreover, in \cite{mardani2017deep}, a framework based on generative adversarial networks is applied for an improved compressive MRI performance, whereas in \cite{sandinodeep}, a convolutional network is trained for faster acquisition and reconstruction for dynamic MRI setting. }

\subsection{Subsampling pattern design}

Generally speaking, the most popular approaches to designing $\Omega$ for compressive MRI make use of {\em random variable-density sampling} according to a non-uniform probability distribution \cite{lustig2007sparse}. The random sampling is done in a manner that favors taking more samples at low frequencies.  Some examples include variable-density Poisson disk sampling \cite{vasanawala2011practical}, multi-level sampling schemes \cite{adcock2013breaking, adcock2015quest}, pseudo-2D random sampling \cite{wang2009pseudo}, and \rev{variable density with continuous and block sampling models \cite{chauffert2014variable, boyer2014algorithm ,bigot2016analysis}.    }

While such variable-density approaches often perform well, they have notable limitations.  First, they typically require parameters to be tuned (e.g., the rate of decay of probability away from the center).   Second, it is generally unclear which particular sampling distribution will be most effective for a given decoding rule and anatomy.  Finally, the very idea of randomization is questionable, since in practice one would like to design a fixed sampling pattern to use across many subjects.

Recently, alternative design methods have been proposed that make use of fully sampled training data (i.e., training signals).  In \cite{knoll2011adapted,zhang2014energy,vellagoundar15robust}, the training data is used to construct a sampling distribution, from which the samples are then drawn randomly.  In \cite{liu2012under,ravishankar2011adaptive}, a single training image is used at a time to choose a row to sample, and in \cite{seeger2010optimization} the rows are chosen based on a mutual information criterion.  Much like the above-mentioned randomized variable-density sampling approaches, these existing adaptive algorithms contain parameters for their mask selection whose tuning is non-trivial. Moreover, to our knowledge, none of these works have provided theoretical \secondrev{justifications of the mask selection method. On the other hand, except for \cite{ravishankar2011adaptive}, these algorithms do not optimize the sampling pattern for a given general decoder. We achieve this via a parallelizable greedy algorithm that implements the given decoder on multiple training images at each iteration of the algorithm until the desired rate is attained.}

A particularly relevant prior work is that of \cite{baldassarre2016learning}, in which we proposed the initial learning-based framework that motivates the present paper.  However, the focus in \cite{baldassarre2016learning} is on a simple linear decoder and the noiseless setting, and the crucial aspects of non-linear decoding and noise were left as open problems.  

An alternative approach to optimizing subsampling based on prior information is given in \cite{weizman2015compressed}.  The idea therein is that if a subject requires multiple similar scans, then the previous scans can be used to adjust both the sampling and the decoding of future scans.  This is done using the randomized variable-density approach, with the probabilities adjusted to favor locations where the previous scans had more energy. \rev{ In \cite{mardani2016tracking}, a proposed informative random sampling approach optimizes the sampling of subsequent frames of dynamic MRI data based on previous frames in real time. In \cite{choi2016implementation}, a highly undersampled pre-scan is used to learn the energy distribution of the image and design the sampling prior to the main scan.}
 
A recent comparative study \cite{zijlstra2016evaluation} showed that the approaches that directly use training data perform better than the purely parametric (e.g., randomized variable-density) methods. 

\rev{Other subsampling design works include the following: In \cite{li2015analysis}, a generalized Rosetta shaped sampling pattern is used for compressive MRI, and in \cite{wang2012smoothed} a random-like trajectory based on higher order chirp sequences is proposed. Radial acquisition designs have been proposed  to improve the performance of compressive MRI in the settings of dynamic MRI \cite{feng2014golden} and phase contrast MRI \cite{hilbert2014accelerated}. In addition, a recent work \cite{lazarus2017sparkling} considered non-Cartesian trajectory design for high resolution MRI imaging at 7T (Tesla). }

\subsection{Theory of compressive sensing}

The theory of CS has generally moved in very different directions to the above practical approaches.  In particular, when it comes to subsampled Fourier measurements, the vast majority of the literature has focused on guarantees for recovering sparse signals with {\em uniform random sampling} \cite{Candes2006a}, which performs very poorly in practical imaging applications.

A recent work \cite{adcock2013breaking} proposed an alternative theory of CS based on {\em sparsity in levels}, along with variable-density random sampling and BP decoding.  \rev{As we outline below, we adopt an entirely different approach that avoids making {\em any} specific structural assumptions, yet can exploit even richer structures beyond sparsity and its variants.}

\subsection{Our contributions}

In this paper, we propose a novel {\em learning-based framework} for designing subsampling patterns, based on the idea of directly maximizing the empirical performance on training data.  We adopt an entirely different theoretical viewpoint to that of the existing CS literature; rather than placing structural assumptions (e.g., sparsity) on the underlying signal, we simply think of the training and test signals as coming from a common {\em unknown} distribution.  Using connections with statistical learning theory, we adopt a learning method that automatically extracts the structure inherent in the signal, and optimizes $\Omega$ specifically for the decoder at hand.

While our framework is suited to general CS scenarios, we focus on the application of MRI, in which we observe several advantages over the above existing approaches:
\begin{itemize}
    \item While our previous work \cite{baldassarre2016learning} exclusively considered a simple linear decoder, in this paper we consider targeted optimization for general non-linear decoders; 
    \item We present a non-trivial extension of our theory and methodology to the noisy setting, whereas \cite{baldassarre2016learning} only considered the noiseless case;
    \item We directly optimize for the performance measure at hand (e.g. PSNR), as opposed to less direct measures such as mutual information.  Similarly, our framework permits the direct optimization of bottom-line costs (e.g., acquisition time), rather than auxiliary cost measures (e.g., number of samples);
    \item We can directly incorporate practical sampling constraints, such as the requirement of sampling entire rows and/or columns rather than arbitrary patterns;
    \item Parameter tuning is not required;
    \item Our learning algorithm is highly parallelizable, rendering it feasible even when the dimension and/or the number of training images is large;
    \item We demonstrate the effectiveness of our approach on several real-world data sets, performing favorably against existing methods.
\end{itemize}

\section{Learning-based Framework} \label{sec:algo}

\subsection{Overview}

Our learning-based framework is outlined as follows:
\begin{itemize}
    \item We have access to a set of fully-sampled training signals $\x_1,\dotsc,\x_m$ that are assumed to be representative of the unknown signal of interest $\x$.  
    \item We assume that the decoder \eqref{eq:phi} is given.  This decoder is allowed to be {\em arbitrary}, meaning that our framework can be used alongside general existing reconstruction methods, and potentially also future methods.
    \item For any subsampling pattern $\Omega$, we can consider its empirical average performance on the training signals:
    \begin{equation}
        \frac{1}{m} \sum_{j=1}^m \eta_{\Omega}(\x_j), \label{eq:erm_overview}
    \end{equation}
    where $\eta_{\Omega}(\x)$ is a performance measure (e.g., PSNR) associated with the signal $\x$ and its reconstruction when the sampling pattern is $\Omega$.  If $\x_1,\dotsc,\x_m$ are similar to $\x$, we should expect that any $\Omega$ such that \eqref{eq:erm_overview} is high will also perform well on $\x$.
    \item While maximizing \eqref{eq:erm_overview} is computationally challenging in general, we can use any preferred method to seek an {\em approximate} maximizer.  We will pay particular attention to a {\em greedy algorithm}, which is parameter-free and satisfies a useful nestedness property.
\end{itemize}
We proceed by describing these points in more detail.  \rev{ For convenience, we initially consider the noiseless setting,
\begin{equation}
\b = \P_{\Omega} \bPsi \x,
\end{equation}
and then turn to the noisy setting in Section \ref{sec:noisy}.}

\subsection{Preliminaries} \label{sec:prelim}

Our broad goal is to determine a good subsampling pattern $\Omega \subseteq \{1,\dotsc,p\}$ for compressive MRI.  To perform this task, we assume that we have access to a set of {\em training signals} $\x_1, \dotsc, \x_m$, with $\x_j \in \C^p$.  The idea is that if an unseen signal has similar properties to the training signals, then we should expect the learned subsampling patterns to generalize well.  

In addition to the reconstruction rule $g$ of the form \eqref{eq:phi}, the learning procedure has knowledge of a {\em performance measure}, which we would like to make as high as possible on the unseen signal.  We focus primarily on PSNR in our experimental section, while also considering the  SSIM index.

For implementation reasons, one may wish to restrict the sampling patterns in some way, e.g., to contain only horizontal and/or vertical lines.  To account for such constraints, we assume that there exists a set $\mathcal{S}$ of subsets of $\{1,\dotsc,p\}$ such that the final sampling pattern must take the form 
\begin{equation}
    \Omega = \bigcup_{j=1}^\ell S_j, \quad S_j \in \mathcal{S}
\end{equation}
for some $\ell > 0$.  If $\mathcal{S} = \{ \{1\}, \dotsc, \{p\} \}$, then we recover the setting of \cite{baldassarre2016learning} where the subsampling pattern may be arbitrary.  However, arbitrary sampling patterns are not always feasible; for instance, masks consisting of only horizontal and/or vertical lines are often considered much more suited to practical implementation, and hence, it may be of interest to restrict $\mathcal{S}$ accordingly.

Finally, we assume there exists a {\em cost function} $c(\Omega) \ge 0$ associated with each subsampling pattern, and that the final cost must satisfy
\begin{equation}
    c(\Omega) \le \Gamma
\end{equation}
for some $\Gamma > 0$.  We will focus primarily on the case that the cost is the total number of indices in $\Omega$ (i.e., we are placing a constraint on the sampling rate), but in practical scenarios on\rev{e} may wish to consider the ultimate underlying cost, such as the scan time.   We assume that $c(\cdot)$ is monotone with respect to inclusion, i.e., if $\Omega_1 \subseteq \Omega_2$ then $c(\Omega_1) \le c(\Omega_2)$.

\subsection{Theoretical motivation via statistical learning theory}

Before describing our main algorithm, we present a theoretical motivation for our learning-based framework.  To do so, we think of the underlying signal of interest $\x$ as coming from a probability distribution $P$.  Under any such distribution, we can write down the indices with the best average performance:
\begin{equation}
    \Omega^* = \argmax_{\Omega \in \mathcal{A}} \mathbb{E}_P\big[ \eta_{\Omega}(\x) \big], \label{eq:Omega_opt}
\end{equation}
where $\mathcal{A}$ is the set of feasible $\Omega$ according to $c(\cdot)$, $\Gamma$, and $\mathcal{S}$, and we define
\begin{equation}
    \eta_{\Omega}(\x) = \eta(\x,\hat{\x})
\end{equation}
with $\hat{\x} = g(\Omega,\b)$ and $\b = \P_{\Omega}\bPsi\x$.  

Unfortunately, the rule in \eqref{eq:Omega_opt} is not feasible in practice, since one cannot expect to know $P$ (e.g., one cannot reasonably form an accurate probability distribution that describes a brain image).  However, if the training signals $\x_1,\dotsc,\x_m$ are also independently drawn from $P$, then there is hope that the {\em empirical average} is a good approximation of the true average.  This leads to the following selection rule:
\begin{equation}
    \hat{\Omega} = \argmax_{\Omega \in \mathcal{A}} \frac{1}{m} \sum_{j=1}^m \eta_{\Omega}(\x_j). \label{eq:erm}
\end{equation}
The rule \eqref{eq:erm} is an instance of {\em empirical risk minimization} in statistical learning theory.  While finding the exact maximum can still be computationally hard, this viewpoint will nevertheless dictate that we should seek indices $\Omega \in \mathcal{A}$ such that $\frac{1}{m} \sum_{j=1}^m \eta_{\Omega}(\x_j)$ is high.

To see this more formally, we consider the following question: If we find a set of indices $\Omega \in \mathcal{A}$ with a good empirical performance $\frac{1}{m} \sum_{j=1}^m \eta_{\Omega}(\x_j)$, does it also provide good performance $\mathbb{E}[\eta_{\Omega}(\x)]$ on an unseen signal $\x$?  The following proposition answers this question in the affirmative using statistical learning theory.

\begin{proposition} \label{prop:guarantee}
     Consider the above setup with a performance measure normalized so that $\eta(\x,\hat{\x}) \in [0,1]$.\footnote{As a concrete example, suppose we are interested in the squared error $\|\x - \hat{\x}\|_2^2$.  If the input is normalized to $\|\x\|_2^2 = 1$, then it can be shown that any estimate $\hat{\x}$ only improves if it is scaled down such that $\|\hat{\x}\|_2^2 \le 1$.  It then follows easily that $\eta(\x,\hat{\x}) = 1 - \frac{1}{4}\|\x - \hat{\x}\|_2^2$ always lies in $[0,1]$.}  For any $\delta \in ( 0, 1 )$, with probability at least $1 - \delta$ (with respect to the randomness of $\x_1, \x_2, \ldots, \x_m$), it holds that
    \begin{equation}
        \left\vert  \frac{1}{m} \sum_{j = 1}^m \eta_{\Omega}(\x_j) - \mathbb{E}_P \left[ \eta_{\Omega}(\x) \right] \right\vert \leq \sqrt{ \frac{1}{2m} \log \left( \frac{2 \abs{ \mathcal{A} }}{\delta} \right) } , \notag
    \end{equation}
    simultaneously for all $\Omega \in \mathcal{A}$.
\end{proposition}

The proof is given in the appendix.  We see that as long as $m$ is sufficiently large compared to $|\mathcal{A}|$, the average performance attained by any given $\Omega \in \mathcal{A}$ on the training data is an accurate estimate of the true performance.   This guarantee is with respect to the \emph{worst case}, regarding all possible probability distributions $P$; the actual performance could exceed this guarantee in practice.

\subsection{Greedy algorithm}

While finding the exact maximizer in \eqref{eq:erm} is challenging in general, we can seek to efficiently find an {\em approximate} solution.  There are several possible ways to do this, and Proposition \ref{prop:guarantee} reveals that regardless of how we come across a mask with better empirical performance, we should favor it.  In this subsection, we present a simple {\em greedy} approach, which is parameter-free \rev{in the sense that no parameter tuning is needed for the mask selection process once the decoder is given (though the decoder itself may still have tunable parameters).} The greedy approach also exhibits a useful nestedness property (described below).

At each iteration, the greedy procedure runs the decoder $g$ with each element of $\mathcal{S}$ that is not yet included in the mask, and adds the subset $S \in \mathcal{S}$ that increases the performance function most on average over the training images, normalized by the cost.  The algorithm stops when it is no longer possible to add new subsets from $\mathcal{S}$ without violating the cost constraint. The details are given in Algorithm \ref{alg:1}.

\begin{algorithm}
\caption{Greedy mask optimization}
\label{alg:1}
\textbf{Input}: Training data $\x_1, \dotsc, \x_m$, reconstruction rule $g$, sampling subsets $\mathcal{S}$, cost function $c$, maximum cost $\Gamma$ \\
\textbf{Output}: Sampling pattern $\Omega$
\begin{algorithmic}[1]
\State $\Omega \leftarrow \emptyset$;
\While{$ c(\Omega) \leq  \Gamma$}

    \For{$S \in \mathcal{S}$ such that  $c(\Omega \cup S) \le \Gamma$}
        \State $\Omega' = \Omega \cup S$
        \State For each $j$, set  $\b_{j} \leftarrow \P_{\Omega'}\bPsi\x_j$, $\hat{\x}_j \leftarrow g(\Omega',\b_j)$ 
        \State $\eta(\Omega') \leftarrow \frac{1}{m}\sum_{j=1}^m \eta(\x_j,\hat{\x}_j)$
     \EndFor
    \State $\Omega \leftarrow \Omega \cup S^*$, where
    \begin{equation*}
        S^* = \argmax_{S\,:\,c(\Omega \cup S) \le \Gamma} \frac{ \eta(\Omega \cup S) - \eta(\Omega) }{ c(\Omega \cup S) - c(\Omega) }
    \end{equation*}
 \EndWhile
 \State {\bf return} $\Omega$
\end{algorithmic}
\end{algorithm}

An important feature of this method is the {\em nestedness} property that allows one to immediately adapt for different costs $\Gamma$ (e.g., different sampling rates).  Specifically, one can record the order in which the elements are included in $\Omega$ during the mask optimization for a high cost, and use this to infer the mask corresponding to lower costs, or to use as a starting point for higher costs. Note that this is not possible for most parametric methods, where  changing the sampling rate requires one to redo the parameter tuning.

We briefly note that alternative greedy methods could easily be used.  For instance:
\begin{itemize}
    \item One could start with $\Omega = \{1,\dotsc,p\}$ (i.e., sampling the entire Fourier space) and then {\em remove} samples until a feasible pattern is attained;
    \item One could adopt a hybrid approach in which samples are both added and removed iteratively until some convergence condition is met.
\end{itemize}
In our experiments, however, we focus in the procedure in Algorithm \ref{alg:1}, which we found to work well.

 \rev{In another related work \cite{ravishankar2011adaptive}, an iterative approach is taken in which only a single nonlinear reconstruction is implemented in each iteration of mask selection, starting with an initial mask whereas we run separate reconstructions for each candidate to be added to the sampling pattern, starting with the empty set.  Moreover, \cite{ravishankar2011adaptive} makes use of several parameters, such as the number of the regions with higher errors to which the samples are moved iteratively, the size of these regions, the power of the polynomial used for a weighting function, etc., which need to be tuned for each experiment. Our greedy algorithm has the advantage of avoiding such heuristics and additional parameters.  While the proposed algorithm requires a larger number of computations for mask selection, these computations can be easily parallelized and performed efficiently. }

\subsection{Parametric approach with learning}

An alternative approach is to generate a number of candidate masks $\Omega_1,\dotsc,\Omega_{L}$ using one or more parametric variable-density methods (possibly with a variety of different choices of parameters), and then to apply the learning-based idea to these candidate masks:  Choose the one with the best empirical performance on the training set.  While similar ideas have already been used when performing parameter sweeps in existing works (e.g., see \cite{knoll2011adapted}), our framework provides a more formal justification to why the empirical performance is the correct quantity to optimize.  The details are given in Algorithm \ref{alg:2}, where we assume that all candidate masks are feasible according to the sampling subsets $\mathcal{S}$ and cost function $c$.

\begin{algorithm}
\caption{Choosing from a set of candidate masks}
\label{alg:2}
\textbf{Input}: Training data $\x_1, \dotsc, \x_m$, reconstruction rule $g$, candidate masks $\Omega_1,\dotsc,\Omega_L$ \\
\textbf{Output}: Sampling pattern $\Omega$
\begin{algorithmic}[1]
\For{$\ell = 1,\dotsc,L$}
        \State For each $j$, set  $\b_{j} \leftarrow \P_{\Omega_{\ell}}\bPsi\x_j$, $\hat{\x}_j \leftarrow g(\Omega_{\ell},\b_j)$ 
        \State $\eta_{\ell} \leftarrow \frac{1}{m}\sum_{j=1}^m \eta(\x_j,\hat{\x}_j)$
 \EndFor
 \State $\Omega \leftarrow \Omega_{\ell^*}$, where $\ell^* = \argmax_{\ell=1,\dotsc,L} \eta_{\ell}$
 \State {\bf return} $\Omega$
\end{algorithmic}
\end{algorithm}

\subsection{Noisy setting} \label{sec:noisy}

So far, we have considered the case that both the acquired signal $\b$ and the training signals $\x_1,\dotsc,\x_m$ are noiseless.  In this subsection, we consider a noisy variant of our setting: The acquired signal is given by
\begin{equation}
    \b = \P_{\Omega}\bPsi\x + \w \label{eq:noisy_model}
\end{equation}
for some noise term $\w \in \mathbb{\rev{C}}^p$, and the learning algorithm does not have access to the exact training signals $\x_1,\dotsc,\x_m$, but instead to noisy versions $\z_1,\dotsc,\z_m$, where
\begin{equation}
    \z_j = \x_j + \v_j, \quad j=1,\dotsc,m
\end{equation}
with $\v_j$ representing the noise.

We observe that the selection rule in \eqref{eq:erm} can no longer be used, since the learning algorithm does not have direct access to $\x_1,\dotsc,\x_m$.  The simplest alternative is to \rev{substitute} the noisy versions of the signals and use \eqref{eq:erm} with $\z_j$ in place of $\x_j$.  It turns out, however, that we can do better if we have access to a {\em denoiser} $\xi(\z)$ that reduces the noise level.  Specifically, suppose that
\begin{equation}
    \xi(\z) = \x_j + \vtil_j
\end{equation}
for some reduced noise $\vtil_j$ such that $\mathbb{E}[ \| \vtil_j \| ] \le \mathbb{E}[ \| \v_j \| ]$.  We then propose the selection rule
\begin{equation}
    \hat{\Omega} = \argmax_{\Omega \in \mathcal{A}} \frac{1}{m} \sum_{j=1}^m \eta(\x_j+\vtil_j, \hat{\x}(\P_{\Omega}\bPsi(\x_j + \v_j)), \label{eq:erm_noisy}
\end{equation}
where $\hat{\x}(\b)$ denotes the decoder applied to $\b$.  Note that we still use the {\em noisy} training signal in the choice of $\b$; by doing so, we are {\em learning how to denoise}, which is necessary because the unseen test signal is itself noisy as per \eqref{eq:noisy_model}.

To understand how well the above rule generalized to unseen signals, we would like to compare the empirical performance on the right-hand side of \eqref{eq:erm_noisy} to the true average performance on an unseen signal, defined as
\begin{equation}
    \etanoisy(\Omega) = \mathbb{E}\big[ \eta(\x, \hat{\x}) \big]
\end{equation}
with $\hat{\x} = g(\Omega,\b)$ and $\b = \P_{\Omega}\bPsi\x + \w$.  The following proposition quantifies this comparison.

\begin{proposition} \label{prop:guarantee_noisy}
     Consider the above noisy setup with $\w$ and $\{\v_j\}_{j=1}^m$ having independent Gaussian entries of the same variance, and a performance measure $\eta(\x,\hat{\x}) \in [0,1]$ that satisfies the continuity assumption $|\eta(\x,\hat{\x}) - \eta(\x',\hat{\x})| \le L \|\x - \x'\|_2$ for all $\x,\x'$ and some $L > 0$.  For any $\delta \in ( 0, 1 )$, with probability at least $1 - \delta$ (with respect to the randomness of the noisy training signals), it holds that
    \begin{multline}
        \left\vert \frac{1}{m} \sum_{j=1}^m \eta(\x_j+\vtil_j, \hat{\x}(\P_{\Omega}\bPsi(\x_j + \v_j)) - \etanoisy(\Omega) \right\vert \\
        \leq L \mathbb{E}[\| \vtil \|_2] + \sqrt{ \frac{1}{2m} \log \left( \frac{2 \abs{ \mathcal{A} }}{\delta} \right) },
    \end{multline}
    simultaneously for all $\Omega \in \mathcal{A}$, where $\vtil_j = \xi(\v_j)$ is the \secondrev{  effective noise remaining in the} $j$-th denoised training signal, and $\vtil$ has the same distribution as any given $\vtil_j$.
\end{proposition}

The proof is given in the appendix.  We observe that the second term coincides with that of the noiseless case in Proposition \ref{prop:guarantee}, whereas the first term represents the additional error due to the residual noise after denoising.  It is straightforward to show that such a term is unavoidable in general.\footnote{For instance, to give an example where the generalization error must contain the $\EE[\|\v\|_2]$ term, it suffices to consider the $\ell_2$-error in the trivial case that $\x = \x_1 = \dotsc = \x_m$ with probability one, and $\hat{\x}$ also outputs the same deterministic signal.}

Hence, along with the fact that more training signals leads to better generalization, Proposition \ref{prop:guarantee_noisy} reveals the intuitive fact that {\em the ability to better denoise the training signals leads to better generalization}.  In particular, if we can do {\em perfect denoising} (i.e., $\|\vtil_j\| = 0$) then we get the same generalization error as the noiseless case.

In Algorithm \ref{alg:3}, we provide the learning-based procedure with an arbitrary denoising function $\xi$.  Note that if we choose the identity function $\xi(\z) = \z$, then we reduce to the case where no denoising is done.

\begin{algorithm}
\caption{Learning-based mask selection with denoising}
\label{alg:3}
\textbf{Input}: Noisy training data $\z_1, \dotsc, \z_m$, reconstruction rule $g$, denoising algorithm $\xi(\z)$, and either the triplet $(\mathcal{S},c,\Gamma)$ or candidate masks $\Omega_1,\dotsc,\Omega_L$ \\
\textbf{Output}: Sampling pattern $\Omega$
\begin{algorithmic}[1]
\State $\x'_j \leftarrow \xi(\z_j)$ for $j=1,\dotsc,m$
 \State Select $\Omega$ using Algorithm \ref{alg:1} or \ref{alg:2} with $\eta(\x'_j, \hat{\x}(\P_{\Omega}\bPsi\z_j))$ replacing $\eta(\x_j, \hat{\x}(\P_{\Omega}\bPsi\x_j)$ throughout.
 \State {\bf return} $\Omega$
\end{algorithmic}
\end{algorithm}


\section{Numerical Experiments} \label{sec:NUMERICAL}

In this section, we provide numerical experiments demonstrating that our learning-based framework provides high-performing sampling patterns for a diverse range of reconstruction algorithms. Our simulation code and data are publicly available online. \footnote{https://lions.epfl.ch/lb-csmri}

\subsection{Implementation details} \label{sec:exp_setup}

{\bf Reconstruction rules.} We consider the decoders described in Section \ref{sec:signal}, which we refer to as BP, TV, BM3D, and NN (i.e., neural network). For BP in \eqref{eq: BP}, we let the sparsifying operator $\bPhi$ be the shearlet transform \cite{kutyniok2016shearlab}, and for both BP and TV, we implement the minimization using NESTA \cite{becker2011nesta} \secondrev{, for which we set the maximum number iterations to 20000, the denoising parameter to $\epsilon=0$, the tolerance value and the smoothing parameter to $\mu=10^{-5}$, and the number of continuation steps to $T=1$.} 

For BM3D, we use the code available in \cite{eksloglu_website}.  We take the observation fidelity parameter $\alpha = 0$, the number of outer iterations $\mathcal{J} = 20$ and the regularization parameters as $\lambda_{\mathrm{max}} = 200 $ and $\lambda_{\mathrm{min}} =0.01$. We also use a varying number of inner iterations between 1 and 10 as described in \cite{eksioglu2016decoupled}.


For the NN decoder, we use the network structure from \cite{schlemper2017deep}, only slightly modifying certain parameters.  We choose depth of the architecture as \rev{$n_d=3$} and depth of the cascade as \rev{$n_c=5$}. We set the mini-batch size for training to $20$.  We use the same training signals for learning indices and tuning the network weights.  Since it is difficult to optimize these jointly, we perform alternating optimization: Initialize the weights, and then alternate between learning indices with fixed weights, and learning weights with fixed indices.  We perform up to three iterations of this procedure, which we found to be sufficient for convergence.

As was done in \cite{schlemper2017deep}, we initialize the network weights using the initialization of He {\em et al.} \cite{he2015delving}, and perform network weight optimization using the Adam algorithm \cite{kingma2014adam} with step size $\alpha=10^{-2}$ and decay rates $\beta_1=0.9$ and $\beta_2=0.999$.  Moreover, we apply an additional $\ell_2$ weight regularization penalty of $10^{-6}$.  Each time we train the network, we run the training for $7000$ epochs (i.e., passes over the training data). We use the Python implementation available in \cite{schlemper2017deep}.  

\secondrev{In principle, it may sometimes be preferable to change the reconstruction parameters as the greedy algorithm adds indices and increases the current sampling rate.  However, we did not find such an approach to provide further benefit in the present setting, so here we stick to the above approach where the reconstruction parameters remain fixed.  }





{\bf Mask selection methods.} In addition to the greedy method in Algorithm \ref{alg:1}, we consider parametric randomized variable-density methods with learning-based optimization according to Algorithm \ref{alg:2}; the details are provided in the relevant subsections below.  Moreover, we consider the following two baselines from the existing literature:
\begin{itemize}
    \item {\em (Coherence-based)} We consider the parametric approach of \cite{lustig2007sparse} with parameters specifying (i) the size of a fully-sampled region at low frequencies; and (ii) the polynomial rate of decay of sampling at higher frequencies.  As suggested in \cite{lustig2007sparse}, we choose the parameters to optimize an {\em incoherence function}, meaning that no training data is used.  The minimization is done using Monte Carlo methods, and we do this using the code used in \cite{lustig2007sparse} available online.
    \item {\em (Single-image)} We consider the approach of \cite{vellagoundar15robust} in which only a single training image is used.  Specifically, this image determines a probability density function where the probability is proportional to energy, and then the samples are randomly selected by drawing from this distribution.
\end{itemize}

{\bf Data sets.} The MRI data used in the following subsections was acquired on a 3T MRI system (Magnetom Trio Scanner, Erlangen, Germany). The protocols were approved by the local ethics committee, and all subjects gave written informed consent. 

The data set used in the first three experiments (subsections) below consists of 2D T1-weighted brain scans of seven healthy subjects, which were scanned with a FLASH pulse sequence and a 12-channel receive-only head coil.  In our experiments, we use 20 slices \secondrev{of sizes 256$\times$256} from five such subjects (two for training, three for testing).  Data from individual coils was processed via a complex linear combination, where coil sensitivities were estimated from an 8$\times$8 central calibration region of $k$-space \cite{bydder2002combination}.  The acquisition used a field of view (FOV) of $220 \times 220 $ mm\textsuperscript{2} and a resolution of 0.9 $\times $ 0.7 mm$^2$. The slice thickness was 4.0 mm. The imaging protocol comprised a flip angle of 70$^{\circ}$, a TR/TE of 250.0/2.46 ms, with a scan time of 2 minutes and 10 seconds.

The data set used in subsection E below consists of angiographic brain scans of five healthy subjects acquired with 12-channel receive-only head coil and 20 slices from each are used in our experiments (two subjects for training, three subjects for testing). \secondrev{The size of the slices is 256$\times$256}.  A 3D TOF sequence was used with FOV of 204$\times$204$\times$51 mm$^3$, 0.8$\times$0.8$\times$0.8 mm$^3$ resolution, flip angle of 18$^{\circ}$, magnetization-transfer contrast, a TR/TE of 47/4.6 ms, and a scan time of 16 min 25 sec.


\subsection{Comparison to baselines}

We first compare to the above-mentioned baselines for a single specific decoder, namely, BP.  We use a conventional method of sampling in which readouts are performed as lines at different {\em phase encodes}, corresponding to a horizontal line in Fourier space.  Hence, our subsampling masks consist of only full horizontal lines, and we let $\mathcal{S}$ in Section \ref{sec:prelim} be the set of all horizontal lines accordingly.

\begin{figure}[ht]
\centering
\includegraphics[width=0.45\textwidth]{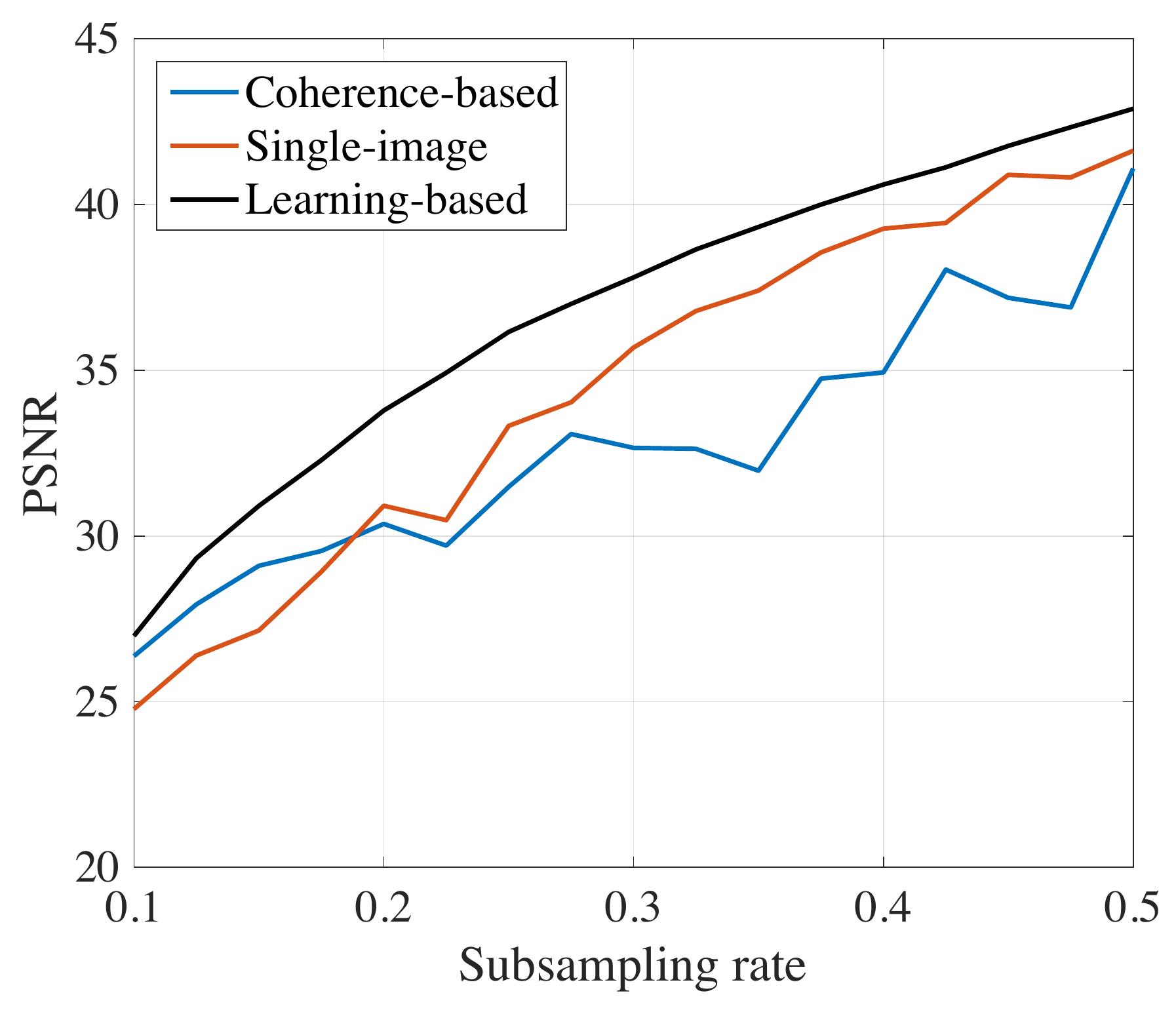} 
\caption{PSNR as a function of subsampling rates with BP reconstruction. }
\label{fig:2D_rates}
\end{figure}

We use our greedy algorithm to find a subset of such lines at a given budget on the total number of samples (or equivalently, the sampling rate).  From the data of the five subjects with 20 slices each, we take the first 2 subjects (40 slices total) as training data.  Once the masks are obtained, we implement the reconstructions on the remaining 3 subjects (60 slices total).   As seen in Figure \ref{fig:2D_rates}, the learning-based approach outperforms the baselines across all sampling rates shown. 

\begin{table}
\centering
\caption{\label{tab:table_2D} PSNR and SSIM performances averaged on 60 test slices at 25\% subsampling rate. The entries where the learning is matched to the decoder and performance measure are shown in bold.   }
\begin{tabular}{|l|||*{4}{c|}}\hline
\backslashbox{Mask}{Decoder}
&\makebox[3em]{TV}&\makebox[3em]{BP}&\makebox[3em]{BM3D} &\makebox[3em]{\rev{NN}} \\\hline \hline 
Coherence-based &  30.76  &   31.48 & 30.04 & 32.02  \\\hline 
Single-image &  32.79   &   33.32 & 32.42 & 33.67 \\\hline \hline 
TV-greedy & {\bf 34.84} & 36.08 & 35.95 & 36.04  \\\hline 
BP-greedy & 34.76 &  {\bf 36.16} & 36.11 & 36.17 \\\hline  
BM3D-greedy &  34.77 &  36.04 & {\bf 36.19}& 35.92 \\\hline 
\rev{NN-greedy} &  34.81 &  36.05 &  36.16 & {\bf 36.36} \\\hline 
\rev{Low Pass} &  31.96  &  32.41 & 32.59 & 32.59 \\\hline
\end{tabular}

\vspace{2mm}

\begin{tabular}{|l|||*{4}{c|}}\hline
\backslashbox{Mask}{Decoder}
&\makebox[3em]{\rev{TV}}&\makebox[3em]{\rev{BP}}&\makebox[3em]{\rev{BM3D}} &\makebox[3em]{\rev{NN}} \\\hline \hline

\rev{Coherence-based}&0.832&0.85&0.822  & 0.798  \\\hline
\rev{Single-image}&0.876&0.889&0.879   & 0.854  \\\hline
\rev{TV-greedy} &{ 0.907}&0.922&0.921  & 0.869  \\\hline
\rev{BP-greedy}&0.906& { 0.923}&0.921  & 0.859  \\\hline
\rev{BM3D-greedy}&0.906&0.922&{0.922}  & 0.909  \\\hline
\rev{NN-greedy}&0.907&0.923&0.923   & {0.925}  \\\hline
\rev{Low Pass}&0.876&0.888&0.893  & 0.893  \\\hline


\end{tabular}

\end{table}

\begin{figure*}[!t]
\centering
\begin{tabular}{ccccccc}
& \hspace{-4mm} \textbf{Mask} & \hspace{-4mm} \textbf{TV decoder} & \hspace{-4mm} \textbf{BP decoder} & \hspace{-4mm} \textbf{BM3D decoder} & \hspace{-5mm} \textbf{NN decoder} &\hspace{-6mm} \textbf{\rev{Ground truth}} \\

\rotatebox{90}{\hspace{3mm} \textbf{Coher. based}} &                         
\hspace{-4mm}\includegraphics[width=.16\textwidth]{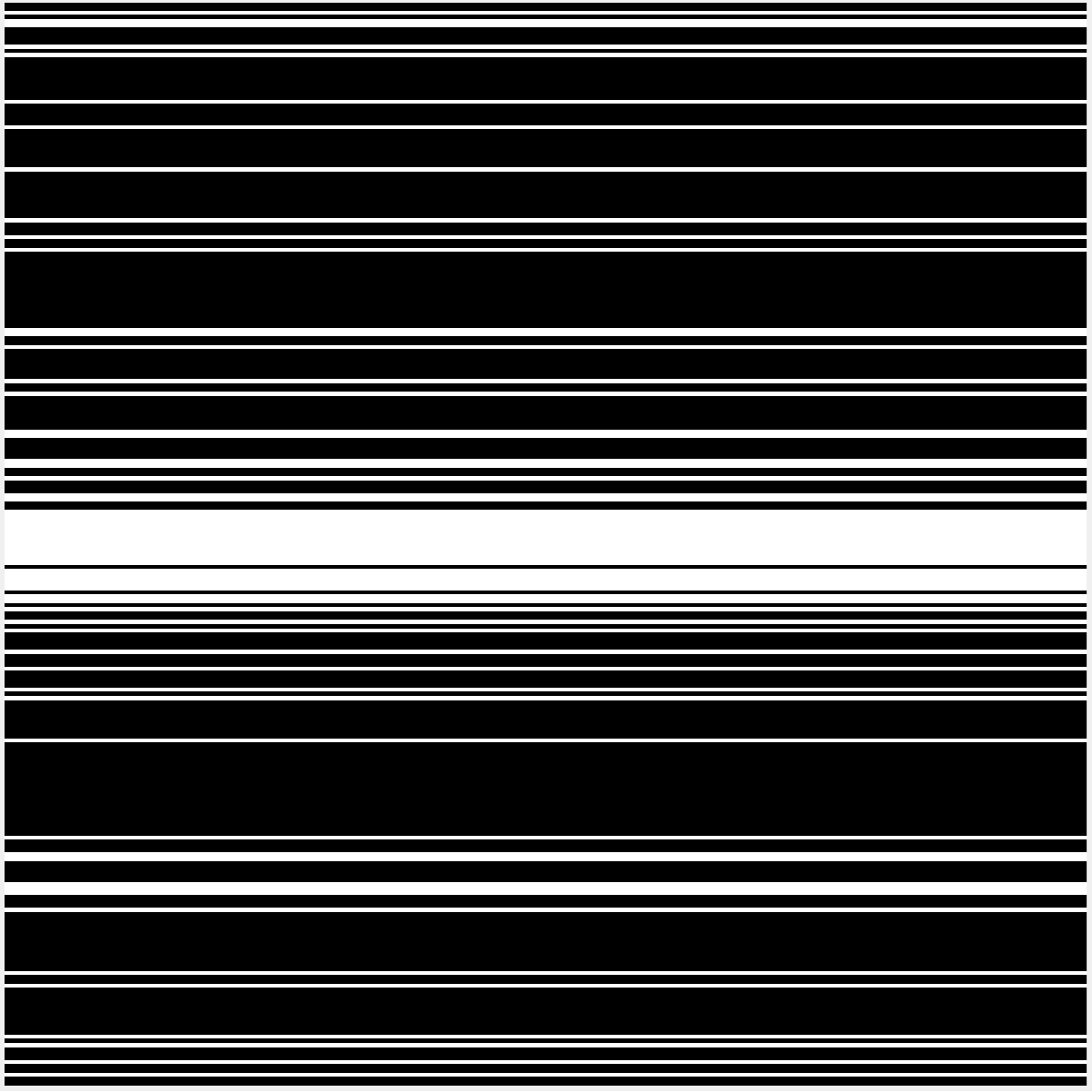} &
\hspace{-4mm}\includegraphics[width=.16\textwidth]{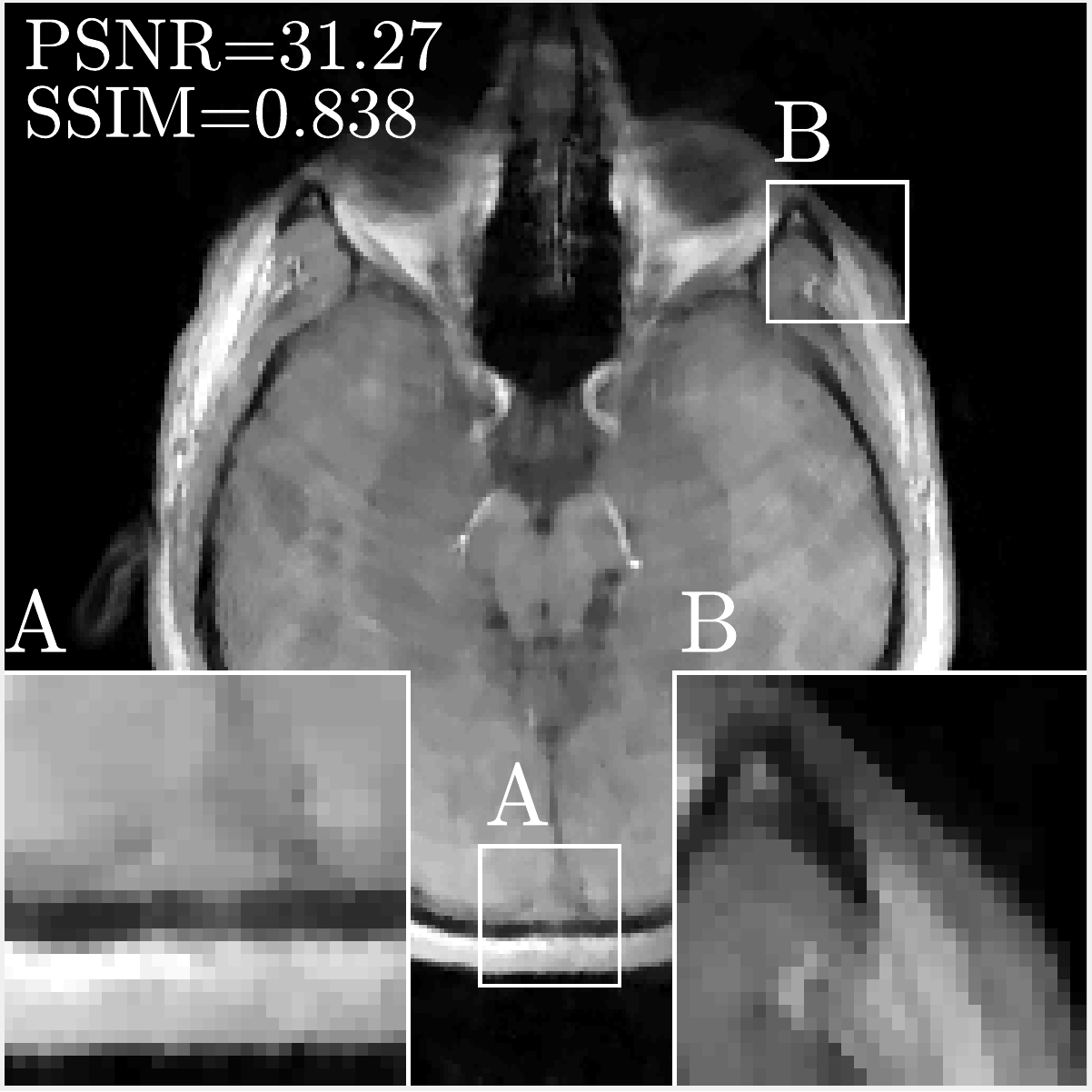} &
\hspace{-4mm}\includegraphics[width=.16\textwidth]{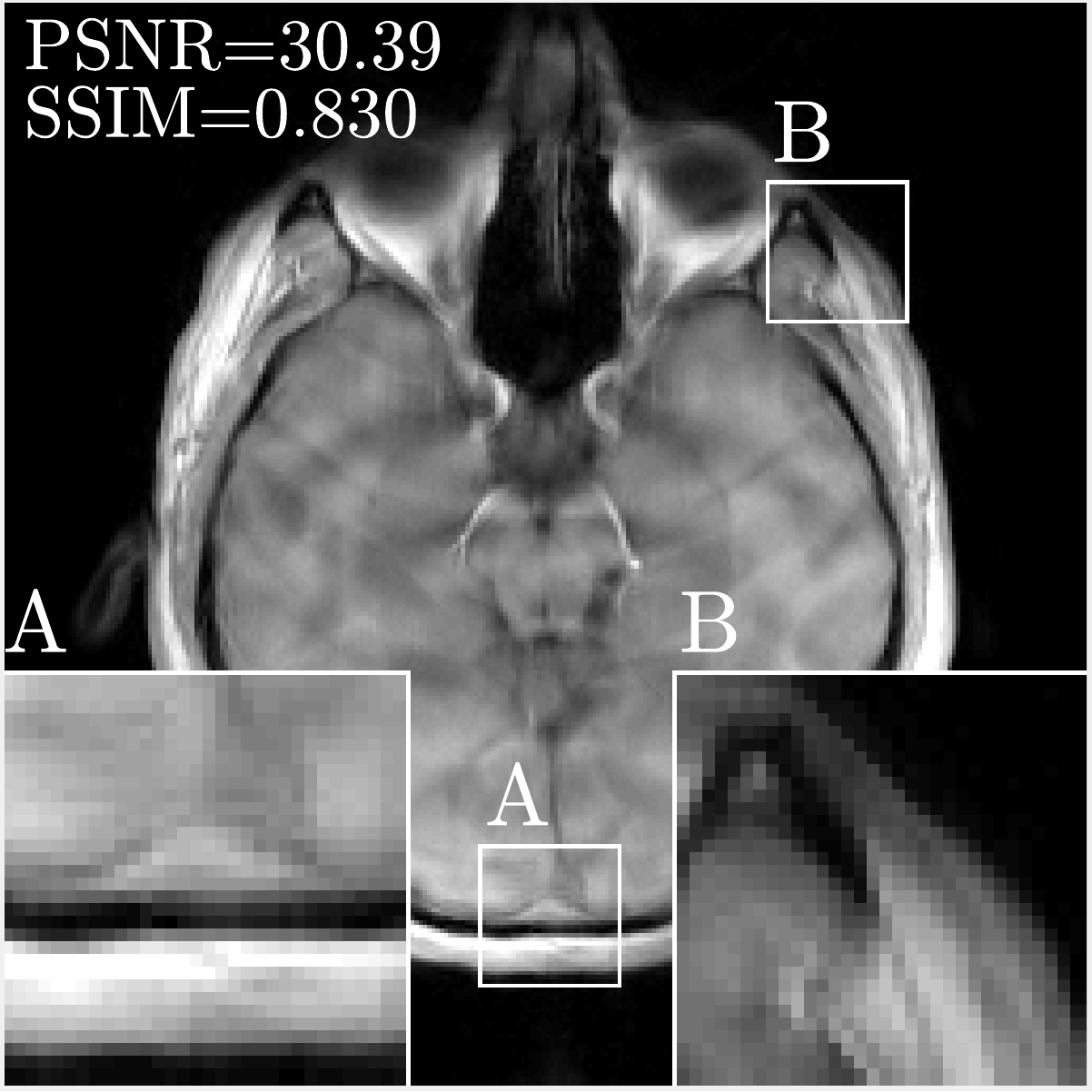} &
\hspace{-4mm}\includegraphics[width=.16\textwidth]{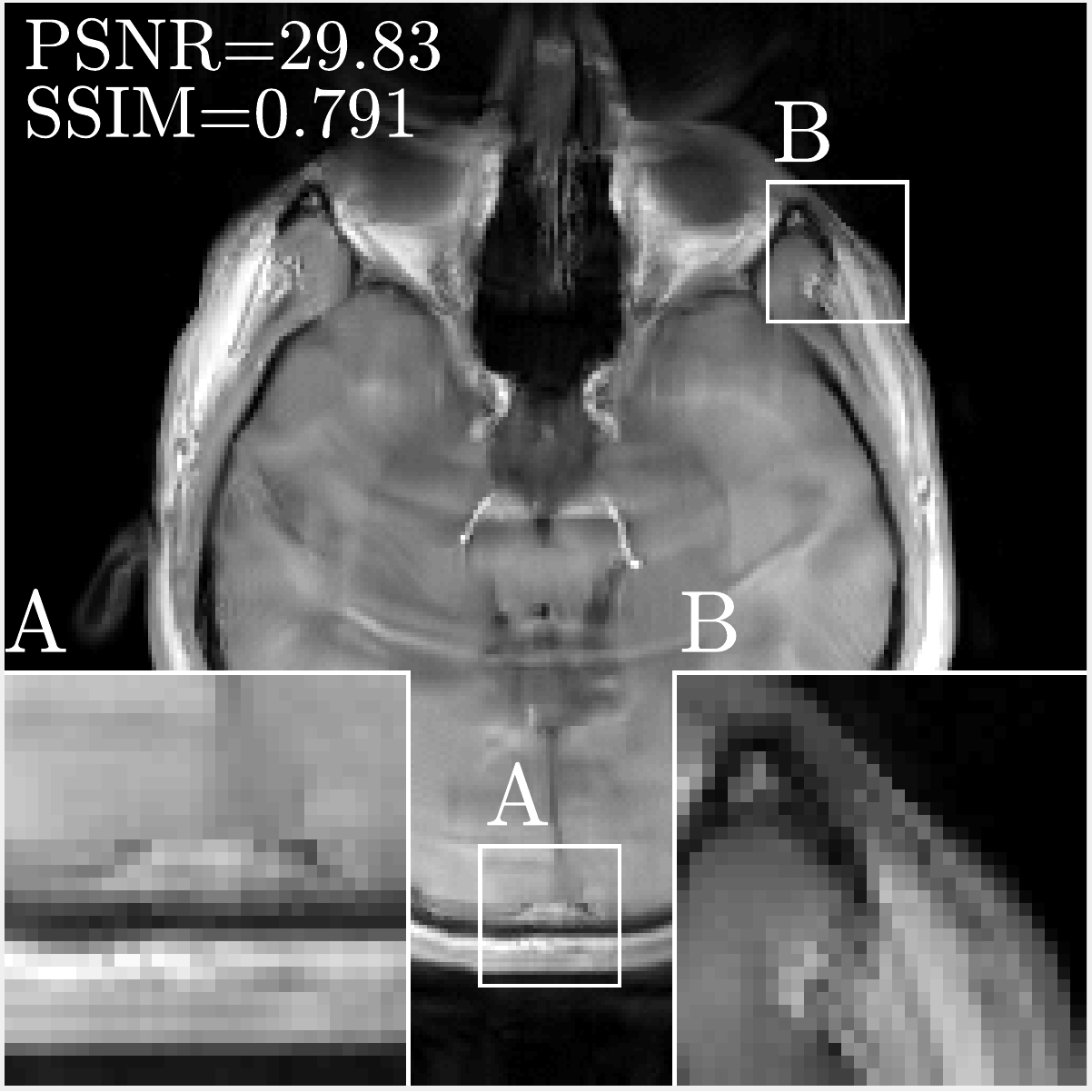} &
\hspace{-4mm}\includegraphics[width=.16\textwidth]{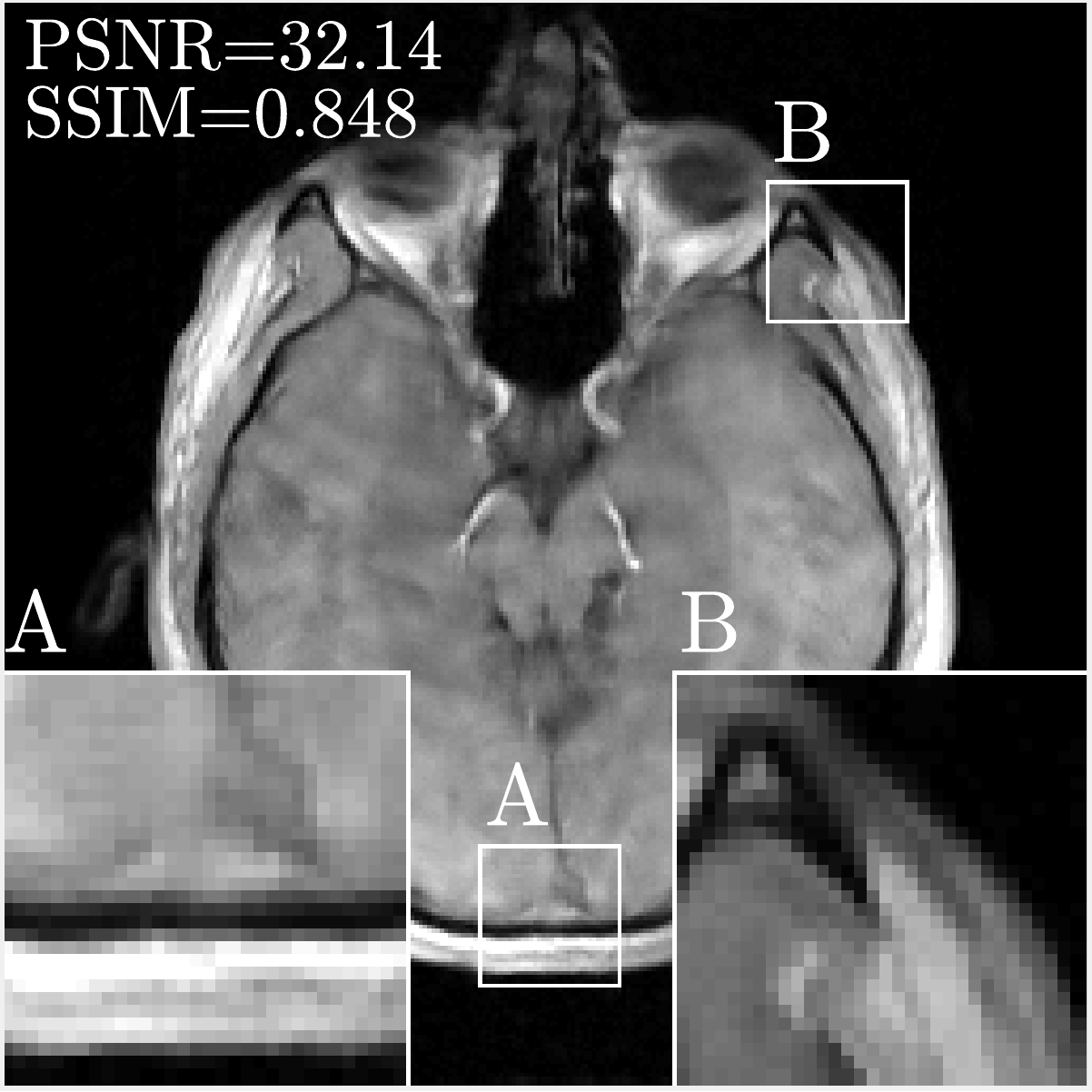}   &
\hspace{-4mm}\includegraphics[width=.16\textwidth]{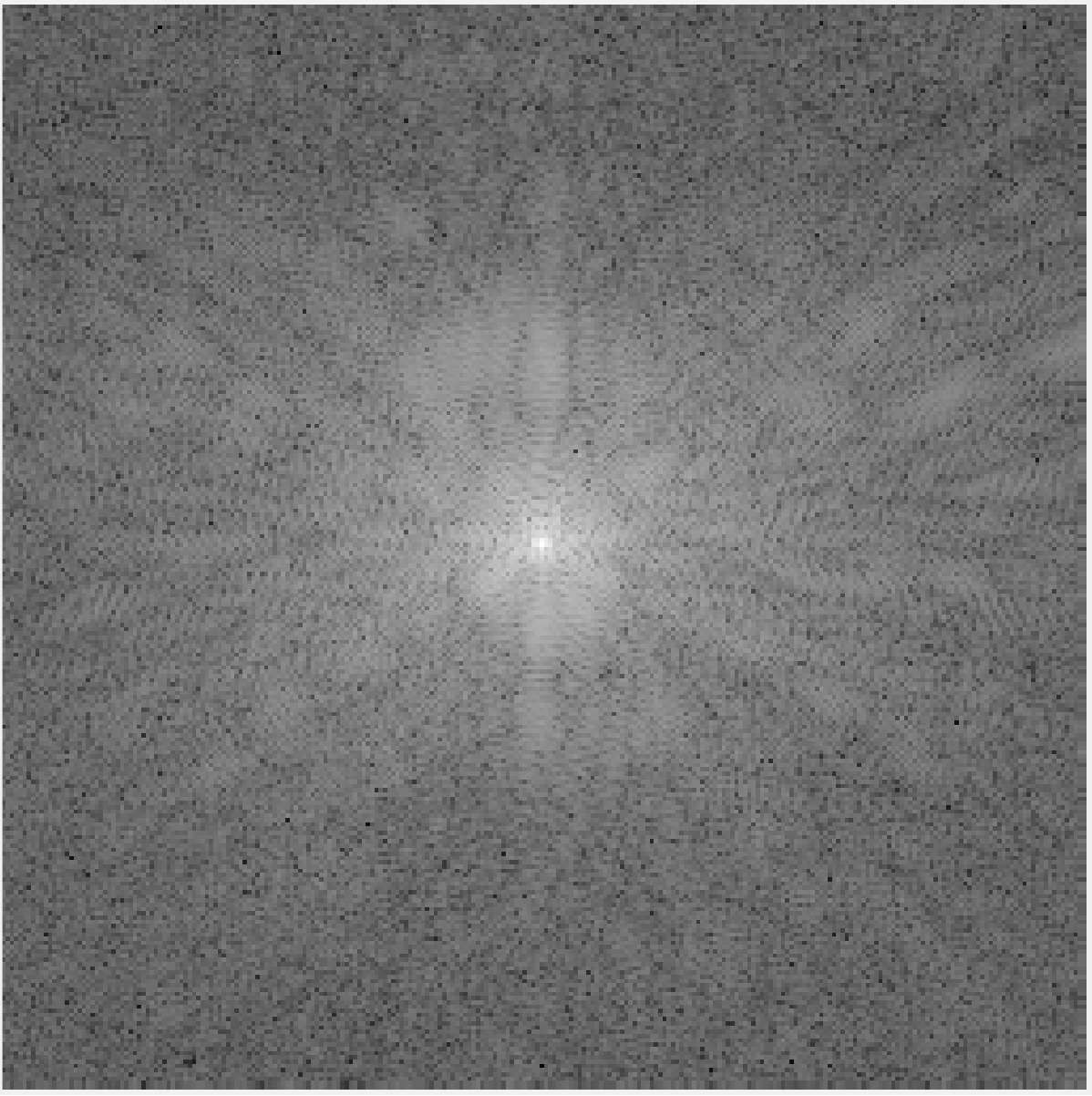}\\ [-1mm]
\rotatebox{90}{\hspace{4mm} \textbf{Single image}} &
\hspace{-4mm}\includegraphics[width=.16\textwidth]{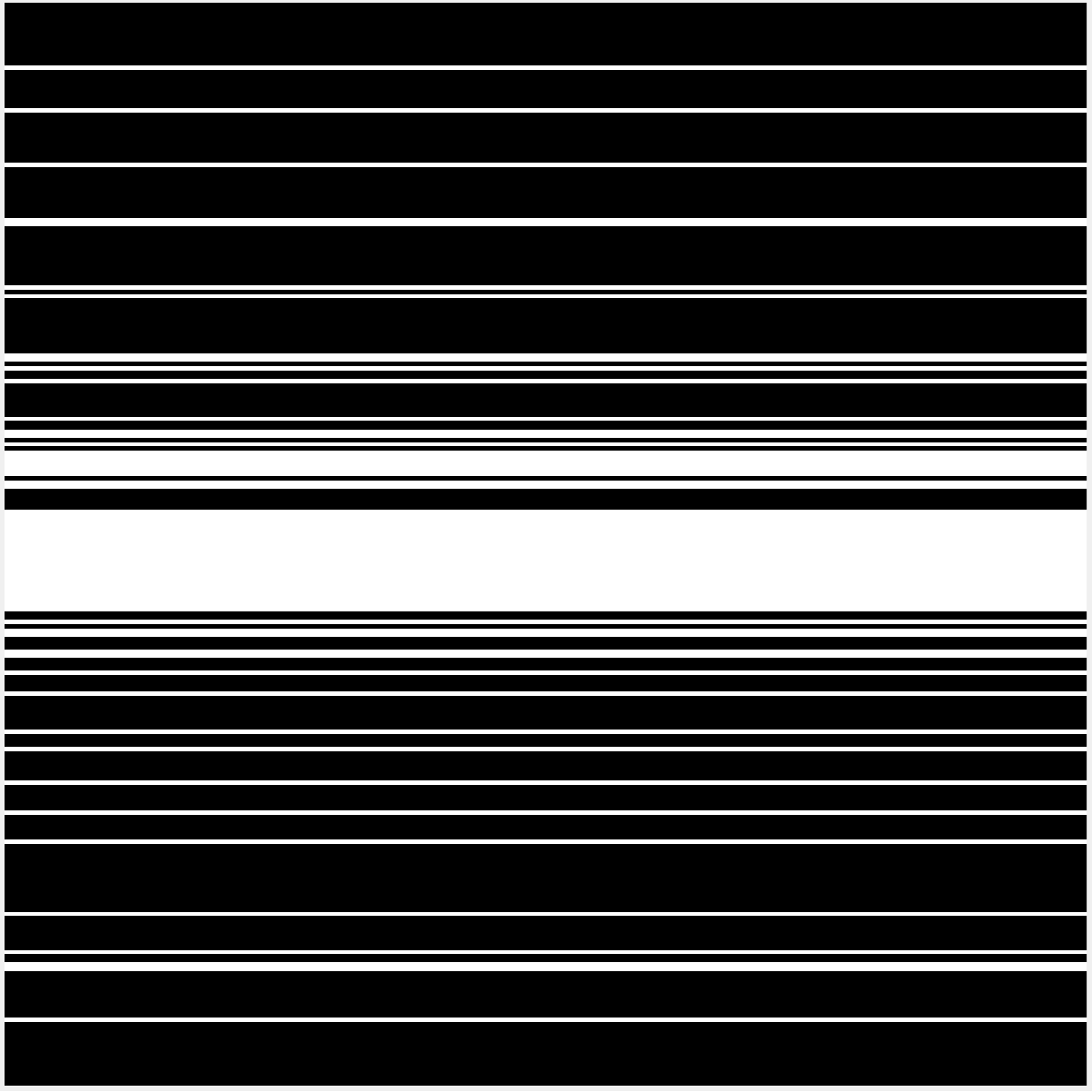} &
\hspace{-4mm}\includegraphics[width=.16\textwidth]{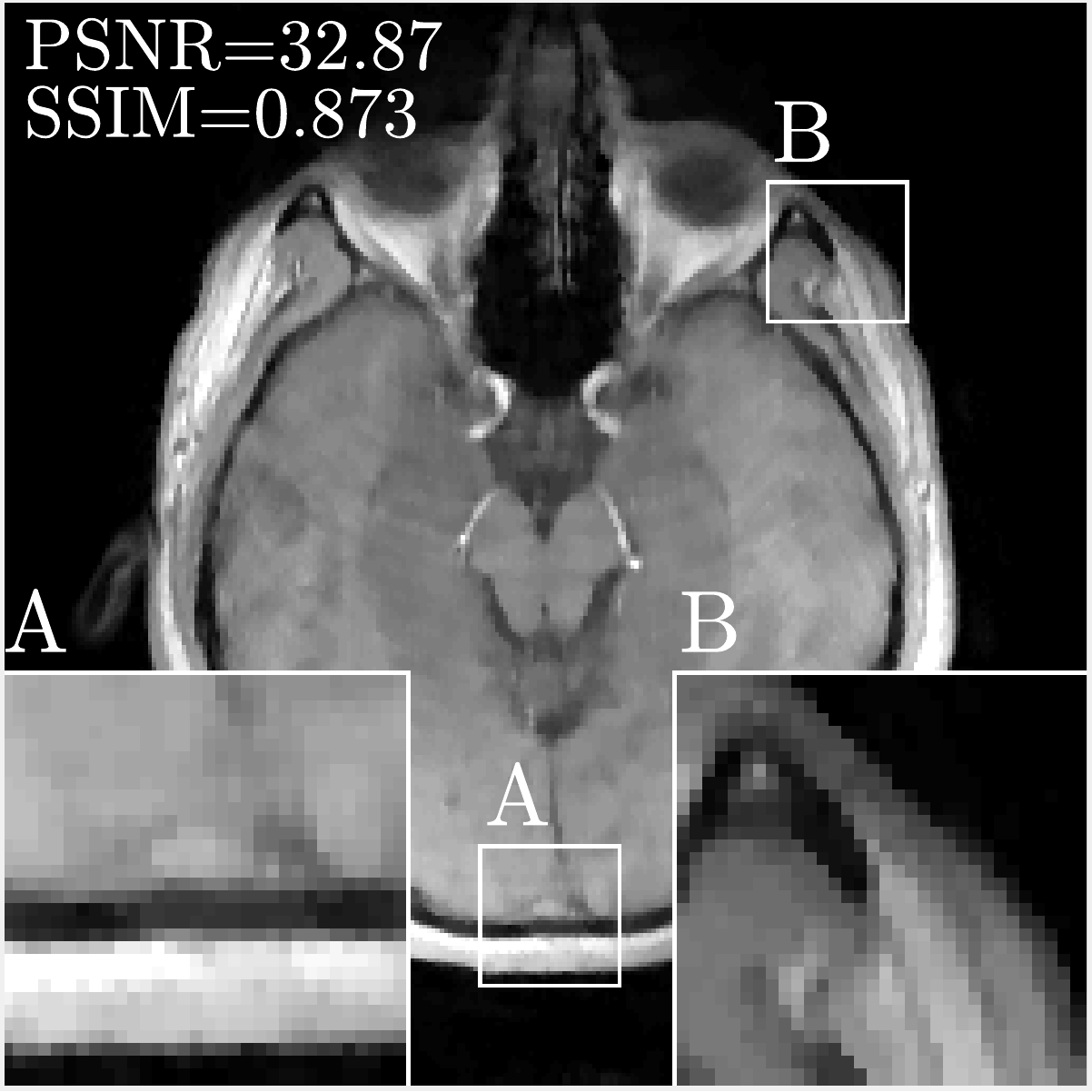} &
\hspace{-4mm}\includegraphics[width=.16\textwidth]{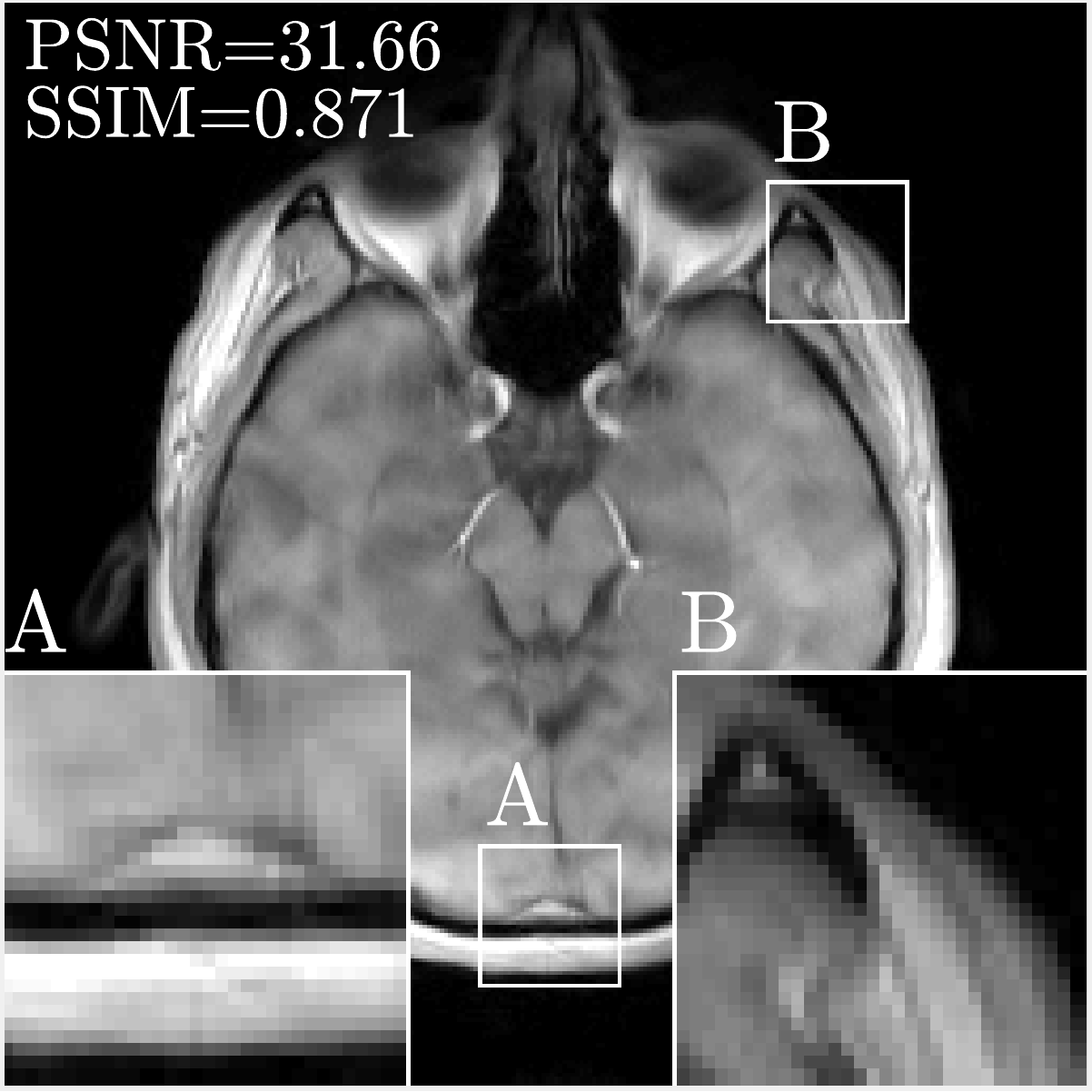} &
\hspace{-4mm}\includegraphics[width=.16\textwidth]{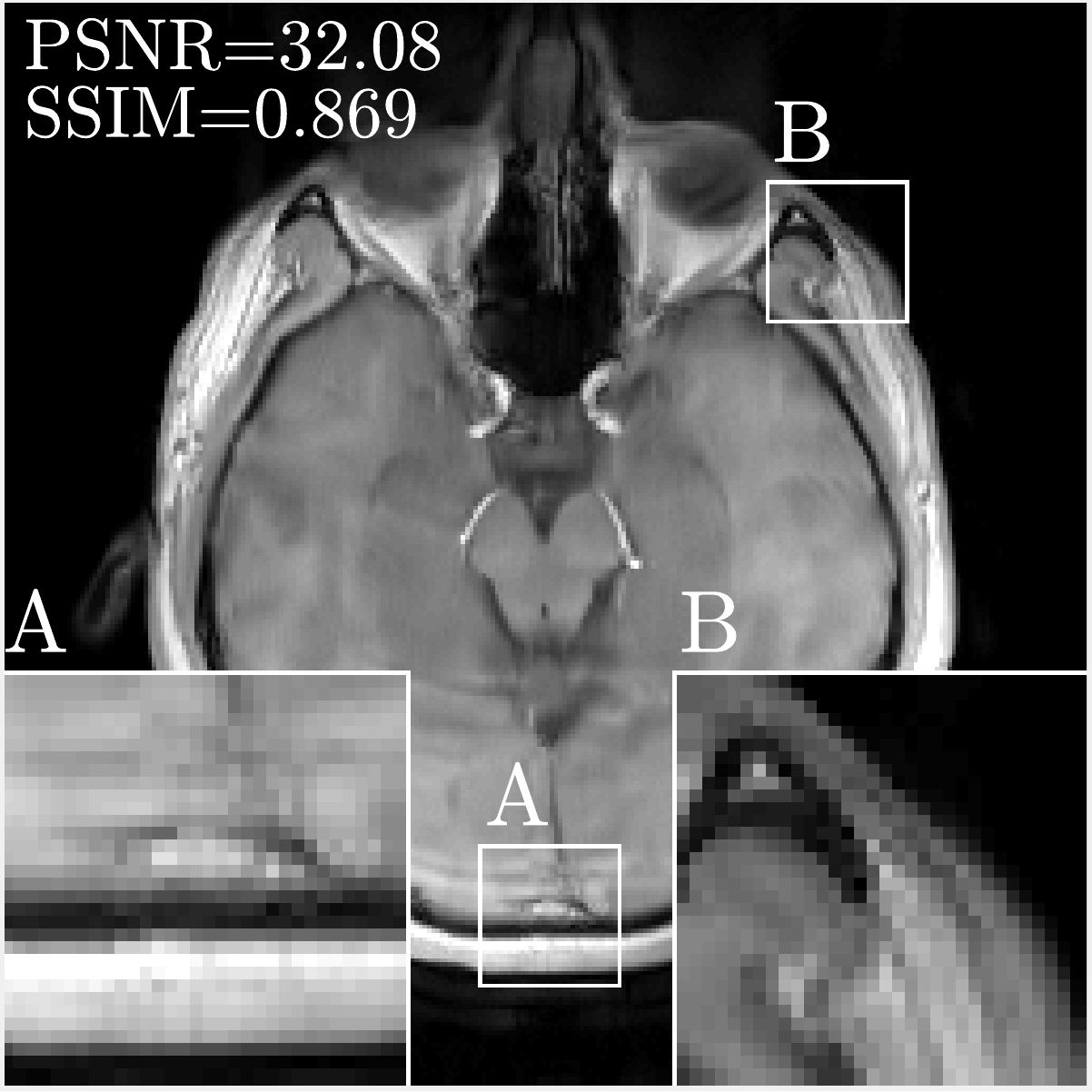} &
\hspace{-4mm}\includegraphics[width=.16\textwidth]{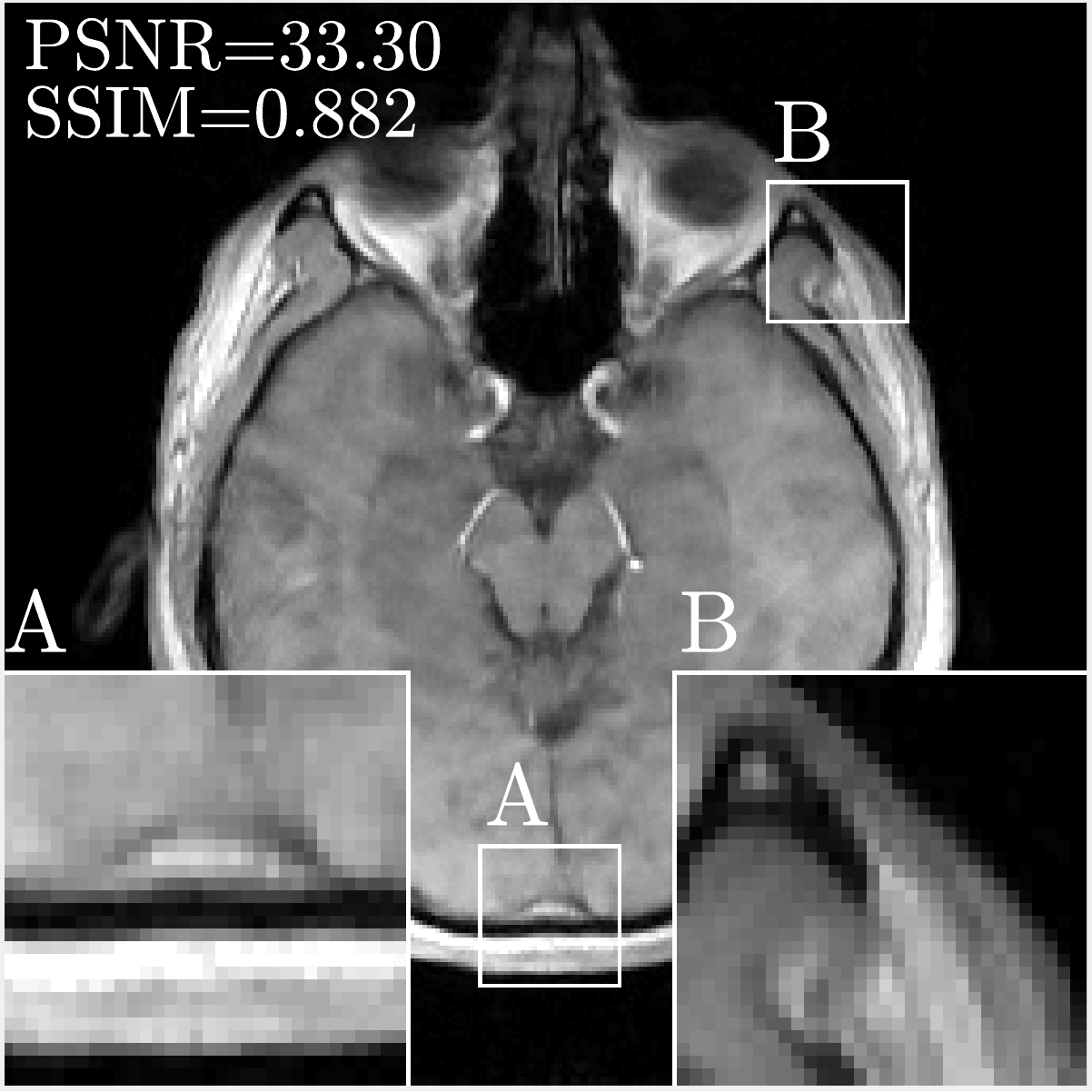}   &
\hspace{-4mm}\includegraphics[width=.16\textwidth]{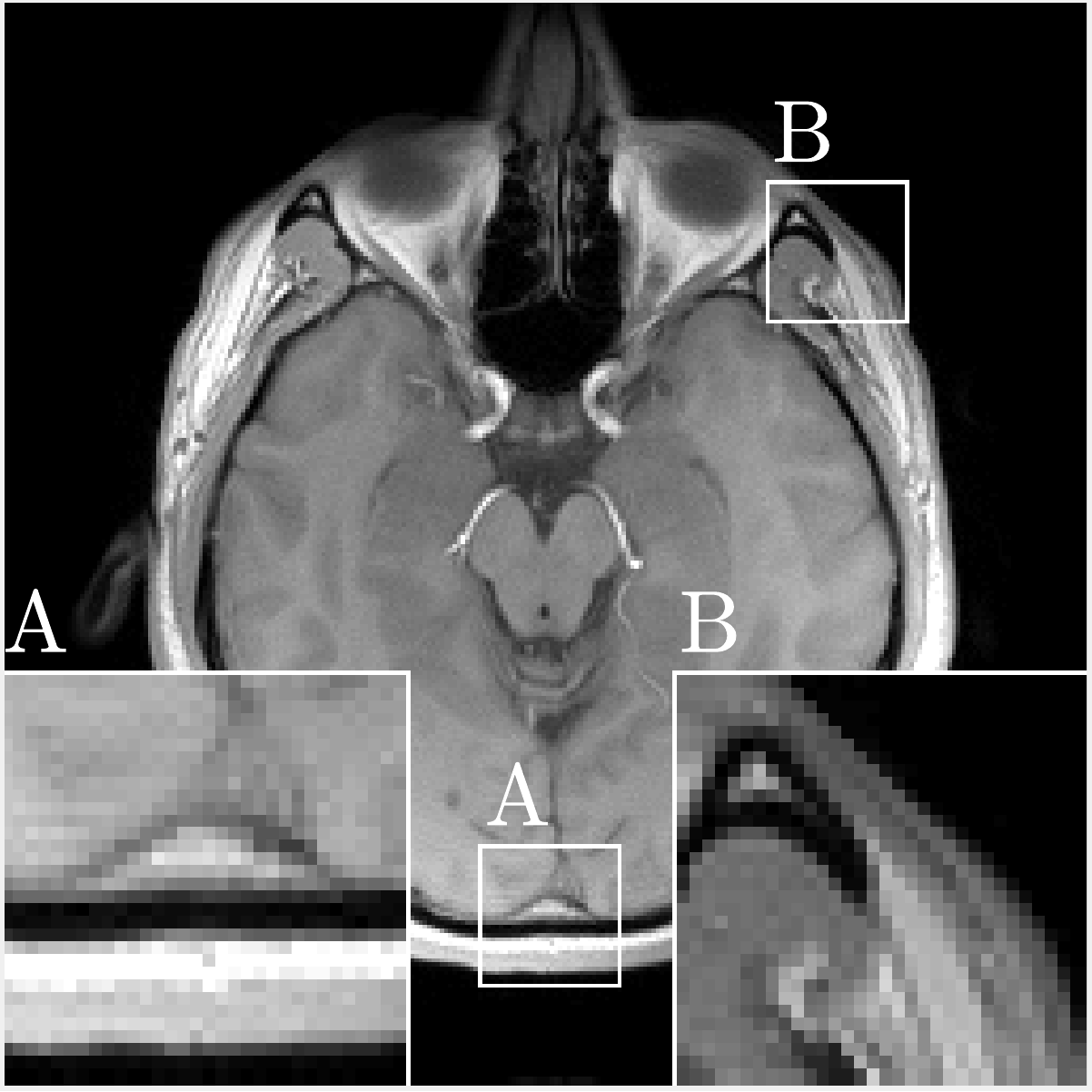}\\ [-1mm]
\rotatebox{90}{\hspace{5mm} \textbf{TV-greedy}} &
\hspace{-4mm}\includegraphics[width=.16\textwidth]{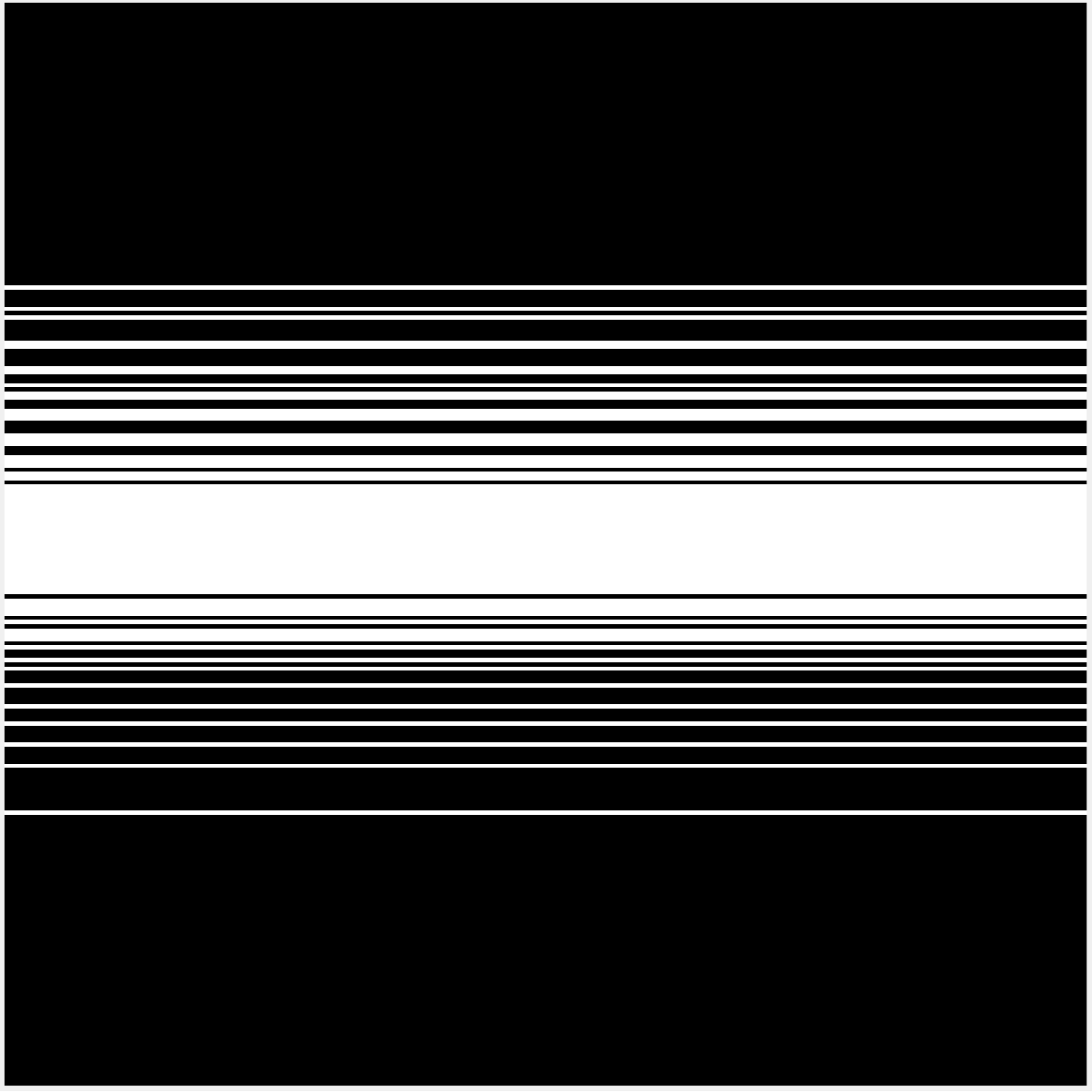} &
\hspace{-4mm}\includegraphics[width=.16\textwidth]{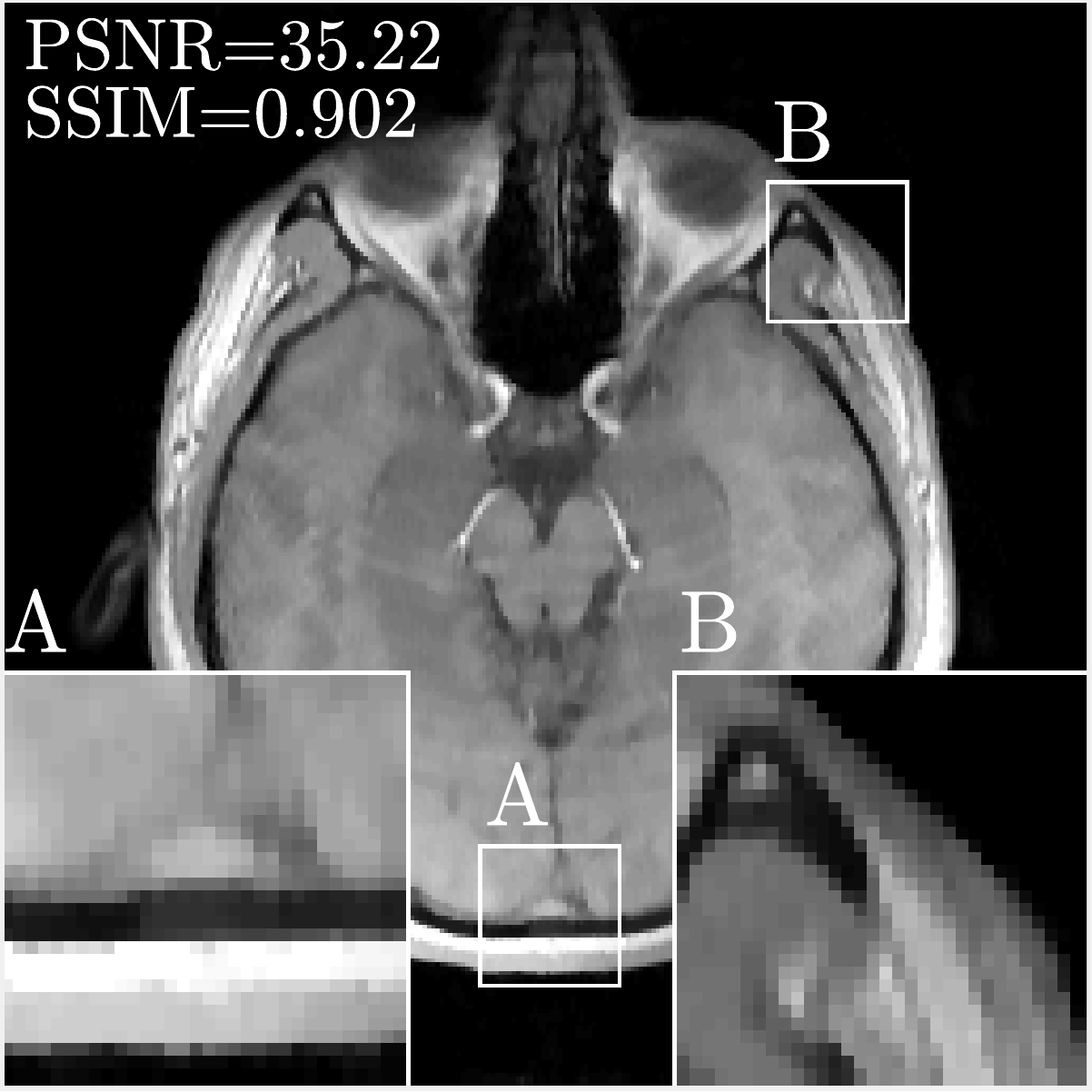} &
\hspace{-4mm}\includegraphics[width=.16\textwidth]{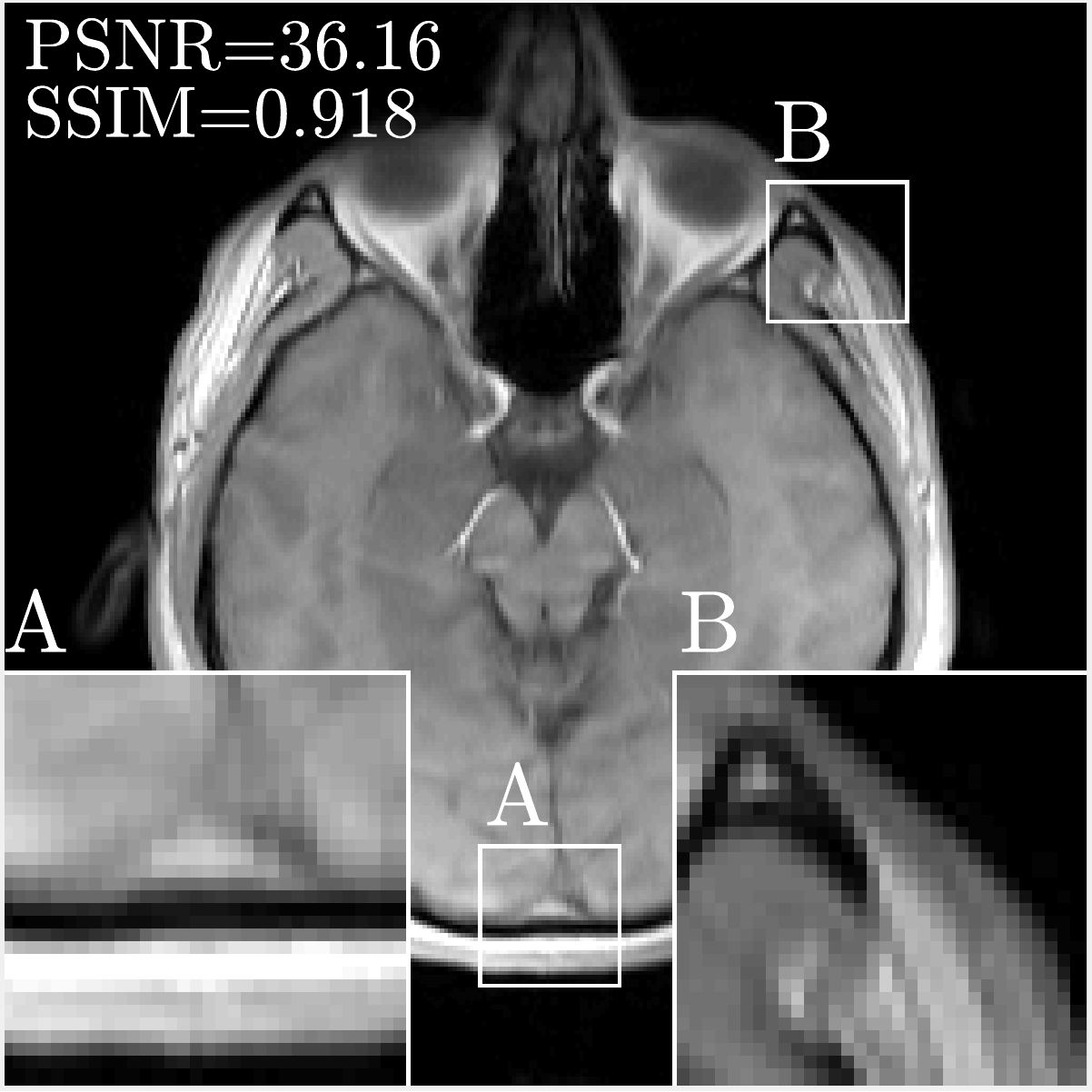} &
\hspace{-4mm}\includegraphics[width=.16\textwidth]{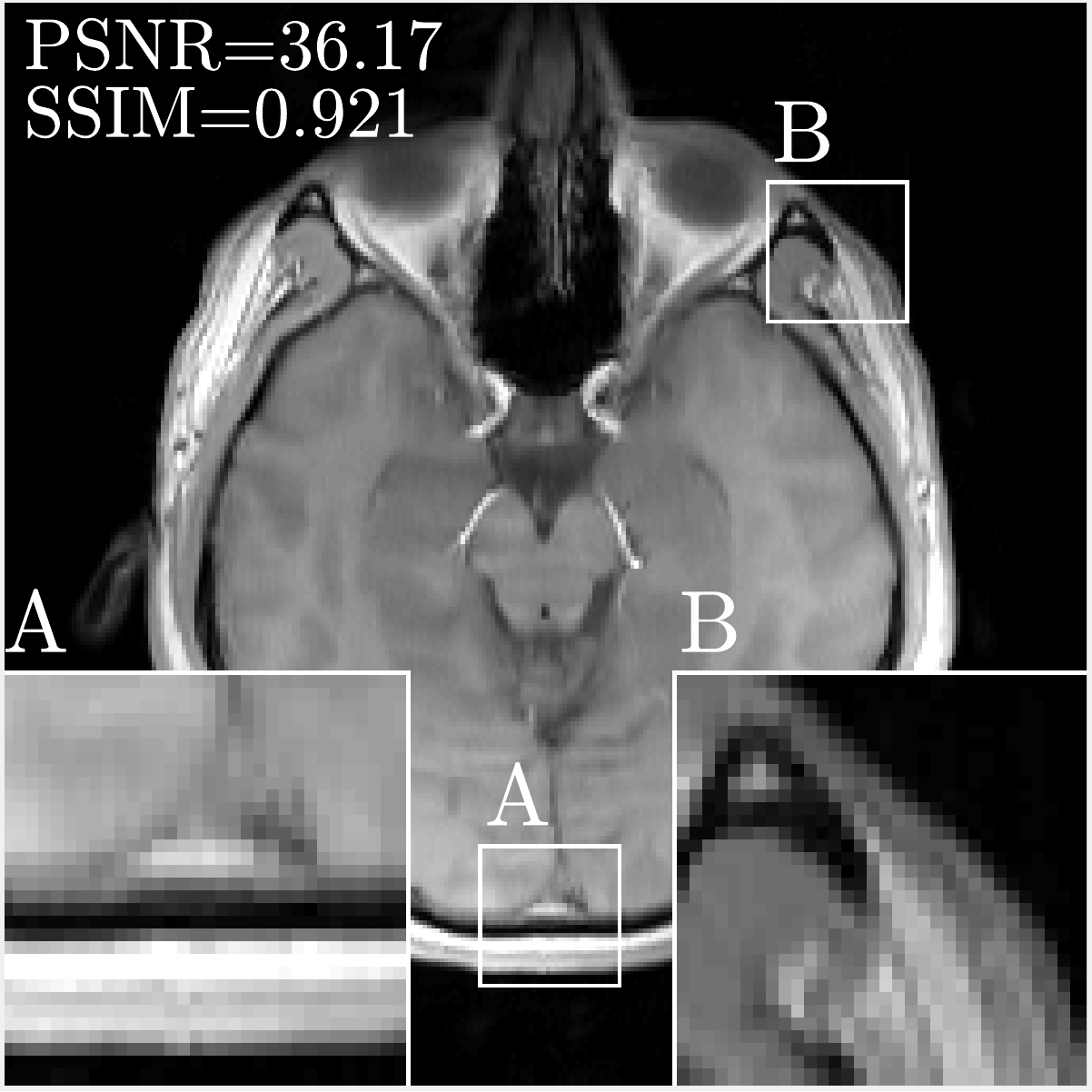} &
\hspace{-4mm}\includegraphics[width=.16\textwidth]{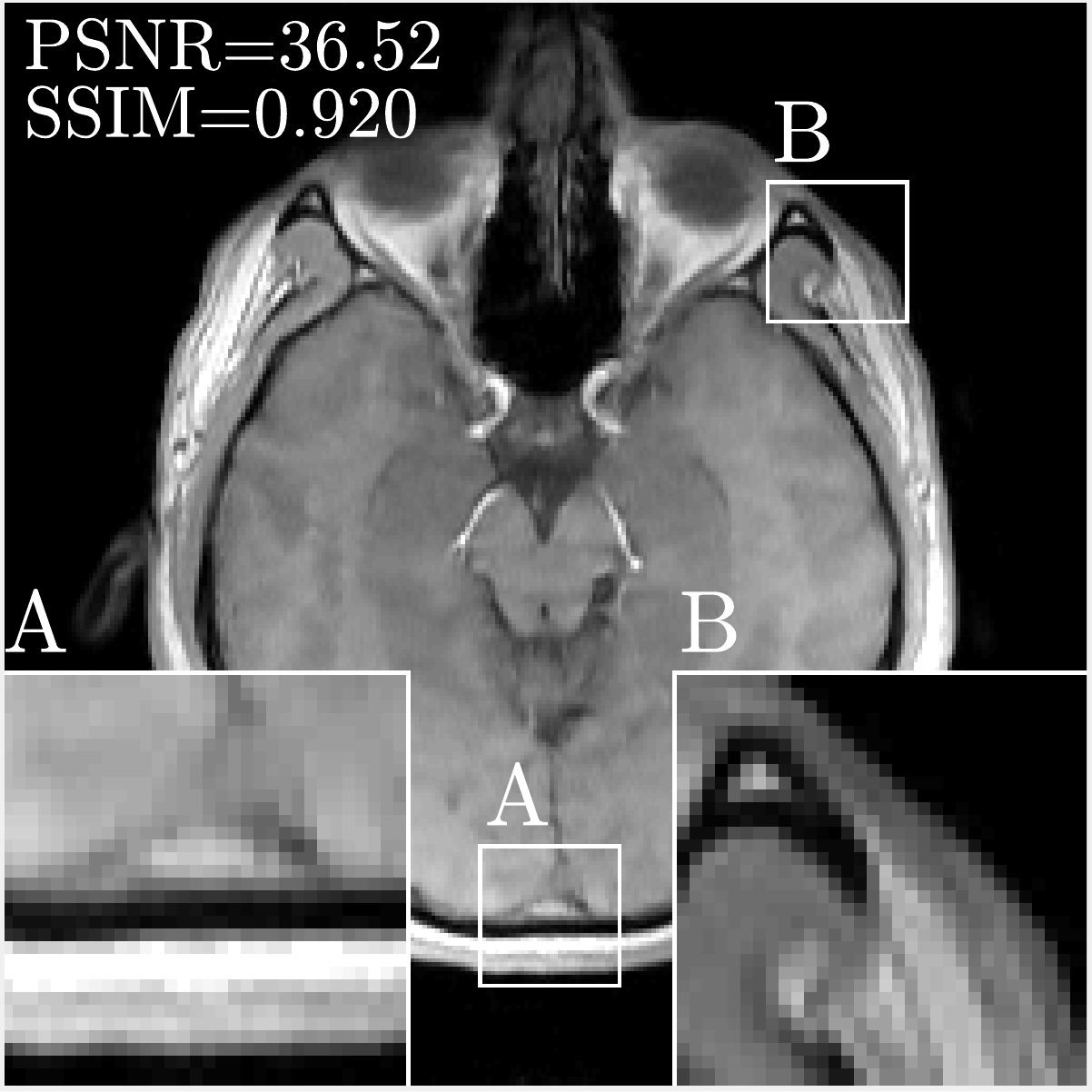}   &
\hspace{-4mm}\includegraphics[width=.16\textwidth]{FIGURES_REV/ABc_original_magnified_3.pdf}\\ [-1mm]

\rotatebox{90}{\hspace{5mm} \textbf{BP-greedy}} &
\hspace{-4mm}\includegraphics[width=.16\textwidth]{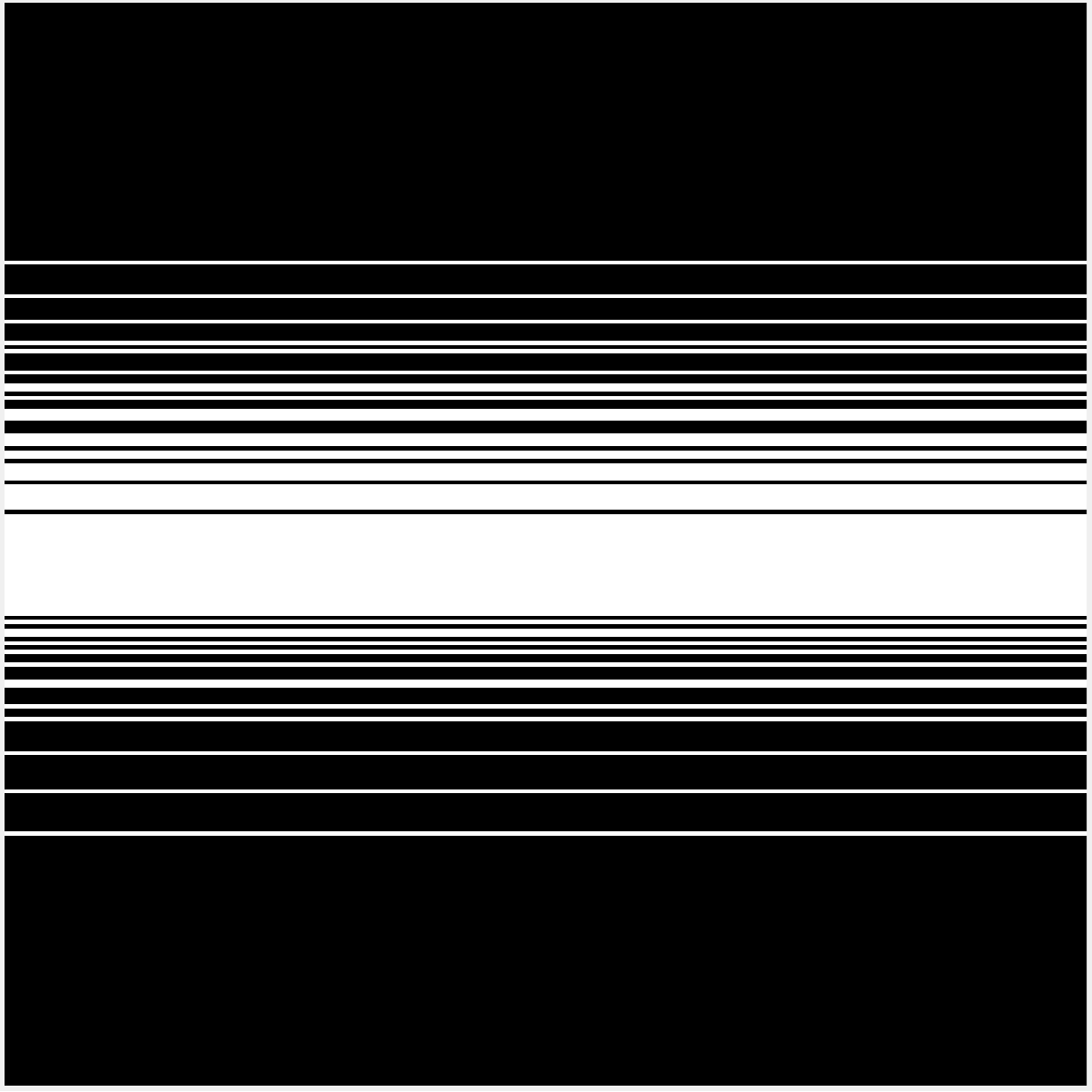} &
\hspace{-4mm}\includegraphics[width=.16\textwidth]{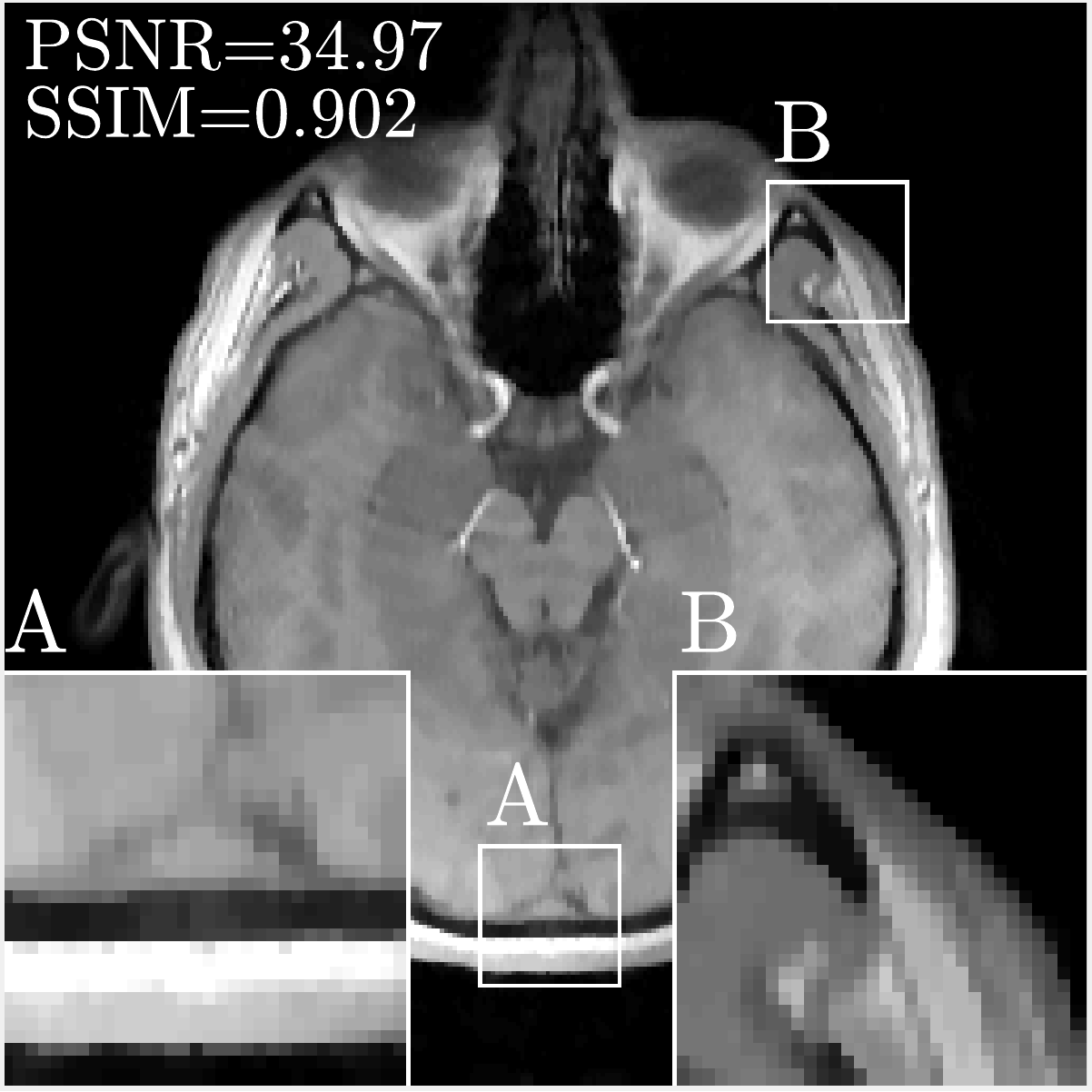} &
\hspace{-4mm}\includegraphics[width=.16\textwidth]{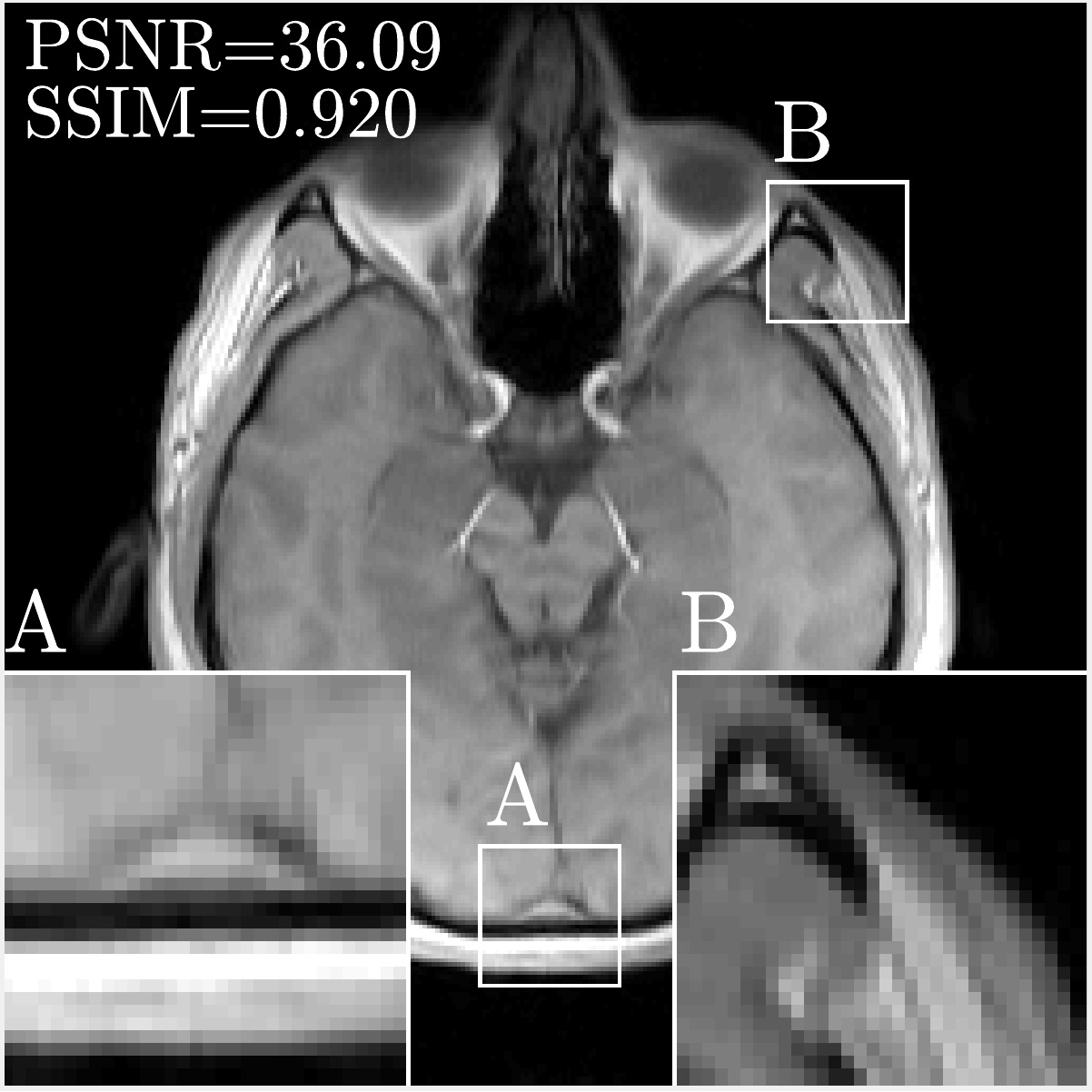} &
\hspace{-4mm}\includegraphics[width=.16\textwidth]{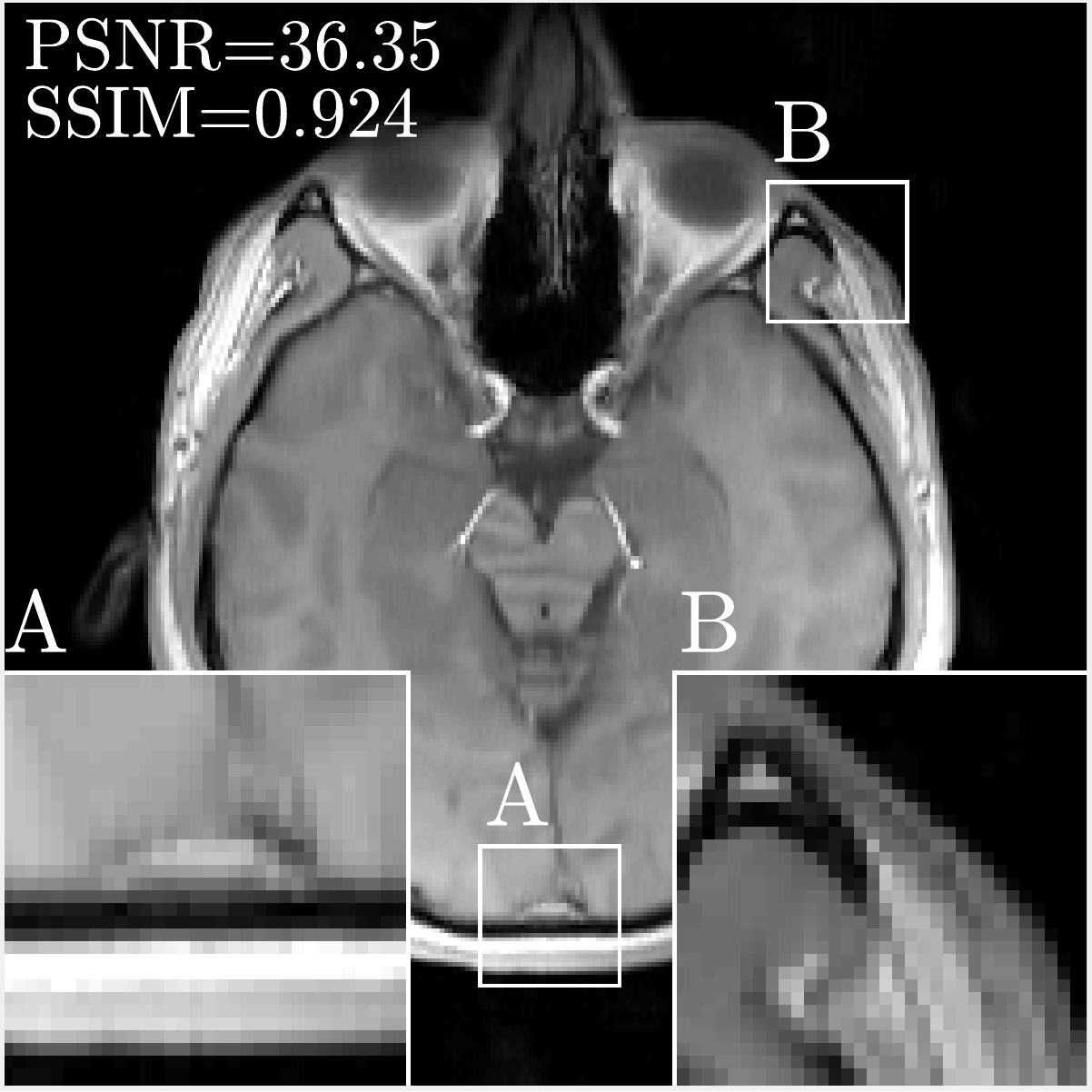} &
\hspace{-4mm}\includegraphics[width=.16\textwidth]{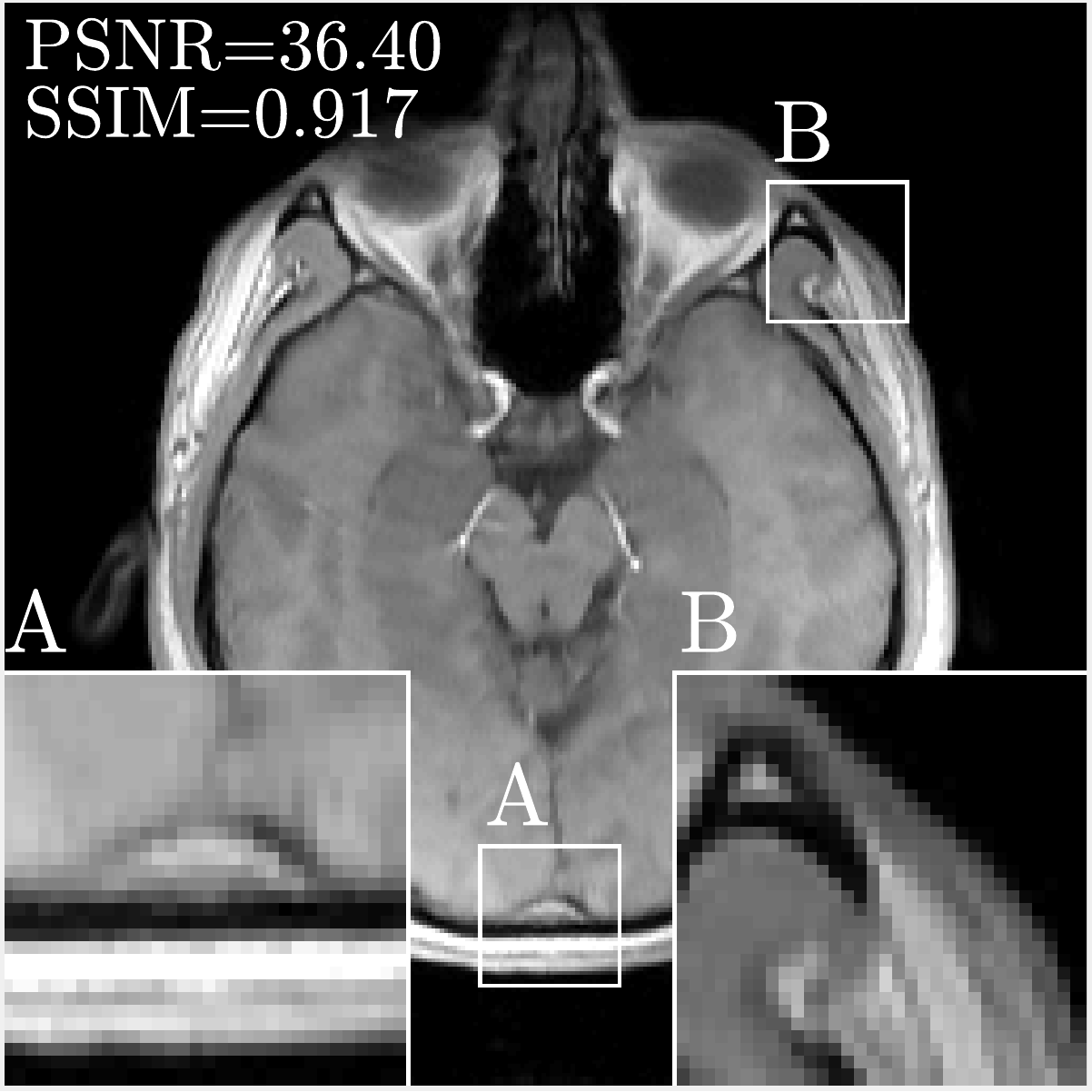}   &
\hspace{-4mm}\includegraphics[width=.16\textwidth]{FIGURES_REV/ABc_original_magnified_3.pdf}\\ [-1mm]
\rotatebox{90}{\hspace{3mm} \textbf{BM3D-greedy}} &
\hspace{-4mm}\includegraphics[width=.16\textwidth]{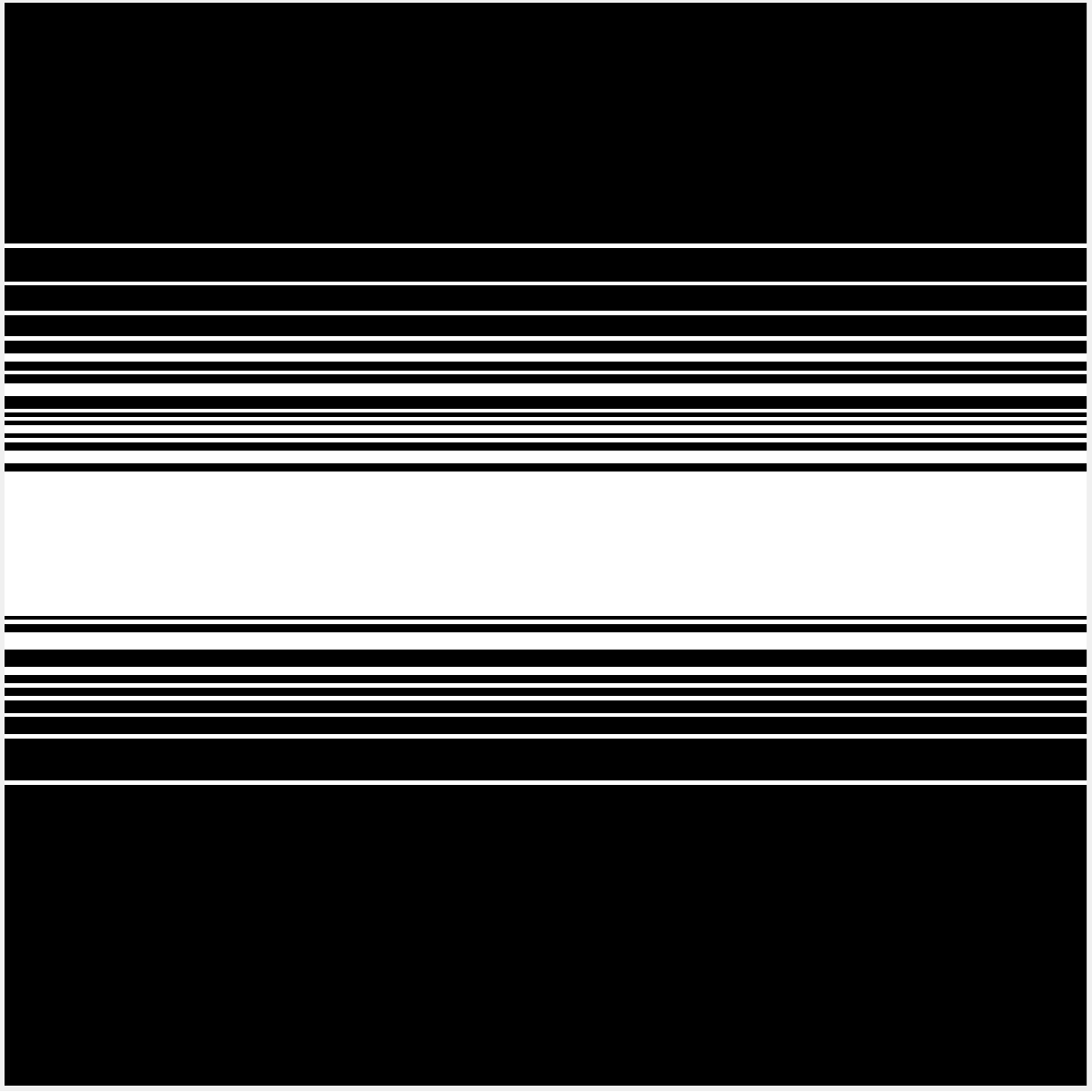} &
\hspace{-4mm}\includegraphics[width=.16\textwidth]{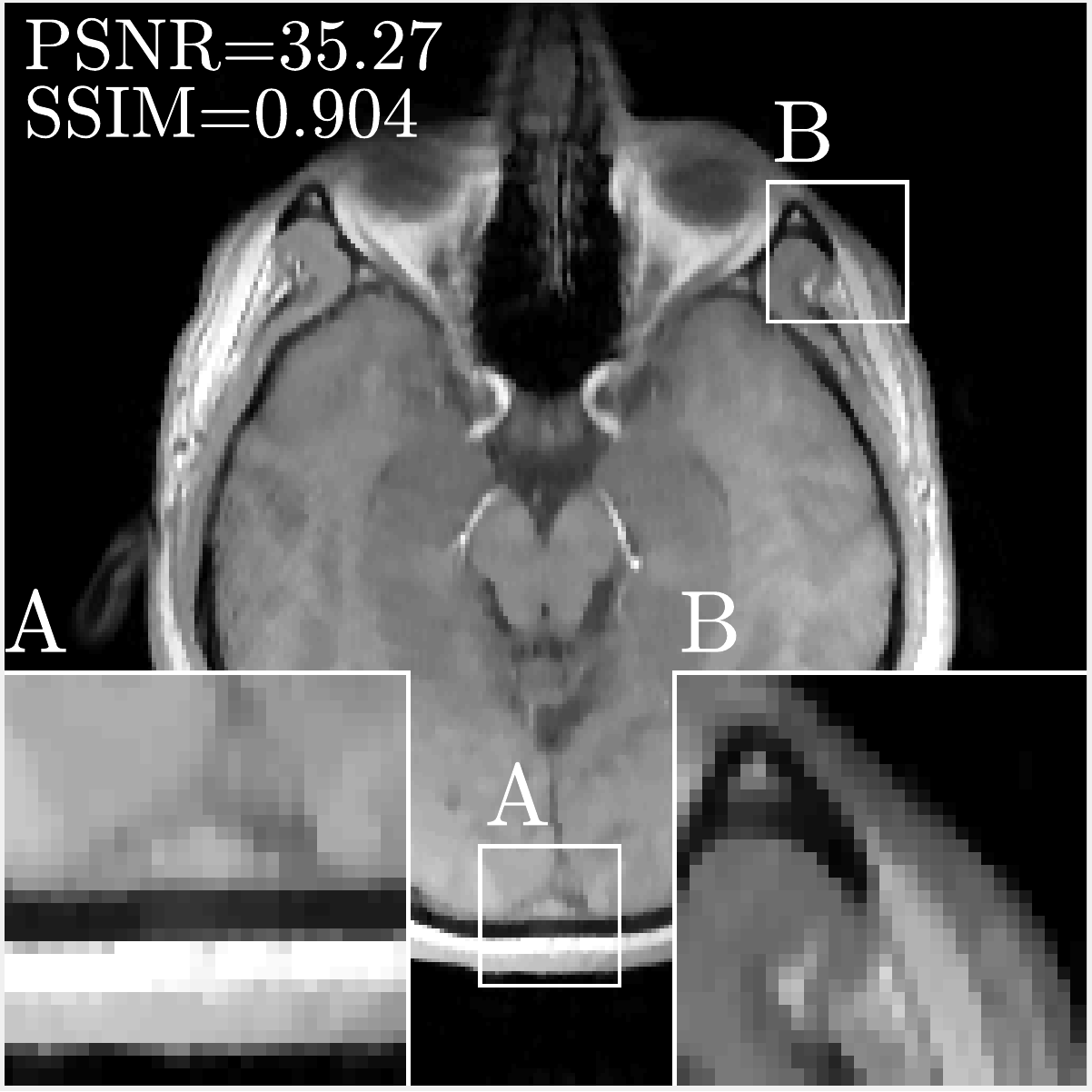} &
\hspace{-4mm}\includegraphics[width=.16\textwidth]{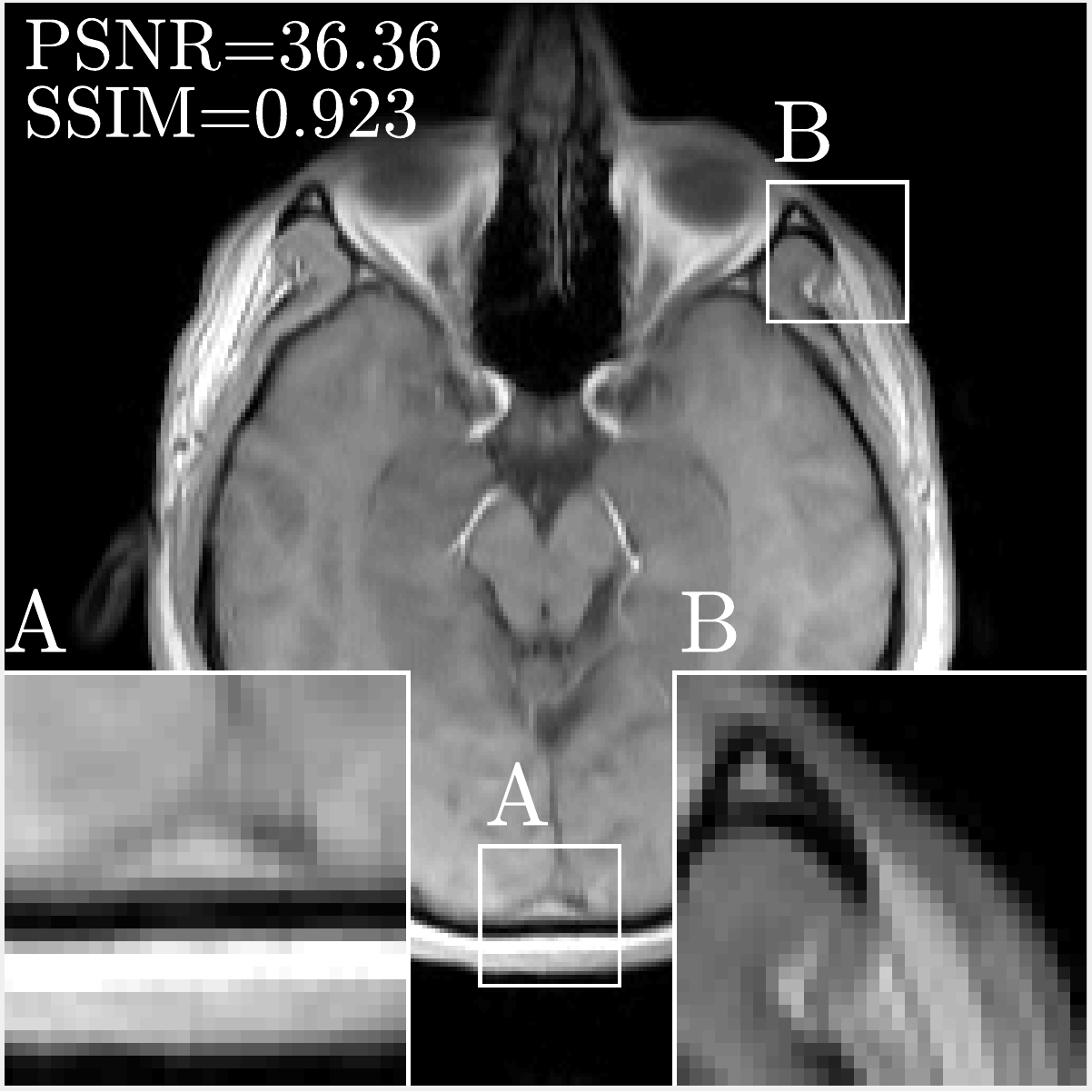} &
\hspace{-4mm}\includegraphics[width=.16\textwidth]{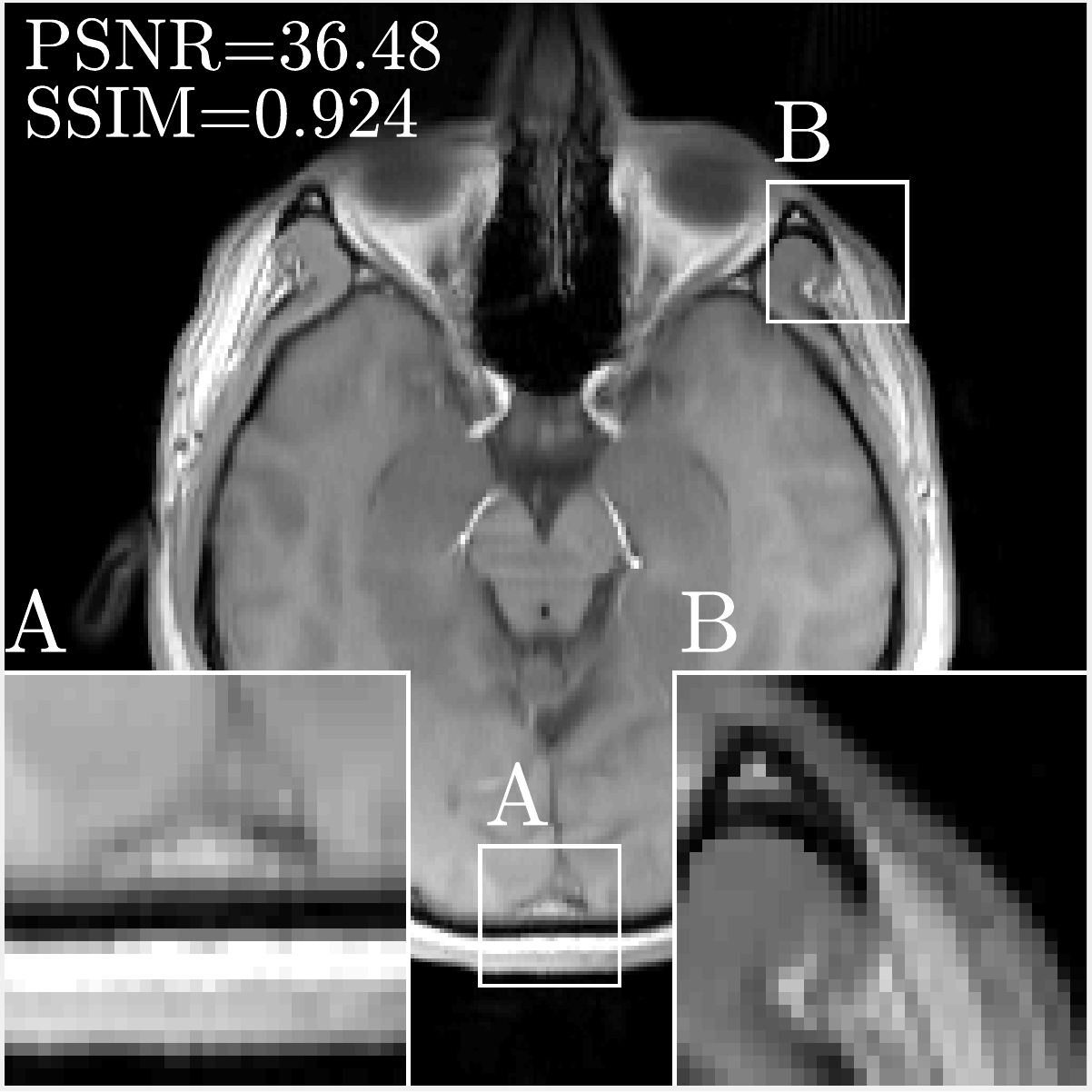} &
\hspace{-4mm}\includegraphics[width=.16\textwidth]{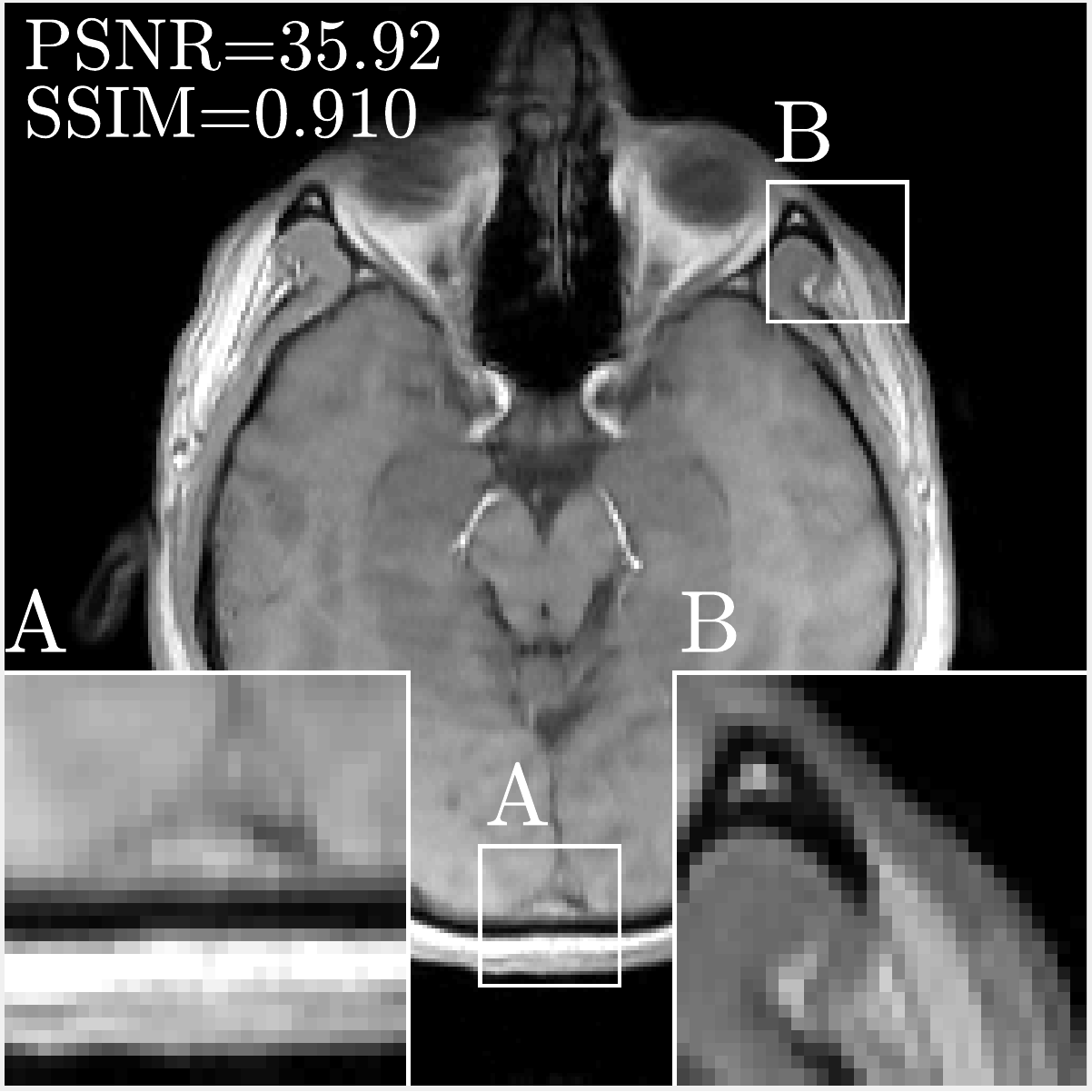}   &
\hspace{-4mm}\includegraphics[width=.16\textwidth]{FIGURES_REV/ABc_original_magnified_3.pdf}\\ [-1mm]
\rotatebox{90}{\hspace{5mm} \textbf{NN-greedy}} &
\hspace{-4mm}\includegraphics[width=.16\textwidth]{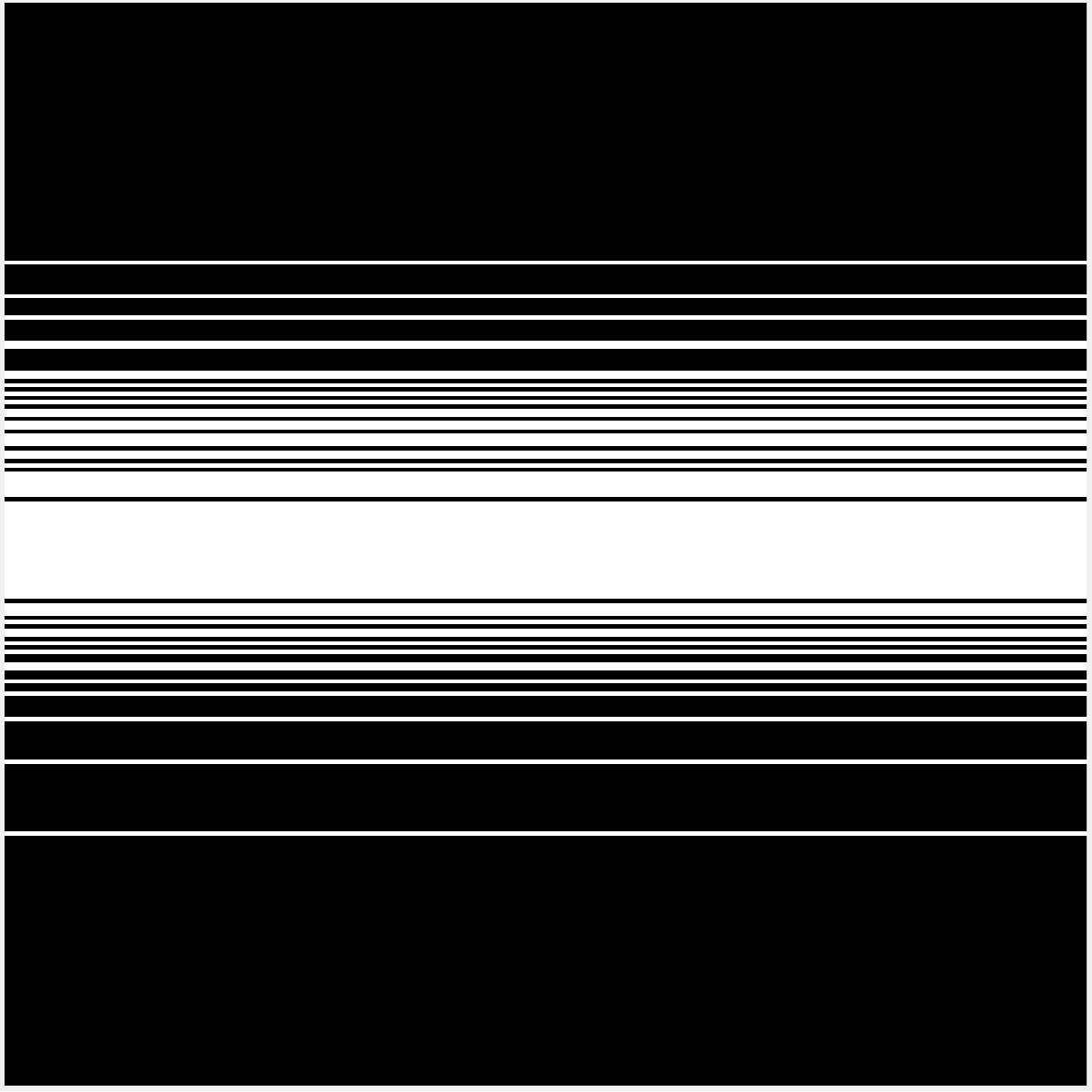} & 
\hspace{-4mm}\includegraphics[width=.16\textwidth]{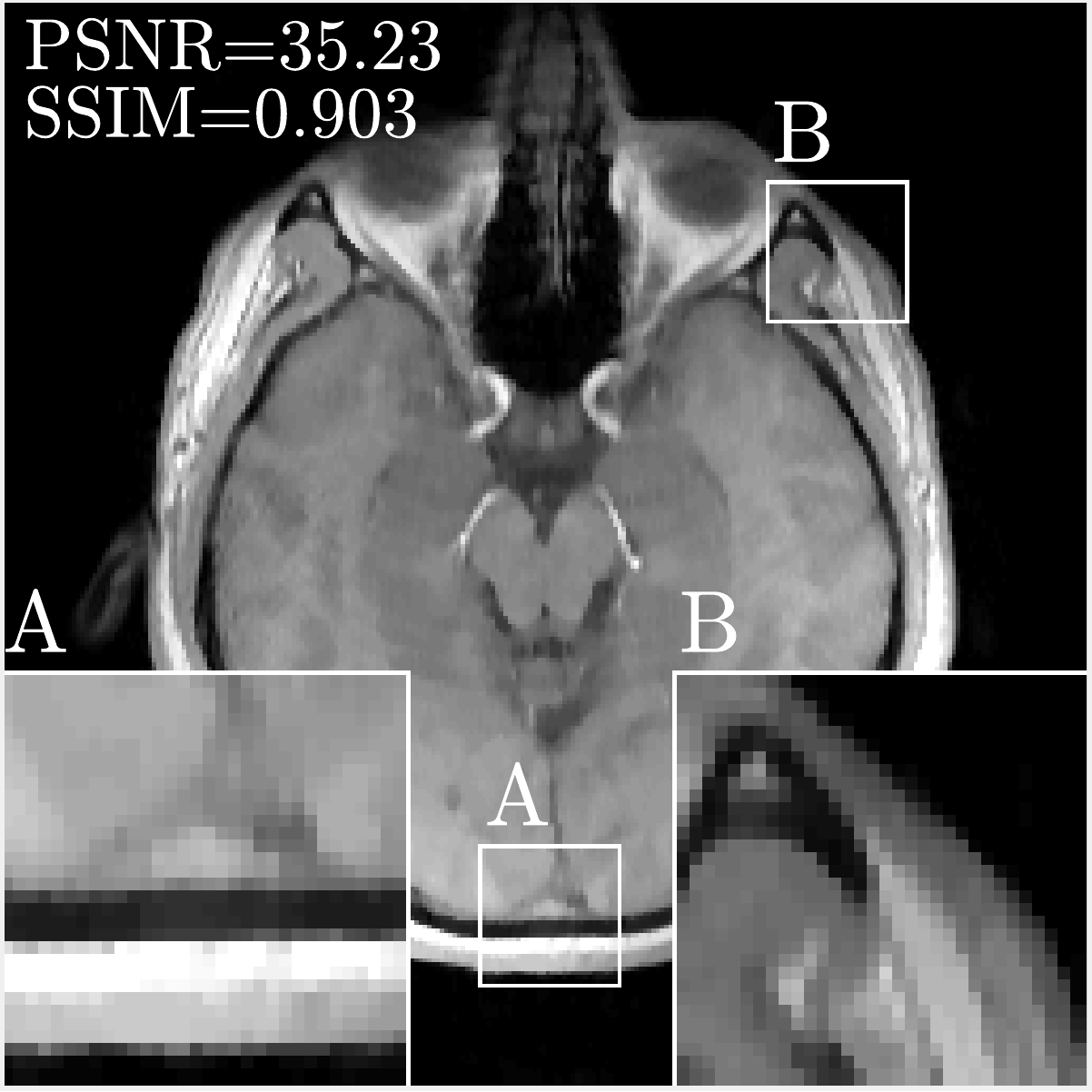} &
\hspace{-4mm}\includegraphics[width=.16\textwidth]{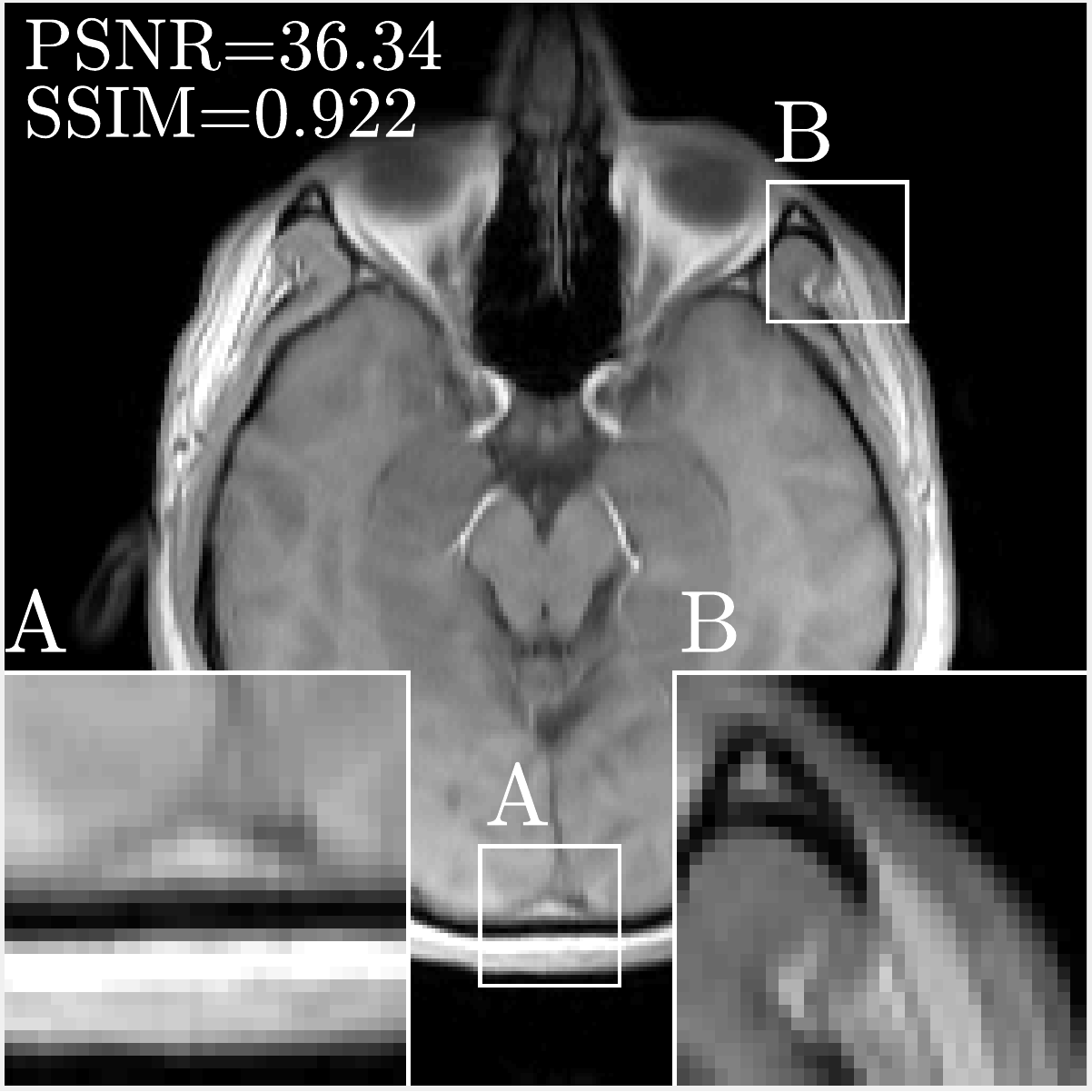} &
\hspace{-4mm}\includegraphics[width=.16\textwidth]{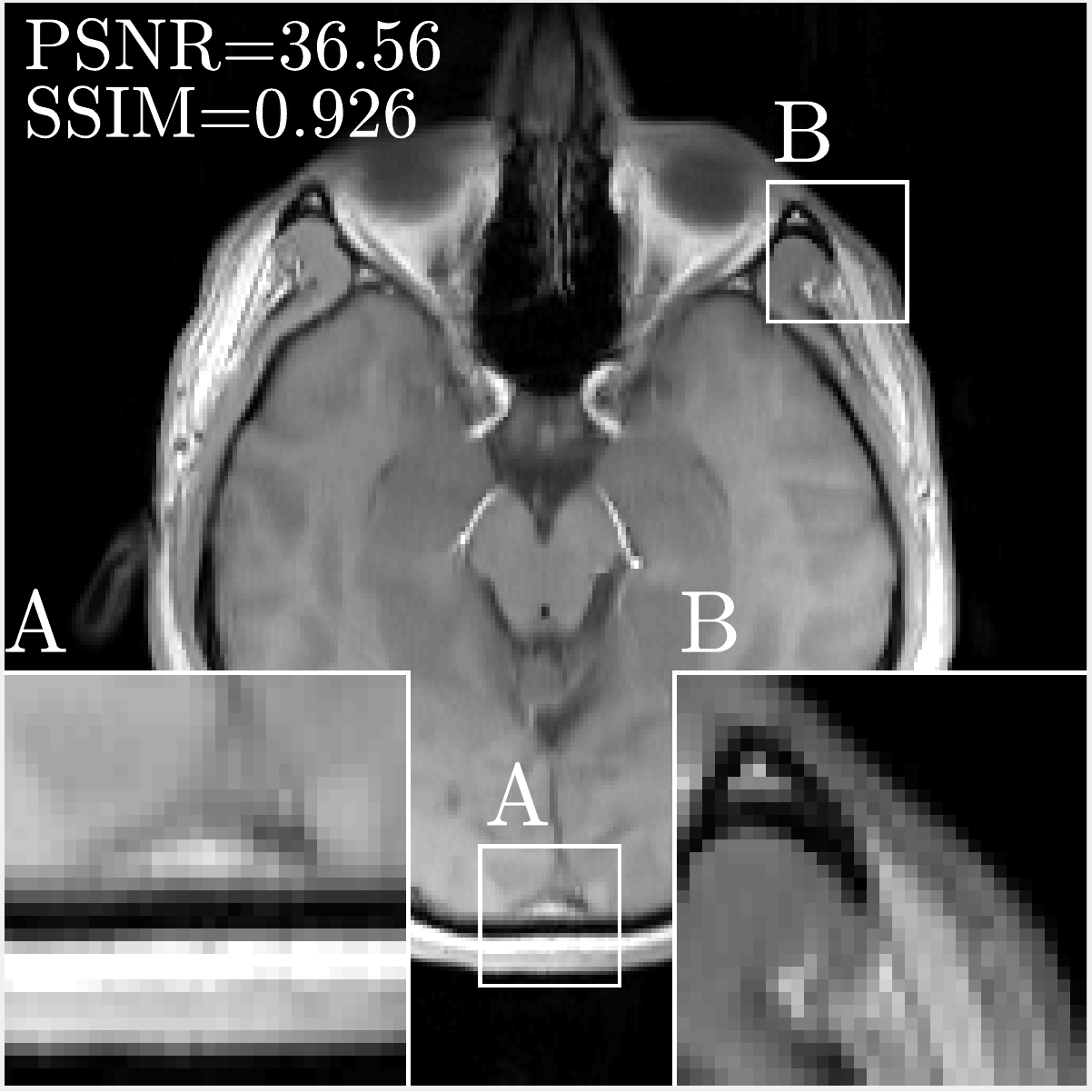} &
\hspace{-4mm}\includegraphics[width=.16\textwidth]{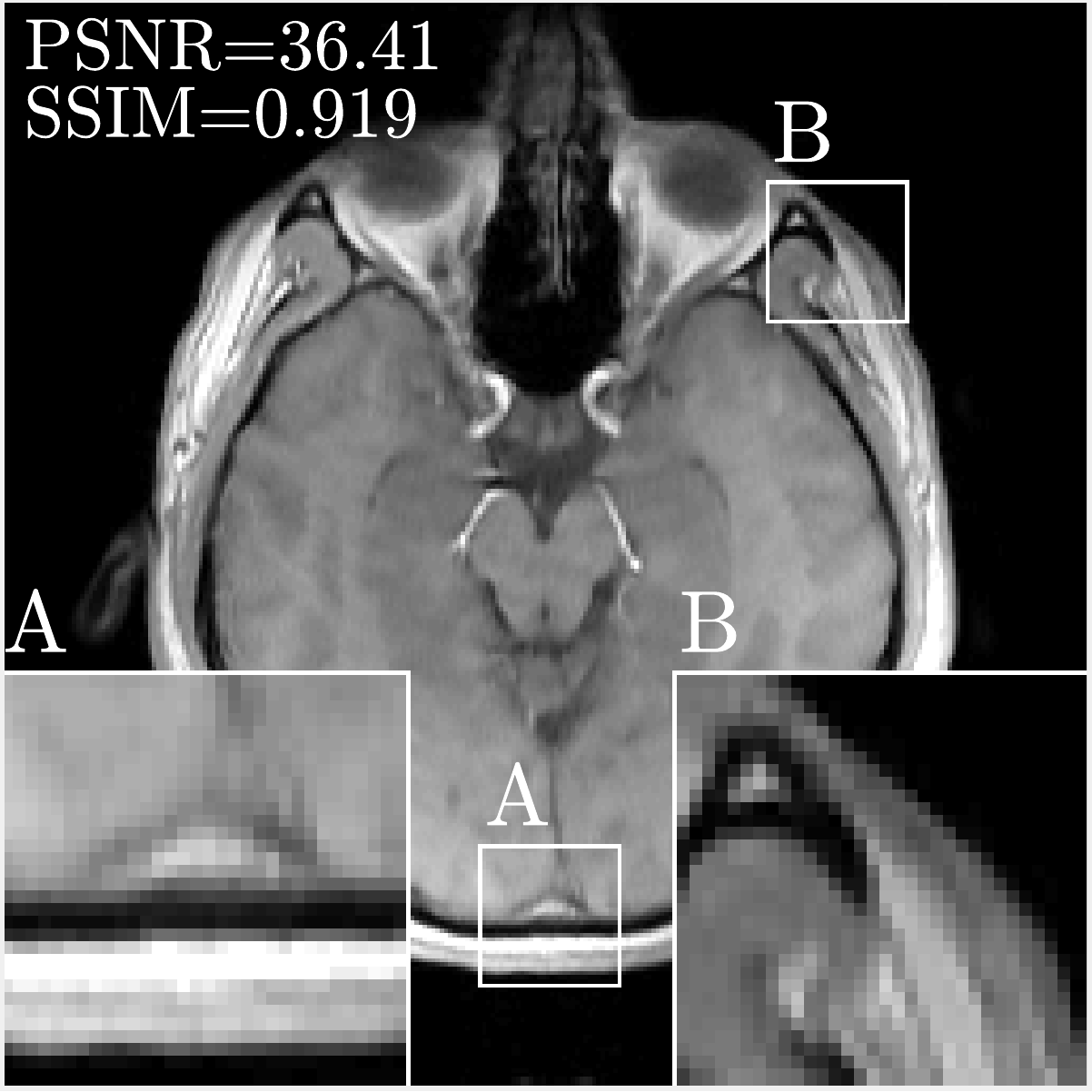}   &
\hspace{-4mm}\includegraphics[width=.16\textwidth]{FIGURES_REV/ABc_original_magnified_3.pdf}\\ [-1mm]
\rotatebox{90}{\hspace{7mm} \textbf{\rev{Low pass}}} &
\hspace{-4mm}\includegraphics[width=.16\textwidth]{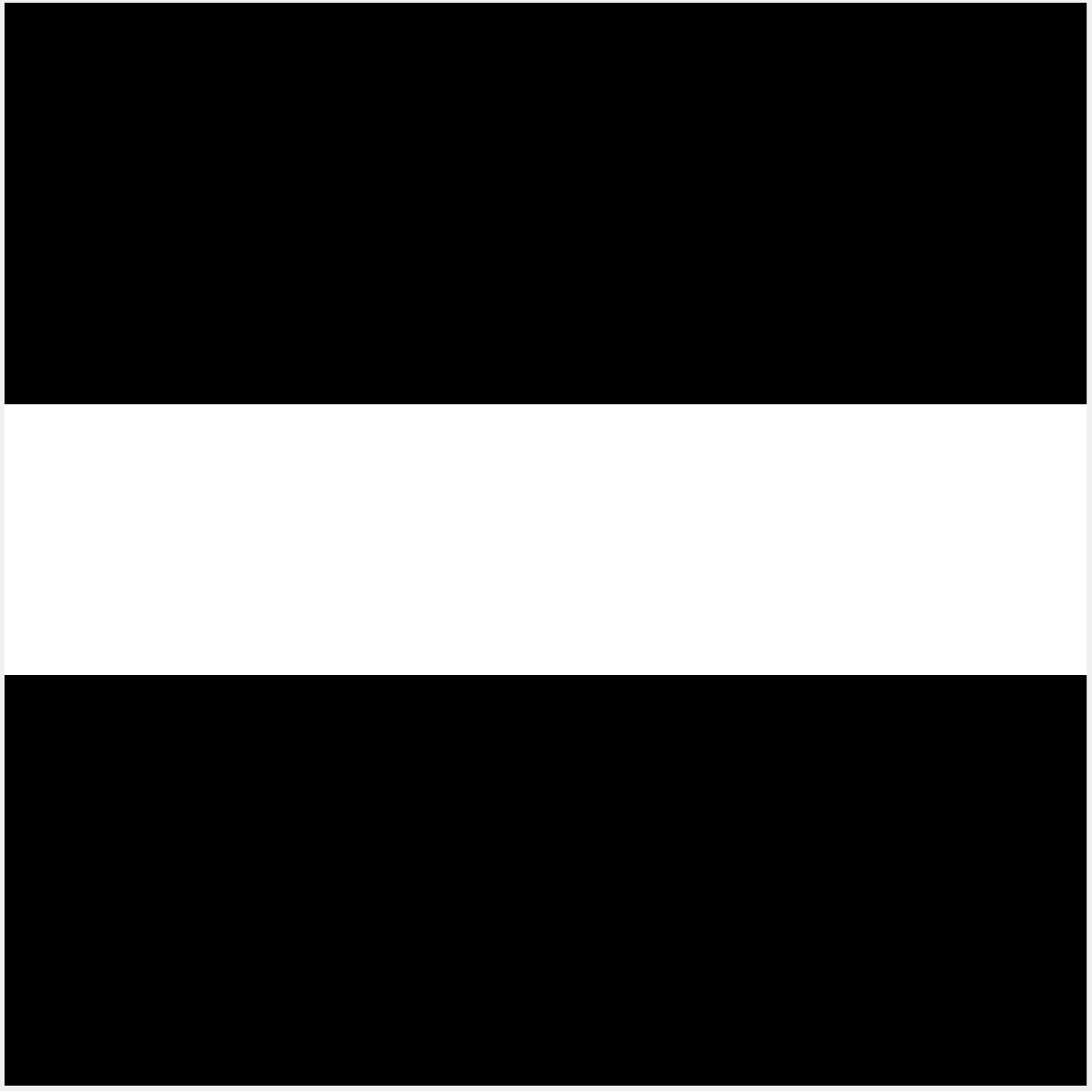} &
\hspace{-4mm}\includegraphics[width=.16\textwidth]{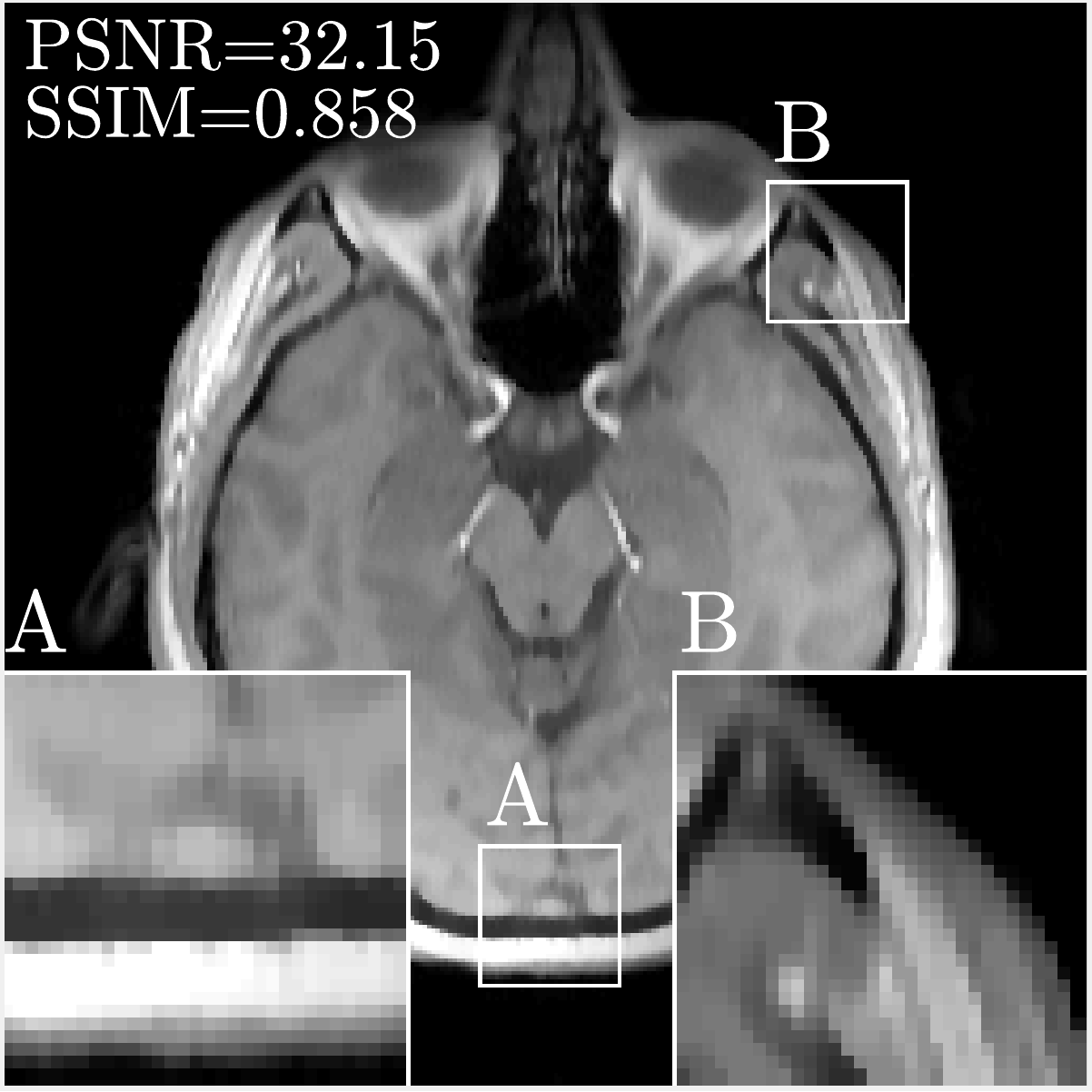} &
\hspace{-4mm}\includegraphics[width=.16\textwidth]{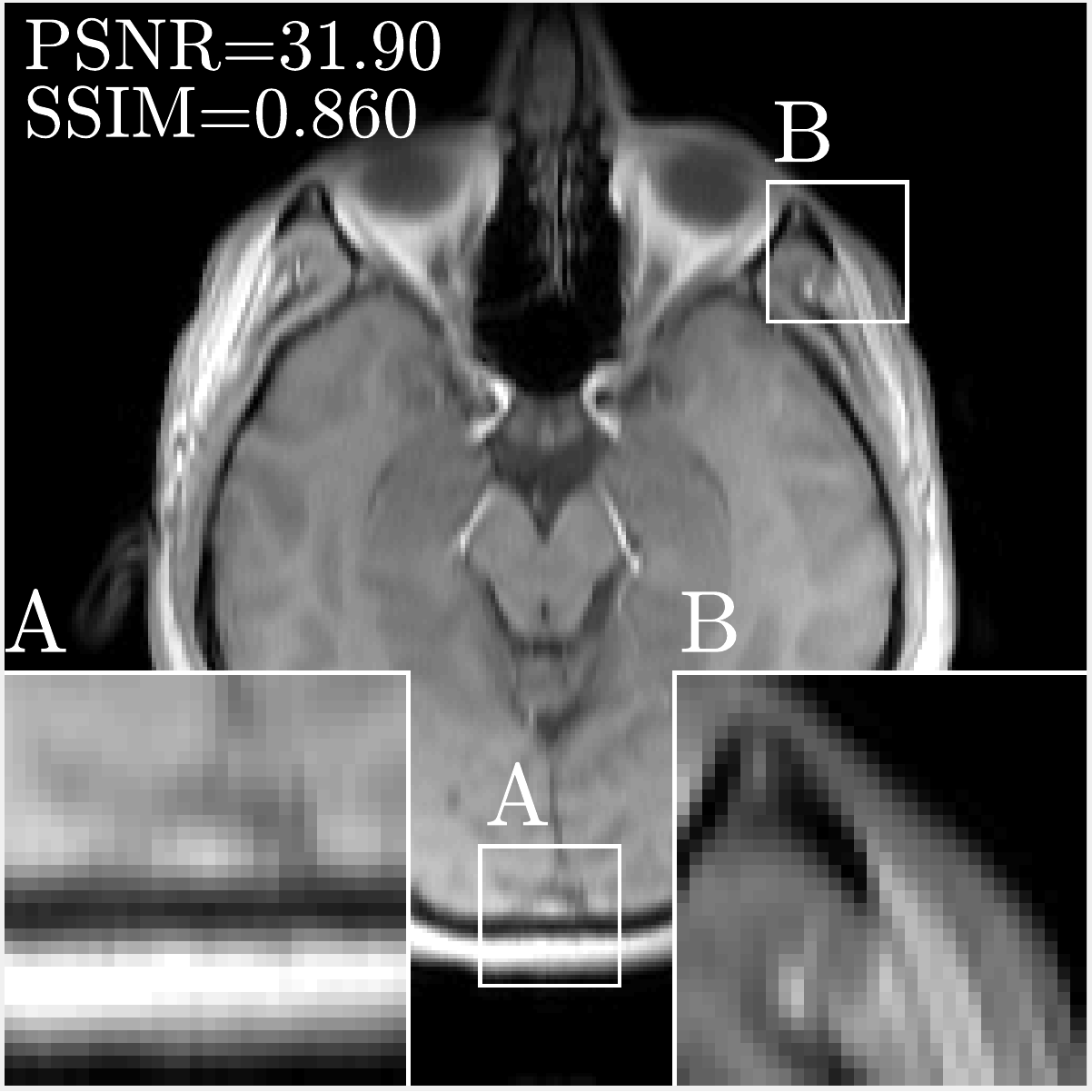} &
\hspace{-4mm}\includegraphics[width=.16\textwidth]{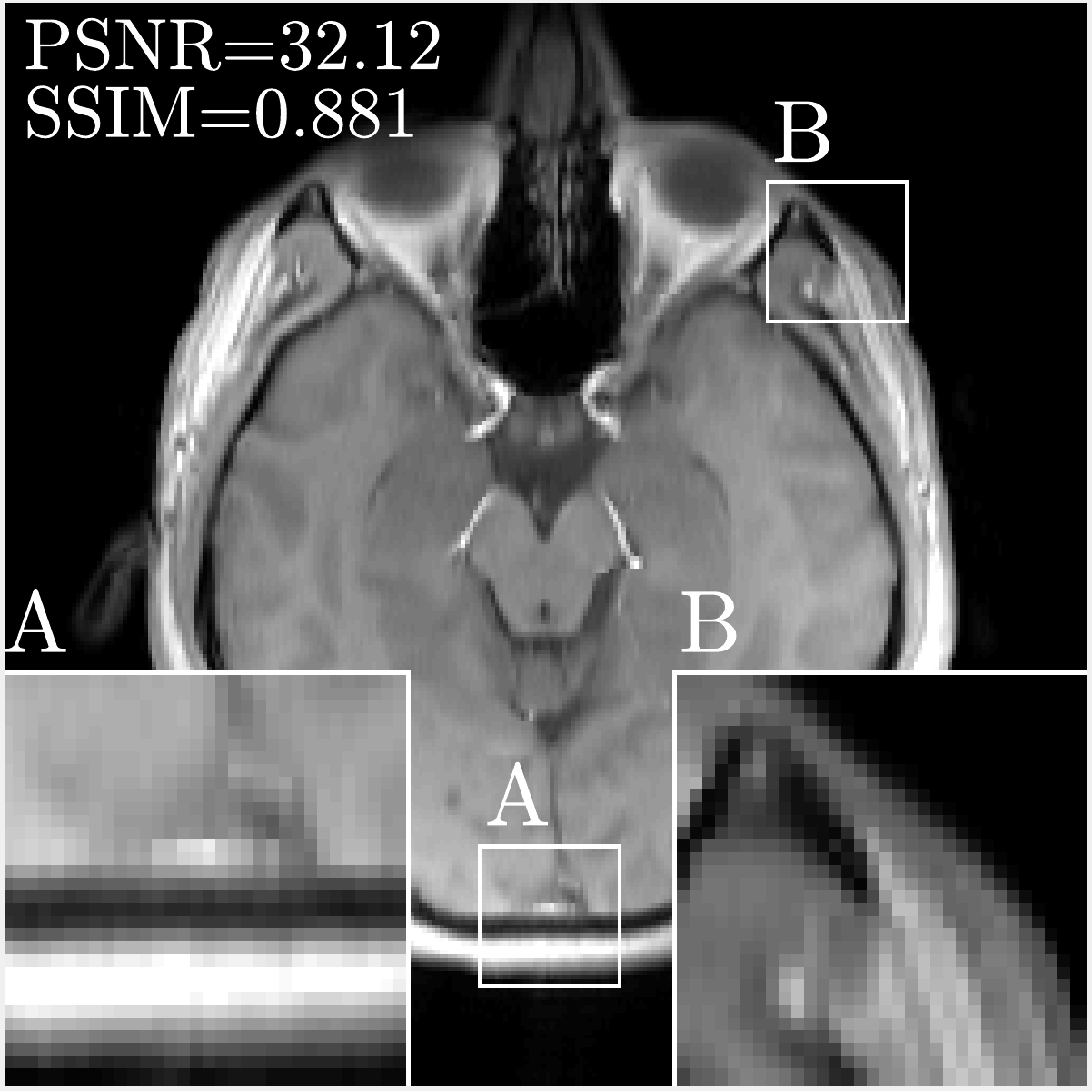} &
\hspace{-4mm}\includegraphics[width=.16\textwidth]{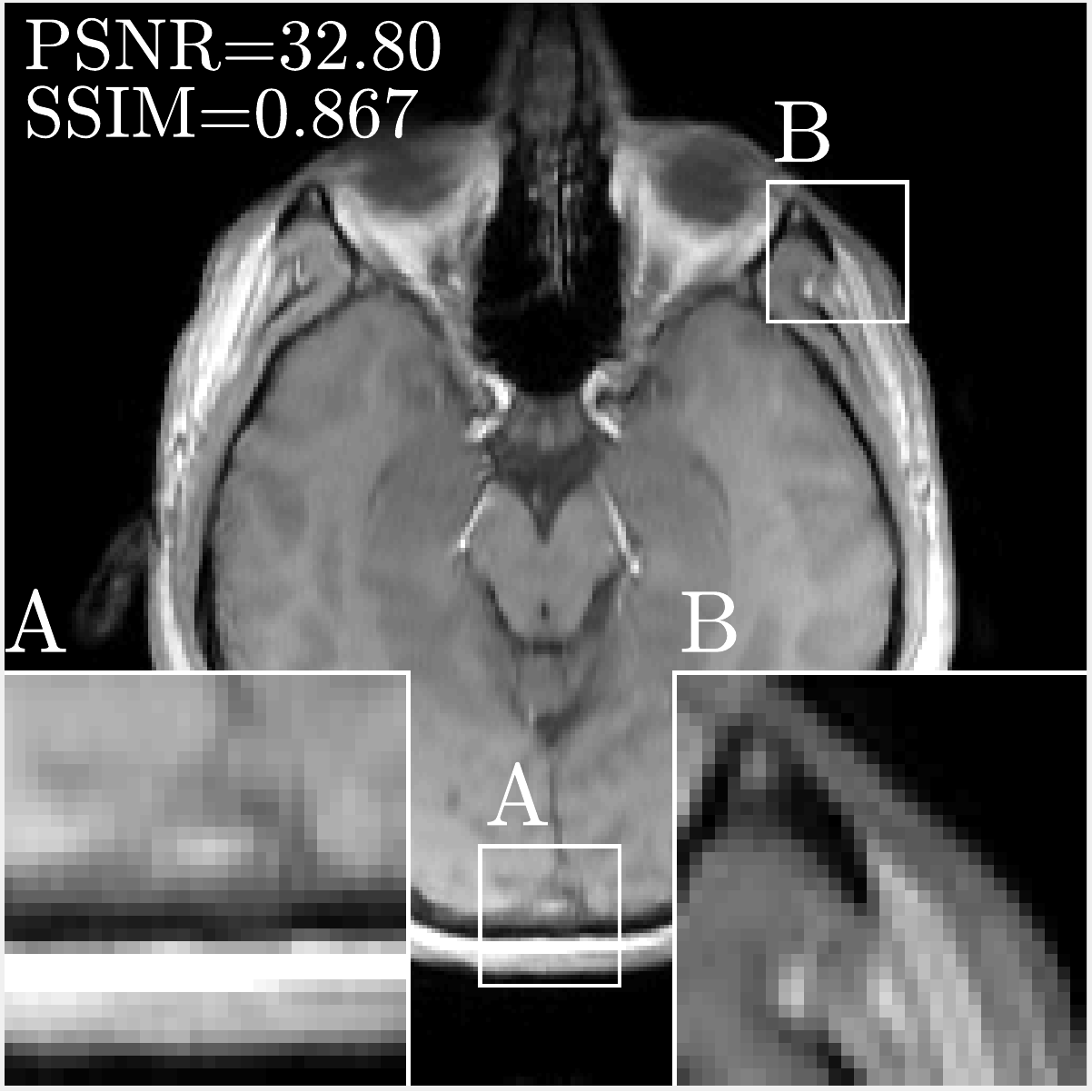}   &
\hspace{-4mm}\includegraphics[width=.16\textwidth]{FIGURES_REV/ABc_original_magnified_3.pdf}  \\ [-1mm]

\end{tabular}
\caption{MRI  reconstruction example for the test subject 2, slice 4 at 25\% subsampling rate. Sampling masks consist of horizontal lines (phase encodes) and are obtained by the baseline methods \cite{lustig2007sparse,vellagoundar15robust} and the greedy method proposed in Algorithm \ref{alg:1} where PSNR is used as the performance measure. PSNR (in dB) and \rev{SSIM} values are shown on the images. \rev{The last row shows the performance of purely low-pass mask with different decoders. We put the ground truth into the each row of the last column for the ease of visual comparison, except for the first row, where we present the k-space of the ground truth image in log-scale.}  }
\label{fig:MRI_recon_table}
\end{figure*}

\subsection{Cross-performances of decoders}

Next continuing in the same setting as the previous subsection, we compare all four decoders (TV, BP, BM3D, and NN), and evaluate how a mask optimized for one decoder performs when applied to a different decoder.  We refer to these as {\em cross-performances}, and the results are shown in Table \ref{tab:table_2D}. \rev{Here we report both the PSNR (top) and SSIM values (bottom), but the training optimizes only the PSNR; see Section IV-E for training with respect to the SSIM.}

Once again, the learning-based approach outperforms the baselines by approximately 2.5-3.5 dB for all decoders considered.  We observe that the greedy method always finds the best performing mask for \secondrev{the given reconstruction algorithms that we use. The performance drop is typically small when the masks optimized for other decoders are used.}  In Figure \ref{fig:MRI_recon_table}, we further illustrate these observations with a single slice from the test data, showing the masks and reconstructions along with their PSNR and SSIM values. \rev{ In this figure, we also compare to a pure low-pass mask given in the bottom row. It can be seen that the greedy masks outperform the the low pass mask as well, in terms of PSNR, SSIM and also visual quality, as they offer sharper images with less aliasing artefacts by balancing between low and high frequency components. On the other hand, as can be seen from the zoomed-in regions, the pure low-pass mask introduces strong blurring, whereas the other baseline masks (coherence-based and single image) cause highly visible aliasing due to suboptimal sampling across low to intermediate frequencies.

}

\rev{Computation times for the greedy mask optimization on a parallel computing cluster depend strongly on the reconstruction algorithm in use, and are as follows for a mask of $ 25 \% $ sampling rate using MATLAB's Parallel Computing Toolbox with 256 CPU nodes: (TV) 2 hours and 41 minutes; (BP with shearlets) 3 hours and 23 minutes;  (BM3D) 5 hours and 24 minutes.  The coherence-based algorithm takes 10 seconds on 256 nodes. The single-image based adaptive algorithm is quite fast, running in 2 seconds on a single node. For the NN decoder, the greedy algorithm takes 2 hours and 19 minutes on 40 GPU nodes using multiprocessing package of Python. 
Note that these computations for mask selection are carried out {\em offline}, and therefore, we contend that the longer computation time for the greedy mask selection should not be considered a critical issue.}

\subsection{Comparison of greedy and parametric methods}

We now perform an experiment comparing the greedy approach (Algorithm \ref{alg:1}) and the parametric approach with learning (Algorithm \ref{alg:2}).  In contrast with the previous experiments, we consider measurements in the 2D Fourier space along both horizontal and vertical lines.  As described in \cite{wang2009pseudo}, this is done via a pulse sequence program that switches between {\em phase encoding} and {\em frequency encoding}, and can provide improvements over the approach of using only horizontal lines.

\begin{table}
\centering
\caption{\label{tab:table_comparison} PSNR and SSIM performances at 25\% subsampling rate averaged on 60 test slices.}
\begin{tabular}{|l|||*{4}{c|}}\hline
\backslashbox{Mask}{Rule}
&\makebox[3em]{TV-PSNR}&\makebox[3em]{\rev{TV-SSIM}}&\makebox[3em]{BP-PSNR} &\makebox[3em]{\rev{BP-SSIM}} \\\hline \hline 
parametric &  35.02   &   0.913 & 36.00 & 0.926 \\\hline  
greedy &  35.89  &   0.927 & 37.37 & 0.942  \\\hline 
\end{tabular}
\end{table}

\begin{figure} 
\centering
\begin{tabular}{ccc}
& \hspace{-5mm} \textbf{Mask} & \hspace{-5mm} \textbf{Reconstruction}  \\
\rotatebox{90}{\hspace{5mm} \textbf{parametric}} &                         
\hspace{-4mm}\includegraphics[width=.16\textwidth]{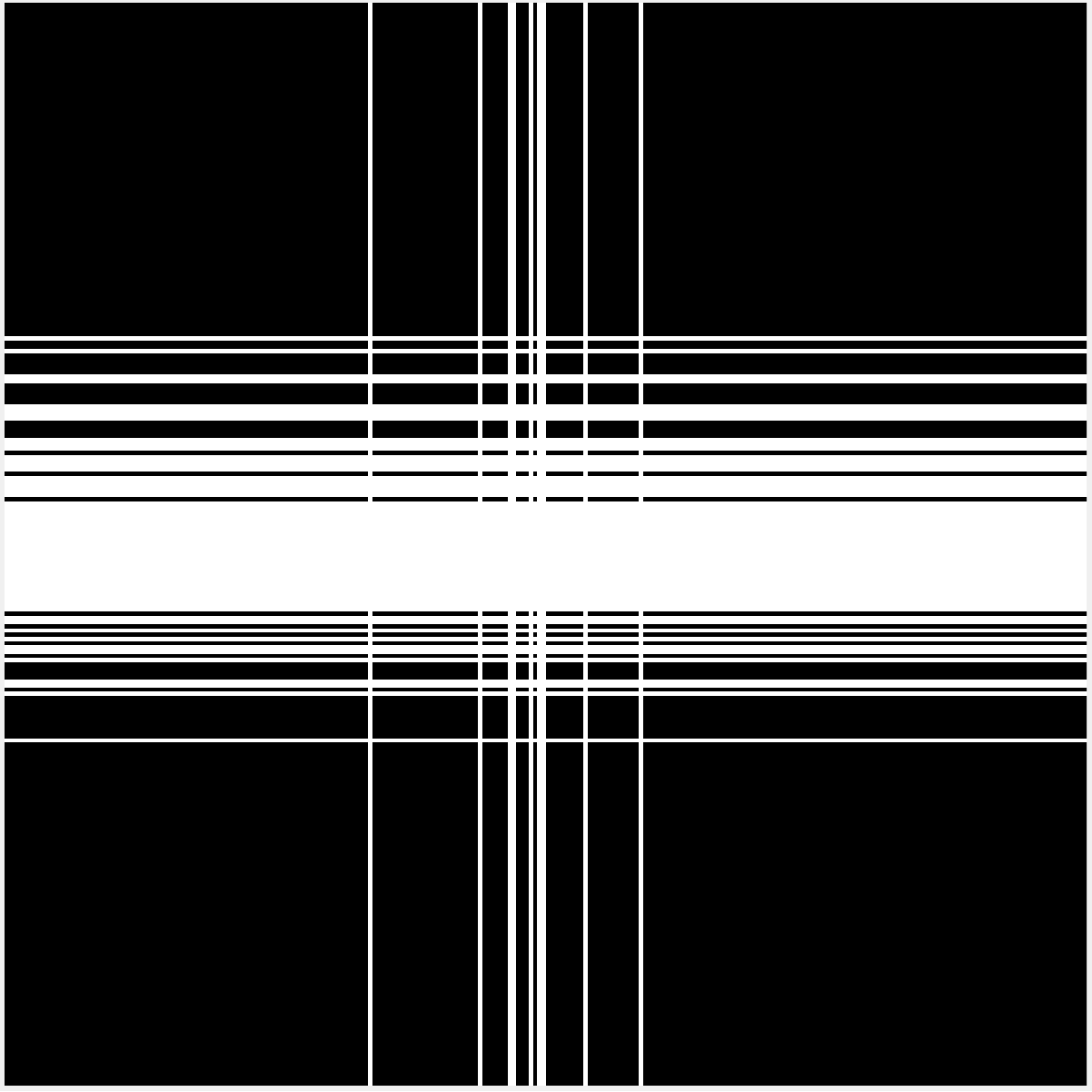} &
\hspace{-4mm}\includegraphics[width=.16\textwidth]{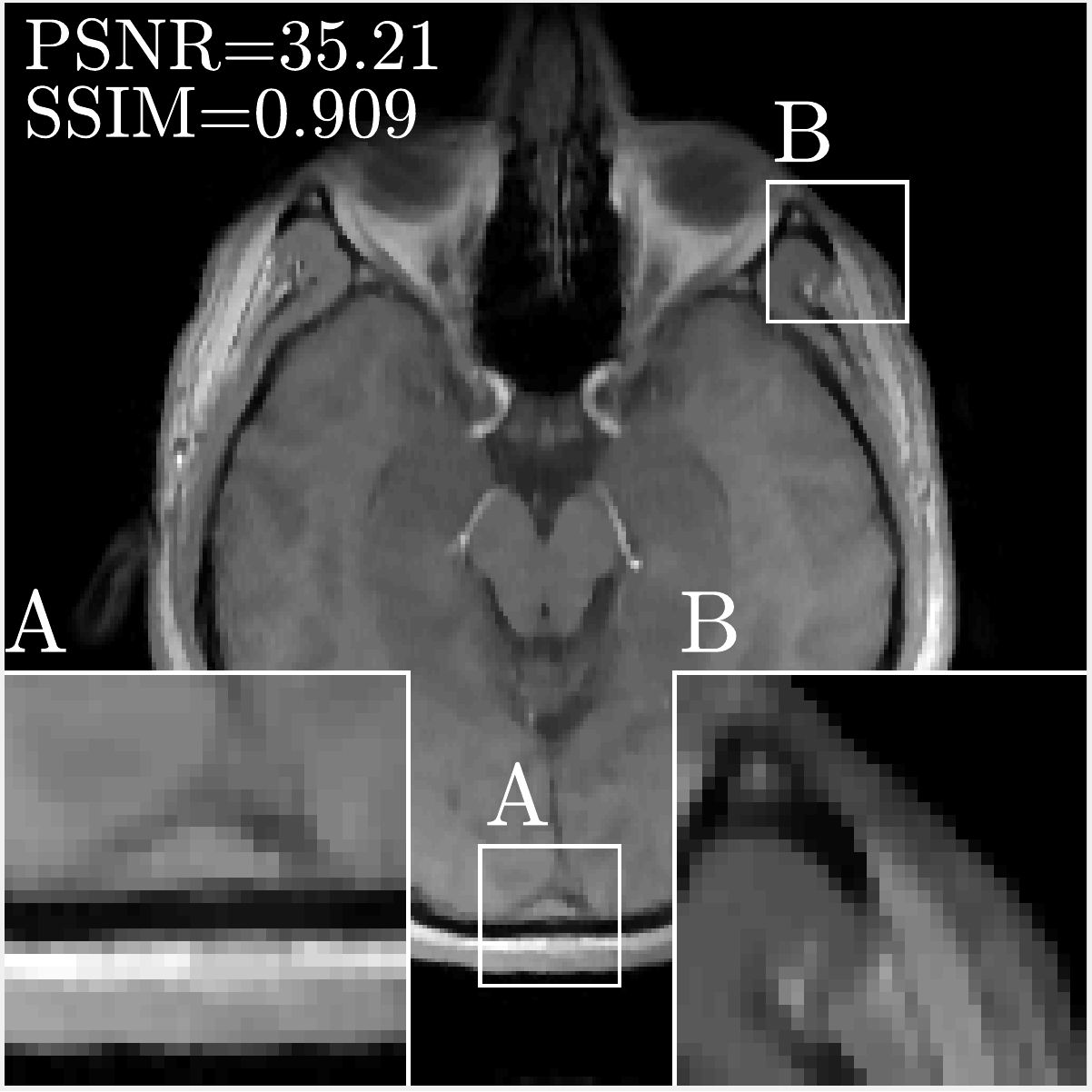}  \\ [-1mm]
\rotatebox{90}{\hspace{9mm} \textbf{greedy}} &
\hspace{-4mm}\includegraphics[width=.16\textwidth]{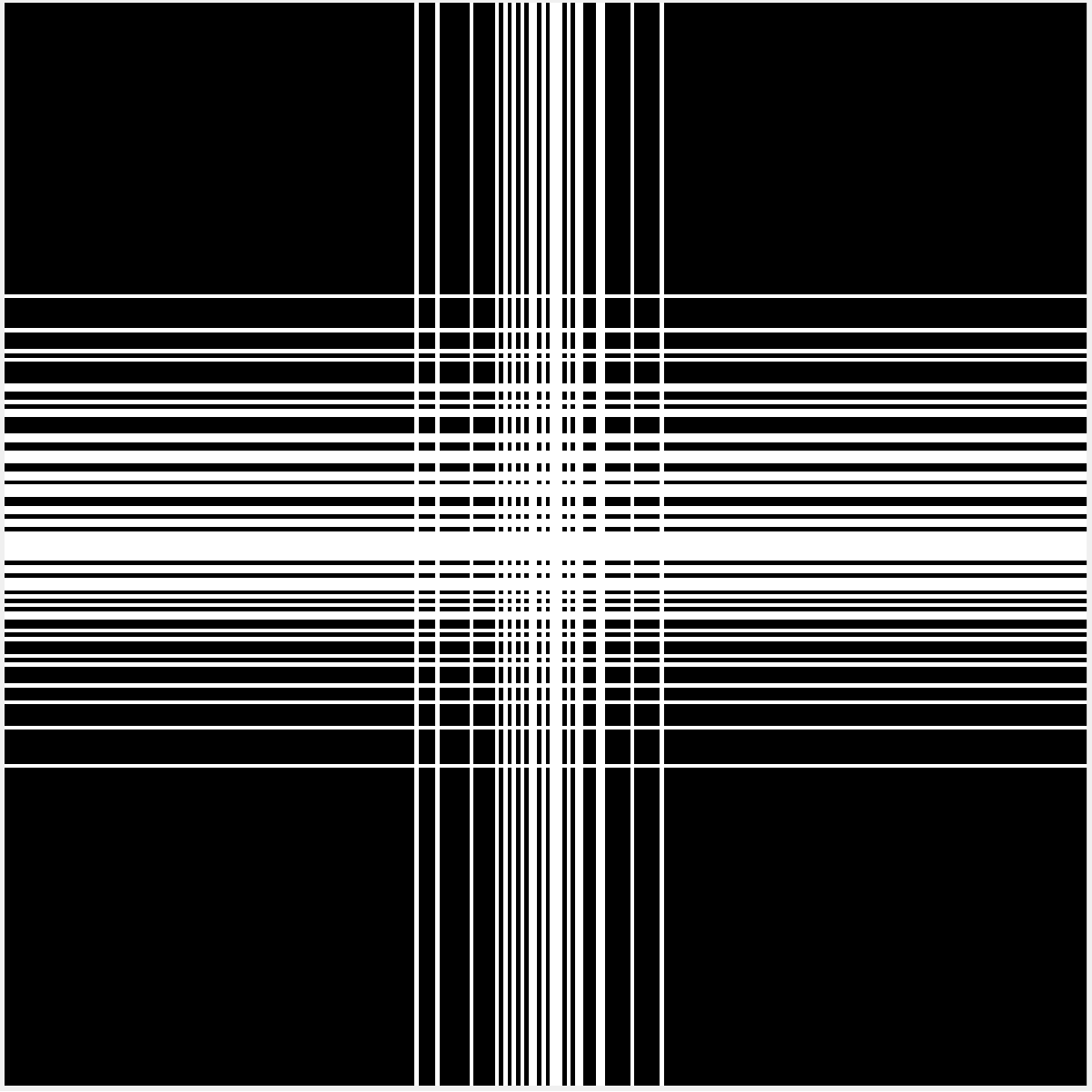} &
\hspace{-4mm}\includegraphics[width=.16\textwidth]{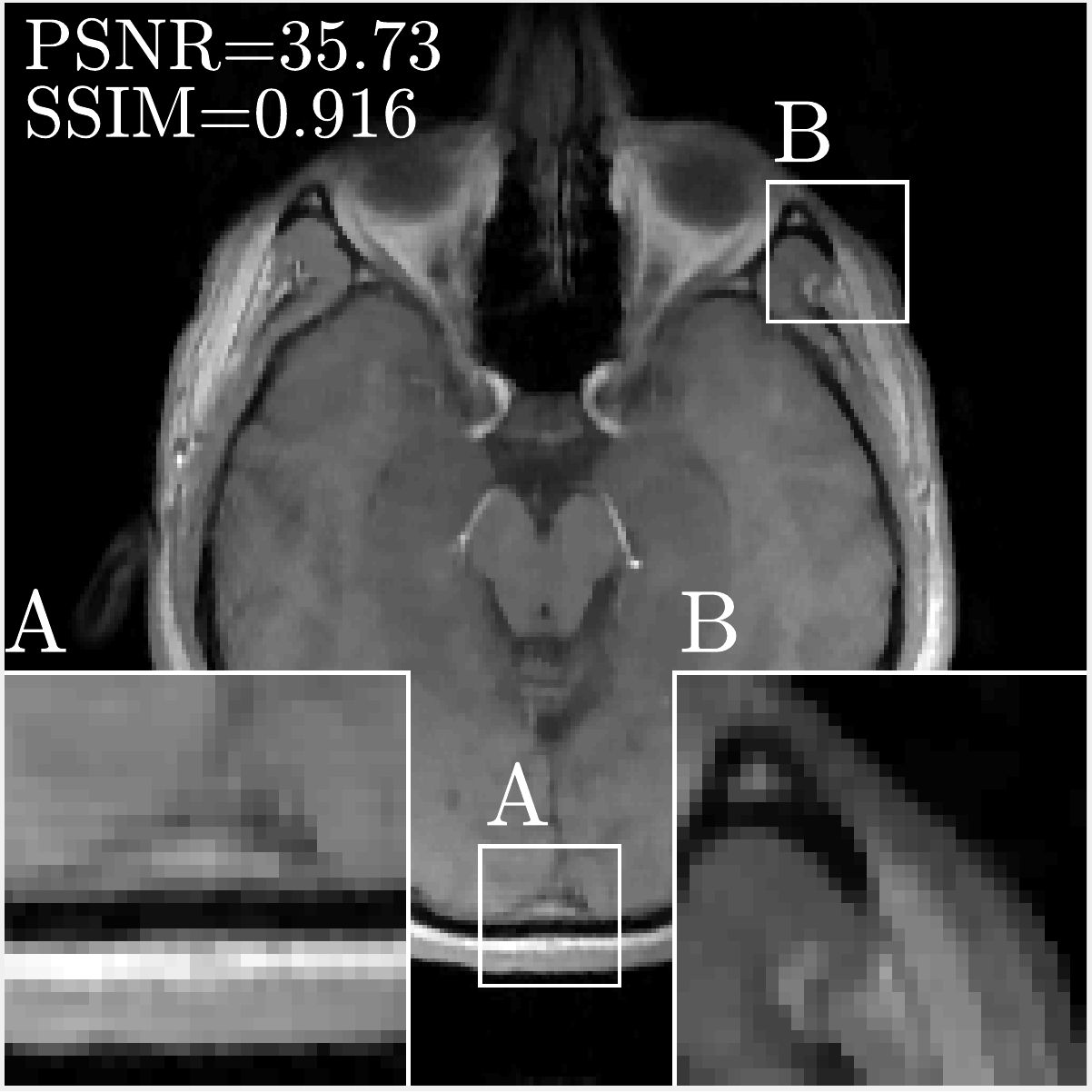}   \\ [-1mm]

\end{tabular}
\caption{Masks obtained and example reconstructions under TV decoding at 25\% sampling rate, for the parametric method of \cite{wang2009pseudo} combined with Algorithm \ref{alg:2}, and the greedy method given in Algorithm \ref{alg:1}.  Both horizontal and vertical lines are permitted. The reconstruction shown is for subject 2, slice 4. \label{fig:TV_reconstructions}}
\end{figure}

\begin{figure} 
\centering
\begin{tabular}{ccc}
&\textbf{Mask} &\hspace{-5mm} \textbf{Reconstruction}  \\                       
\rotatebox{90}{\hspace{5mm} \textbf{parametric}} &  
\hspace{-4mm}\includegraphics[width=.16\textwidth]{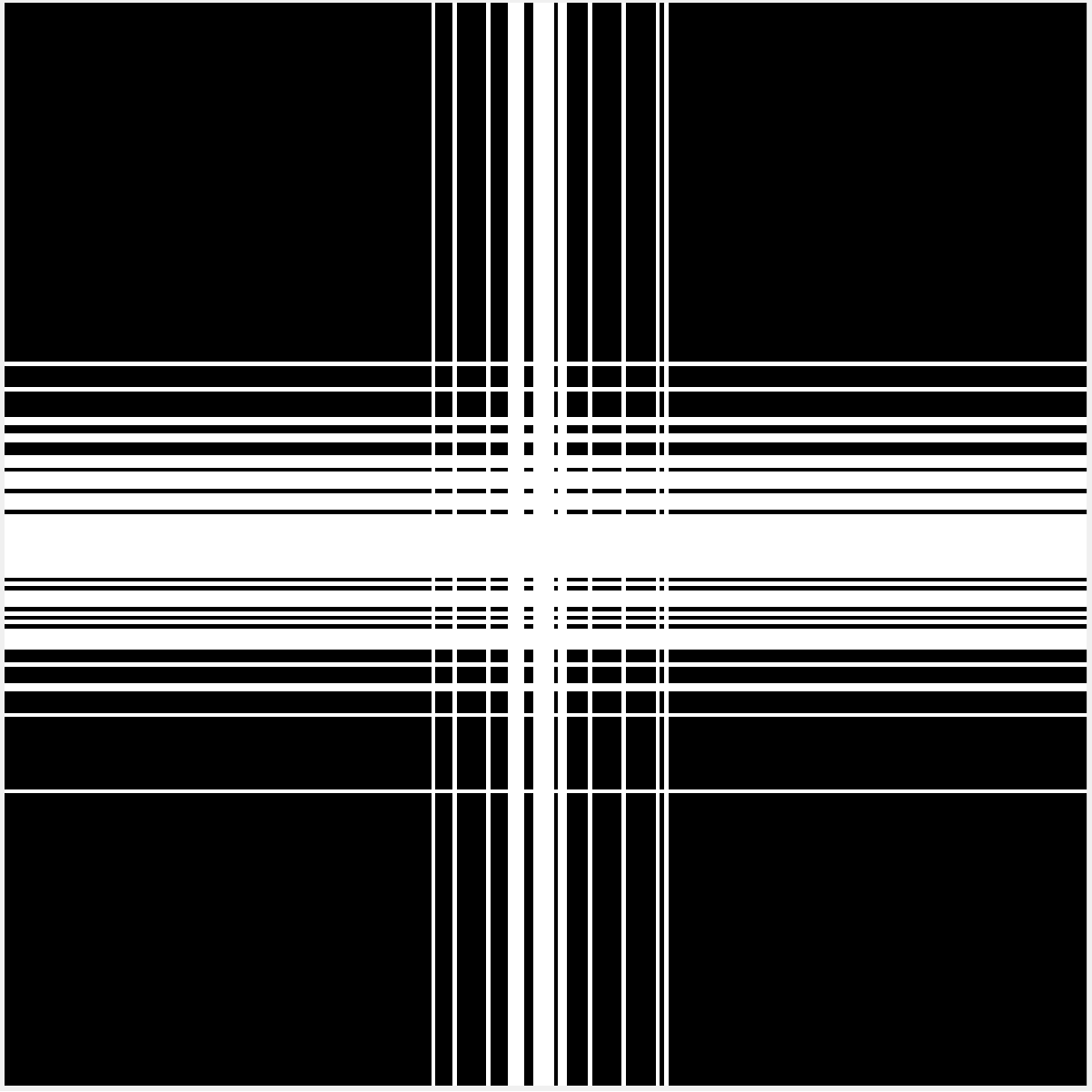} &
\hspace{-4mm}\includegraphics[width=.16\textwidth]{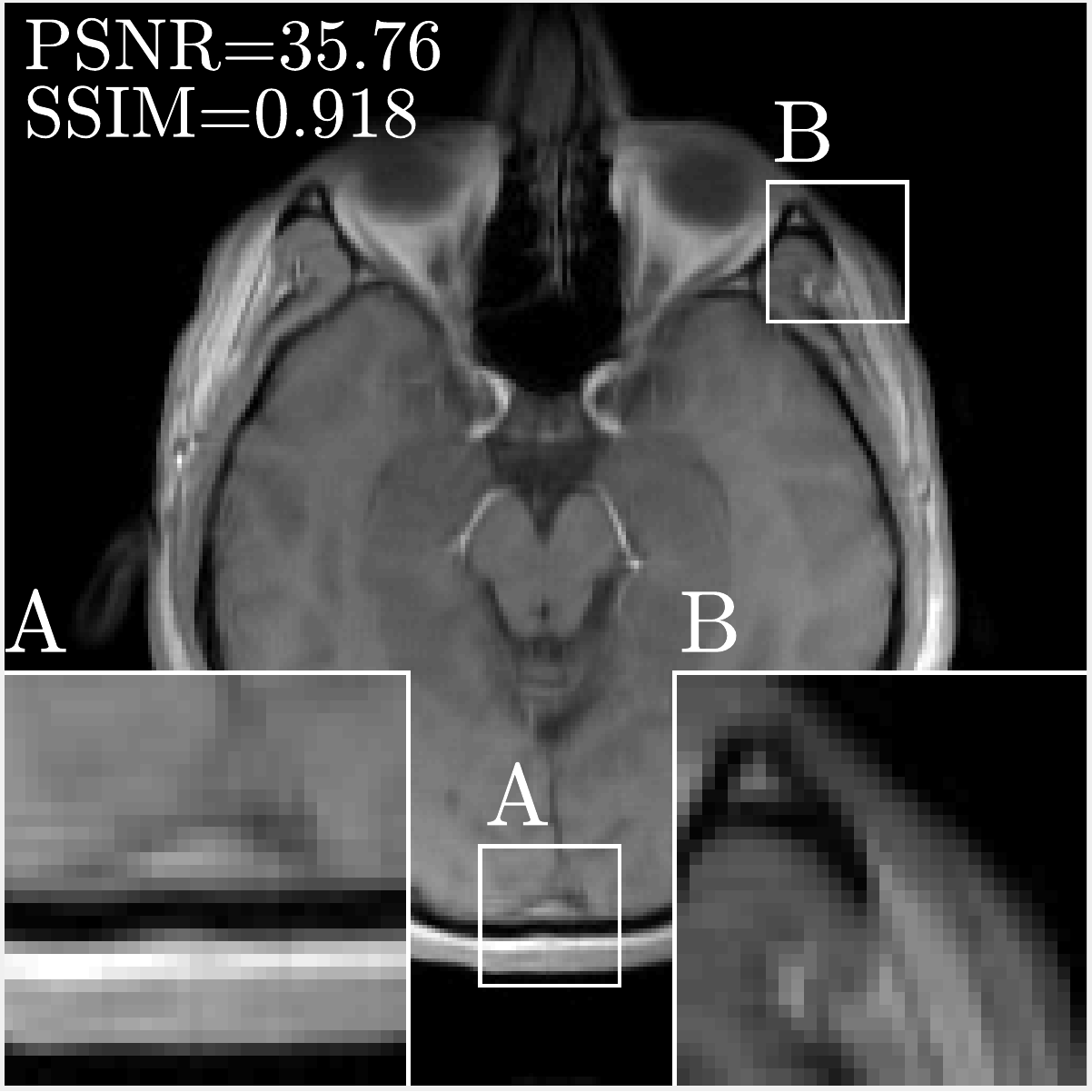}  \\ [-1mm]
\rotatebox{90}{\hspace{9mm} \textbf{greedy}} &
\hspace{-4mm}\includegraphics[width=.16\textwidth]{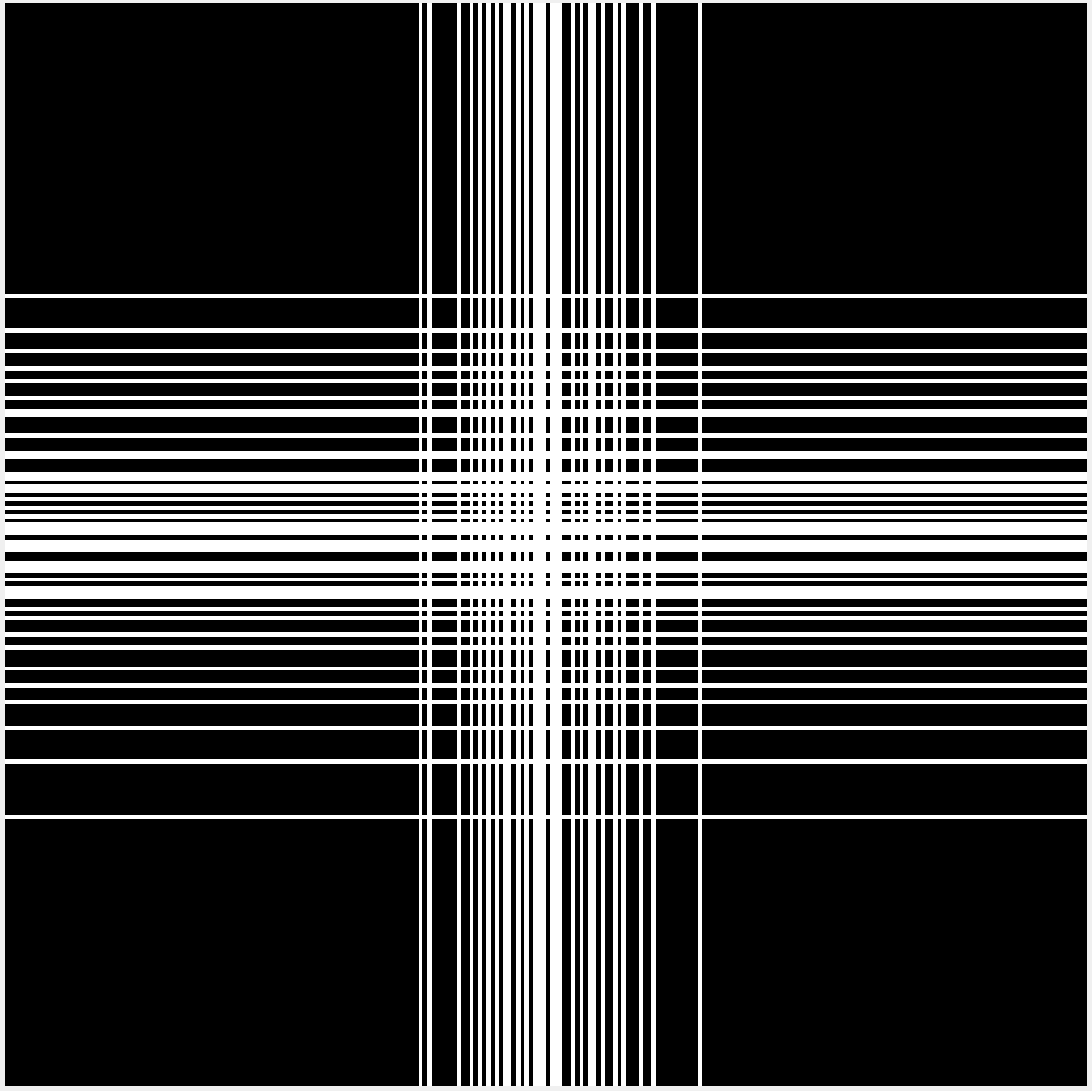} &
\hspace{-4mm}\includegraphics[width=.16\textwidth]{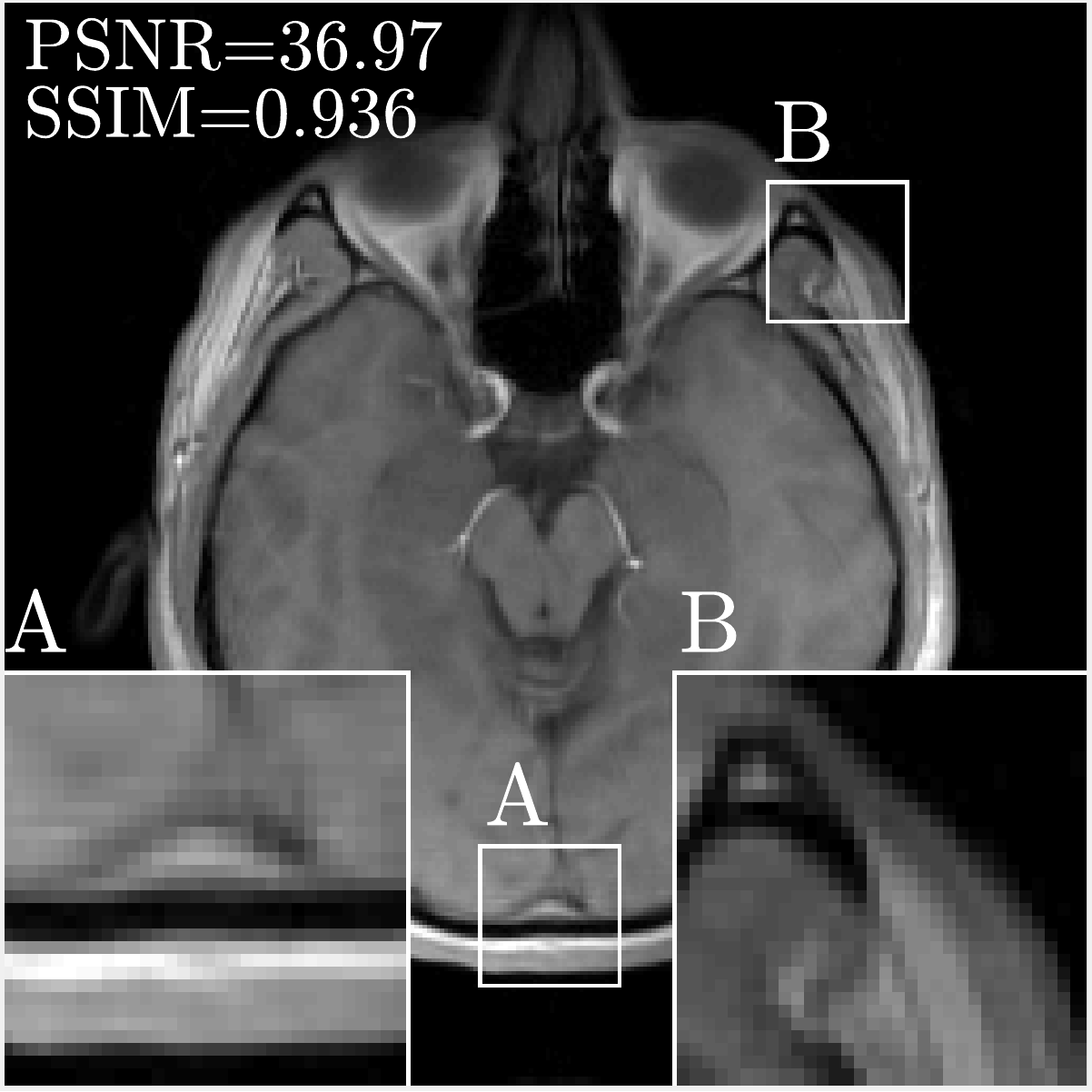}   \\ [-1mm]

\end{tabular}
\caption{Masks obtained and example reconstructions under BP decoding at 25\% sampling rate, for the parametric method of \cite{wang2009pseudo} combined with Algorithm \ref{alg:2}, and the greedy method given in Algorithm \ref{alg:1}. Both horizontal and vertical lines are permitted.  The reconstruction shown is for subject 2, slice 4. \label{fig:BP_reconstructions}}
\end{figure}

We tune the parameters of \cite{wang2009pseudo} on the training data using Algorithm \ref{alg:2}. The first two of the three parameters are $d_x$ and $d_y$, which are the sizes of the fully sampled central regions in horizontal and vertical directions. We sweep this for $d_x, d_y \in \{2, 4, \ldots, d_{\max}\}$ where $d_{\max}$ is maximum feasible fully sampled region size for a given subsampling rate. The last parameter $D$ is the degree of the polynomial that defines the probability distribution function from which random masks are drawn. We sweep over $D \in \{1,3,5, \ldots, 13\}$. We then randomly draw 5 masks for each choice of parameters, and we use the mask that gives the best average PSNR on the training data, as per Algorithm \ref{alg:2}.

As seen in Table \ref{tab:table_comparison}, the greedy approach outperforms the parametric approach for both the TV and BP reconstruction algorithms.  Interestingly, the masks obtained are also visually rather different (\emph{cf.}, Figures \ref{fig:TV_reconstructions} and \ref{fig:BP_reconstructions}, which also show the reconstructions for a single slice), with the greedy masks being more ``spread'' rather than taking a continuum of rows at low frequencies. \rev{ It can be also noticed that both methods choose more horizontal lines than vertical lines, due to the fact the the energy in $k$-space is distributed relatively more broadly across the horizontal direction, as can be seen on the top-right corner of Figure \ref{fig:MRI_recon_table}. }

%
%
%

\subsection{Cross-performances of performance measures}

In the previous experiments, we focused on the PSNR performance measure.  Here we show that considering different measures can lead to different optimized masks, and that it is important to learn a pattern targeted to the correct performance measure.  Specifically, we consider both the PSNR and the structural similarity index (SSIM) \cite{wang2004image}.  Also different from the previous experiments, we use the data set of angiographic brain scans instead of T1-weighted scans (see Section \ref{sec:exp_setup} for details).  We return to the method of taking horizontal lines only in the sampling pattern.

\rev{Table \ref{tab:cross_recon} gives the PSNR and SSIM performances for the TV and BP decoders, under the masks obtained via the greedy algorithm (\emph{cf.}, Algorithm \ref{alg:1}) with the two different performance measures and the decoders at 30\% sampling rate. These results highlight the fact that certain decoders are often better suited to certain performance measures.  Here, TV is suited to the PSNR measure, as both tend to prefer concentrating the sampling pattern at low frequencies, whereas BP is better suited to SSIM, with both preferring a relatively higher proportion of high frequencies. Note also that in some columns, the performance is not highest on the rows where the training is matched to the decoder and the performance measure (shown in bold), but slightly lower than the highest values, which is most likely either due to limited training data or the suboptimality of the greedy algorithm. 
}

\begin{figure}[t] 
\centering
\begin{tabular}{cccc}
& \hspace{-5mm} \textbf{Mask} & \hspace{-5mm} \textbf{Reconstruction} & \hspace{-6mm} \textbf{\rev{MIP}}   \\
\hspace{-7mm} \rotatebox{90}{\hspace{6mm} \textbf{\rev{Original}}} &                         
\hspace{-5mm}\includegraphics[width=.16\textwidth]{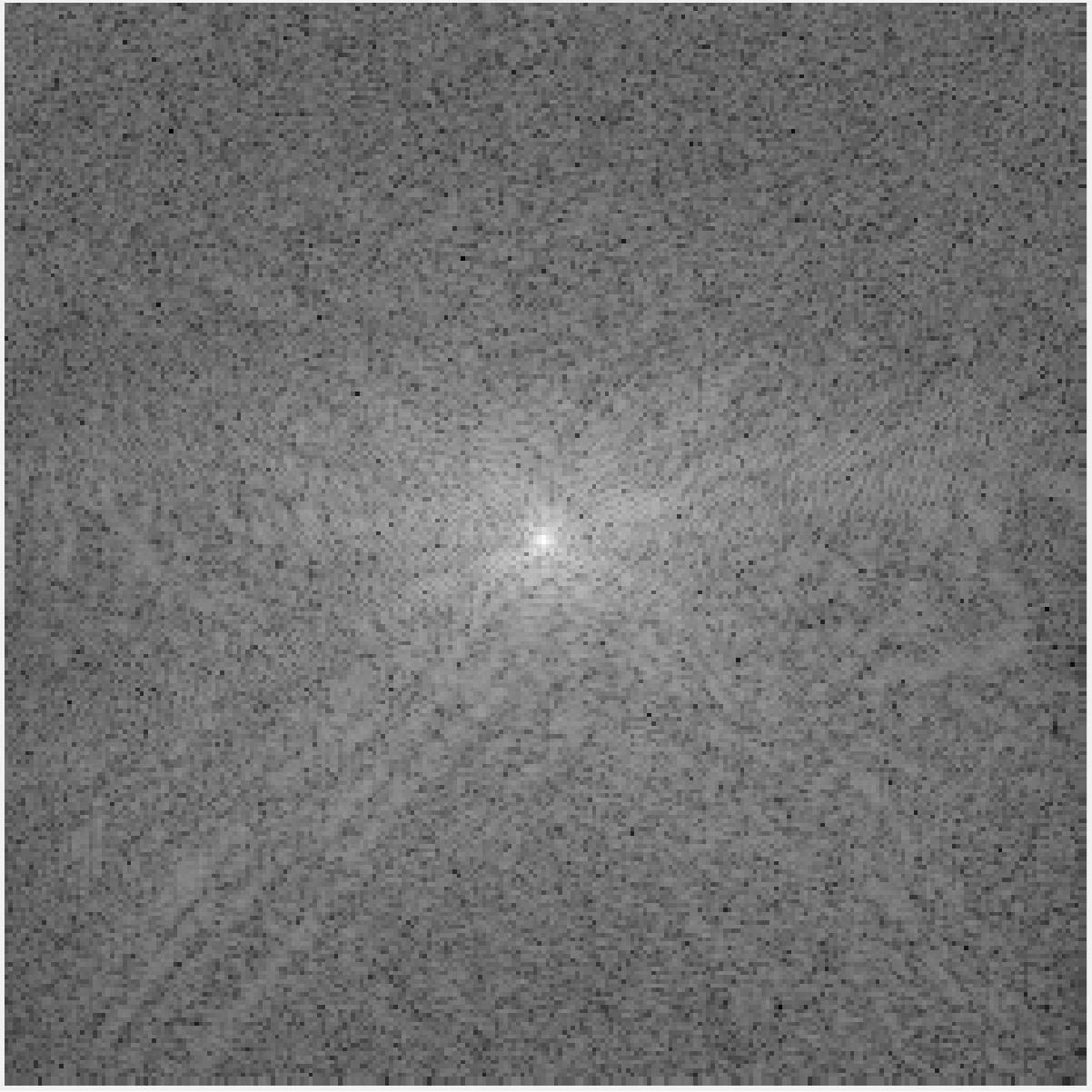} &
\hspace{-4mm}\includegraphics[width=.16\textwidth]{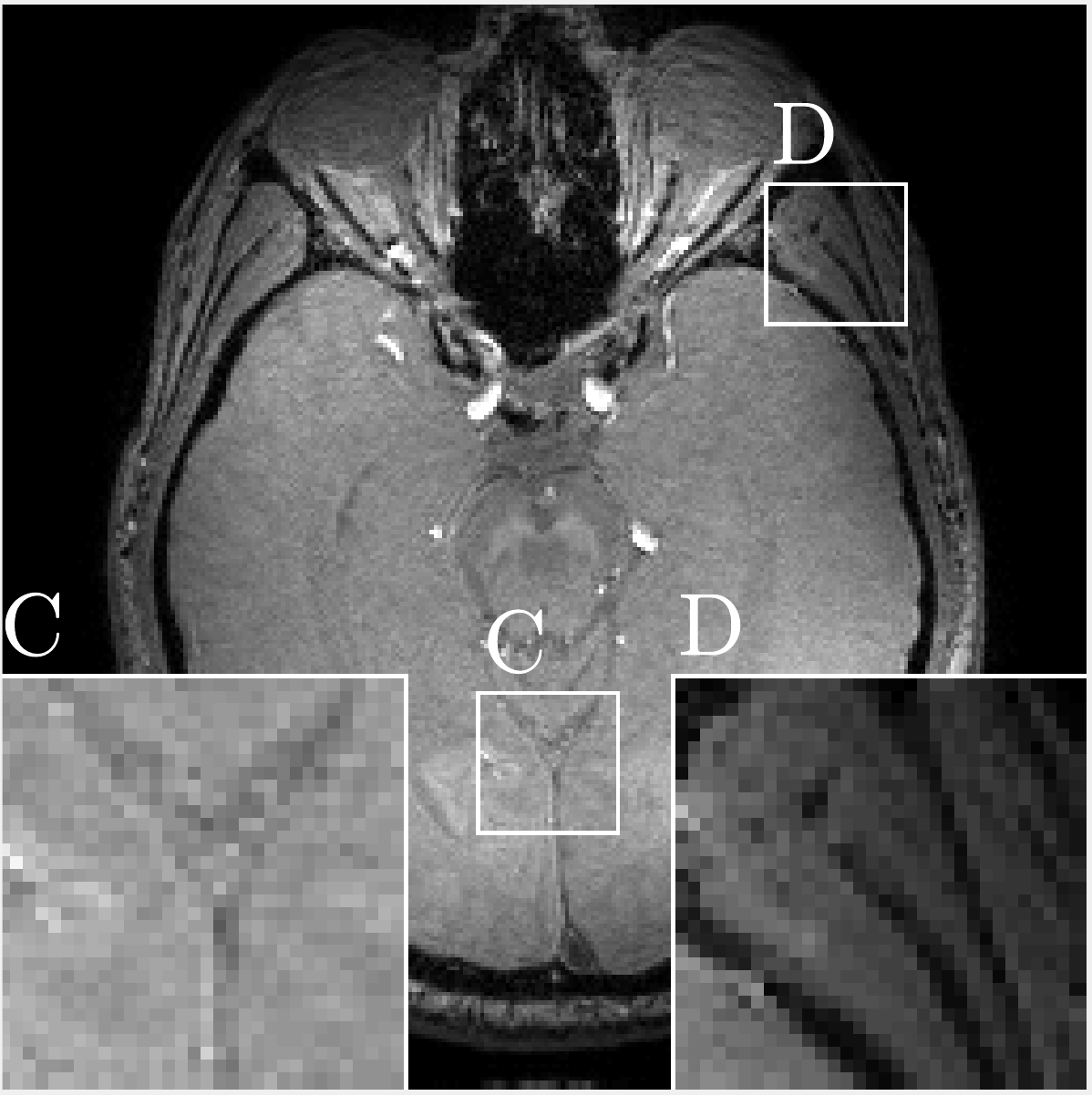}  & 
\hspace{-4mm}\includegraphics[width=.16\textwidth]{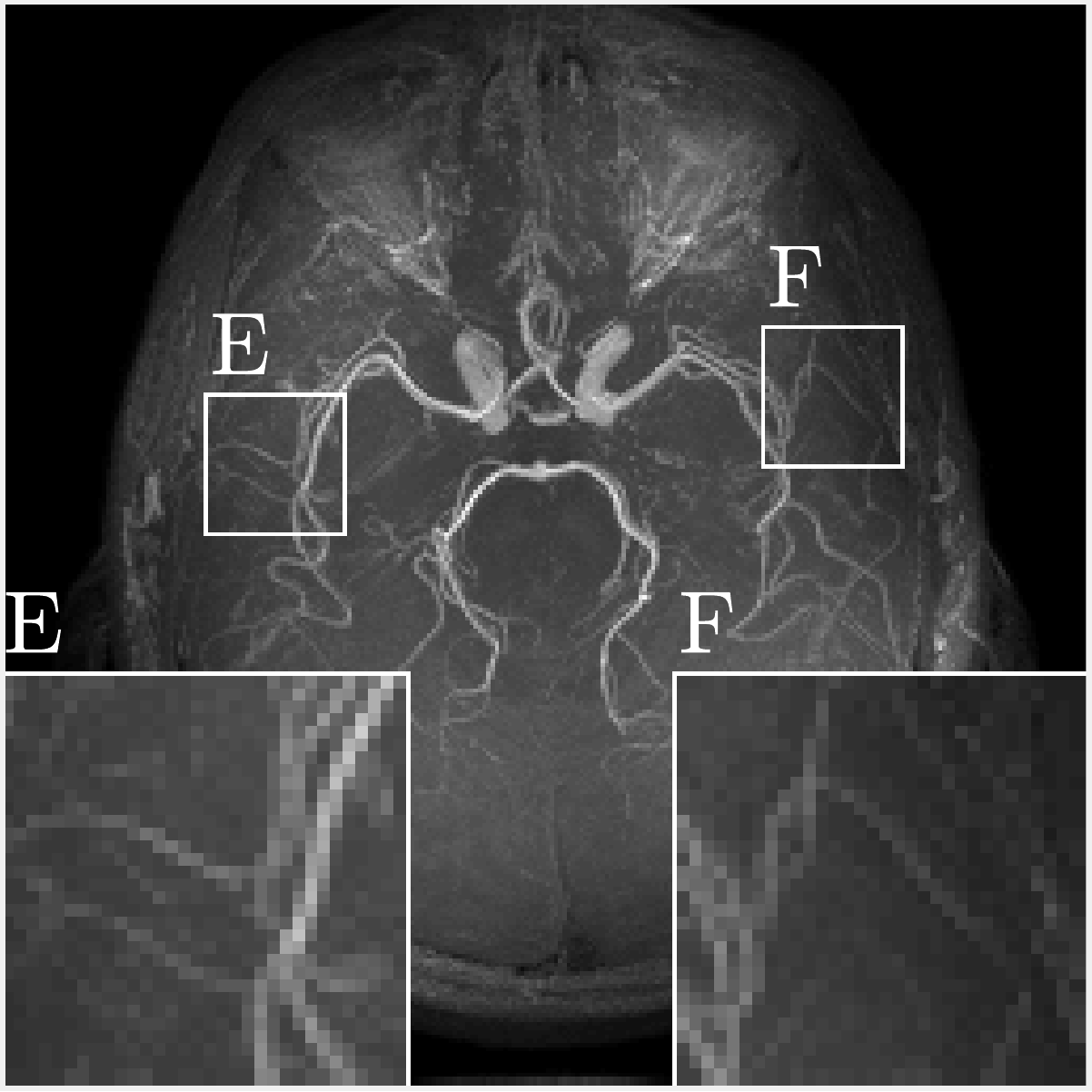}  \\ [-1mm]

\hspace{-7mm} \rotatebox{90}{\hspace{3mm} \textbf{\rev{Coher. based}}} &                         
\hspace{-5mm}\includegraphics[width=.16\textwidth]{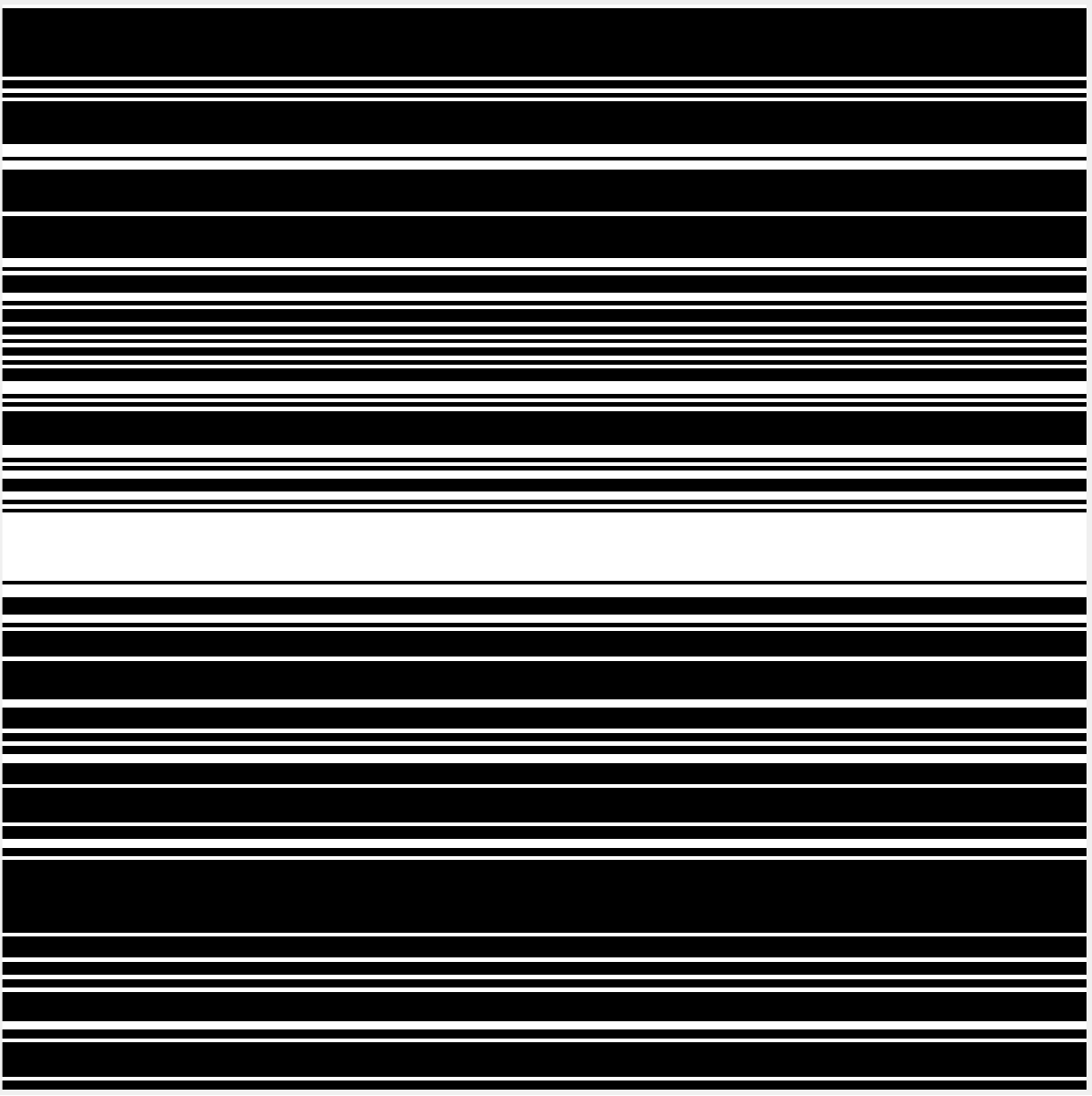} &
\hspace{-4mm}\includegraphics[width=.16\textwidth]{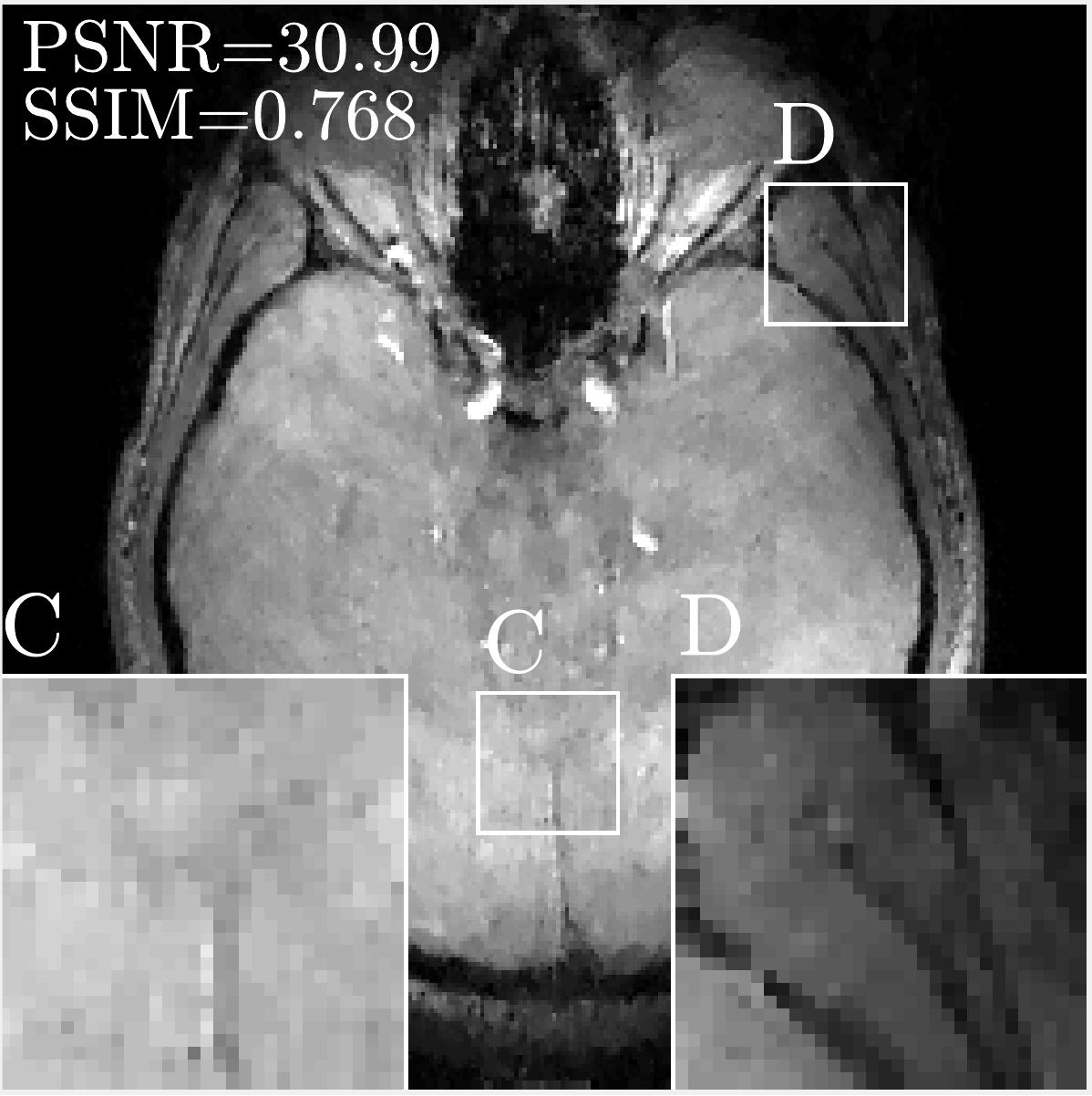}  & 
\hspace{-4mm}\includegraphics[width=.16\textwidth]{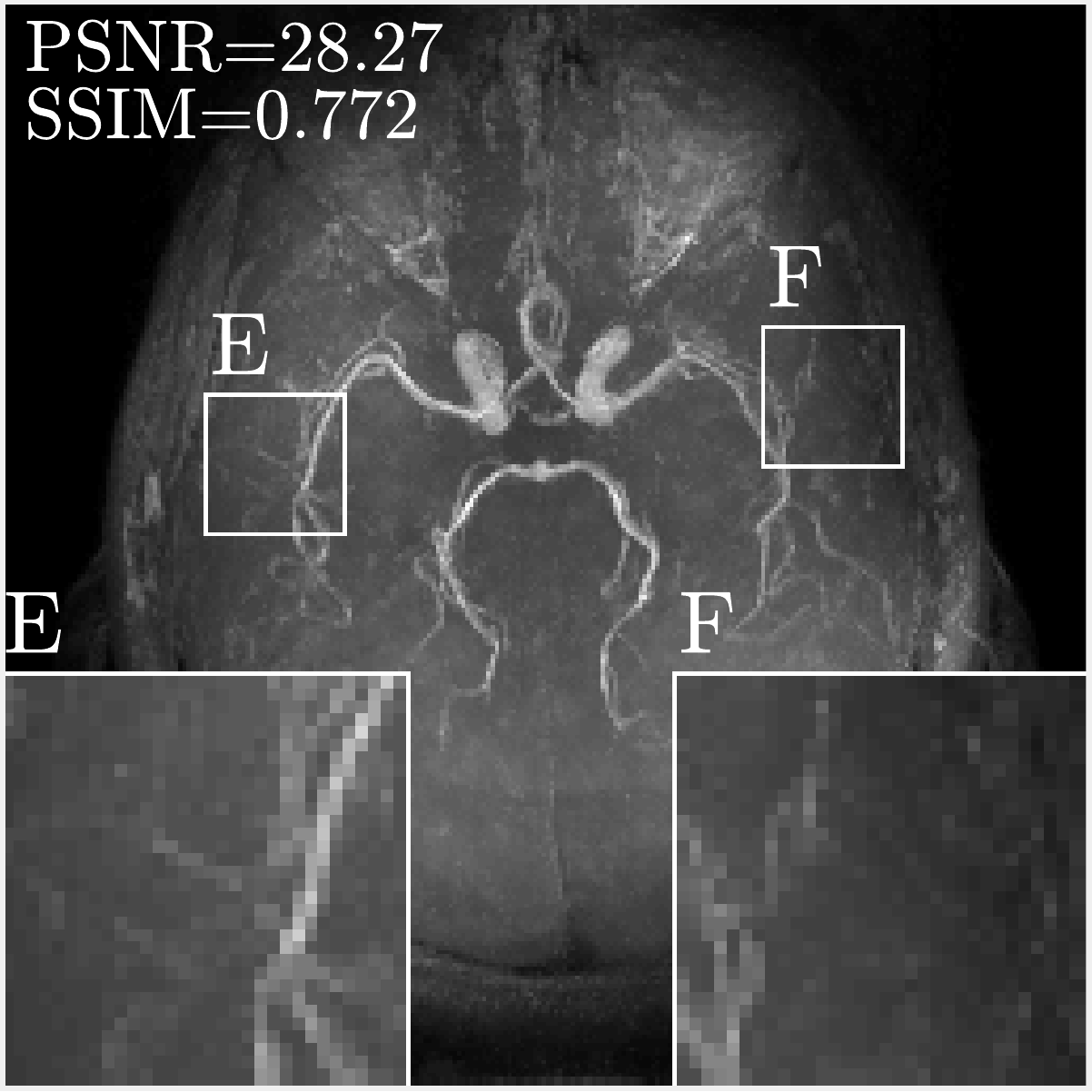}  \\ [-1mm]
\hspace{-7mm} \rotatebox{90}{\hspace{4mm} \textbf{\rev{Single image}}} &                         
\hspace{-5mm}\includegraphics[width=.16\textwidth]{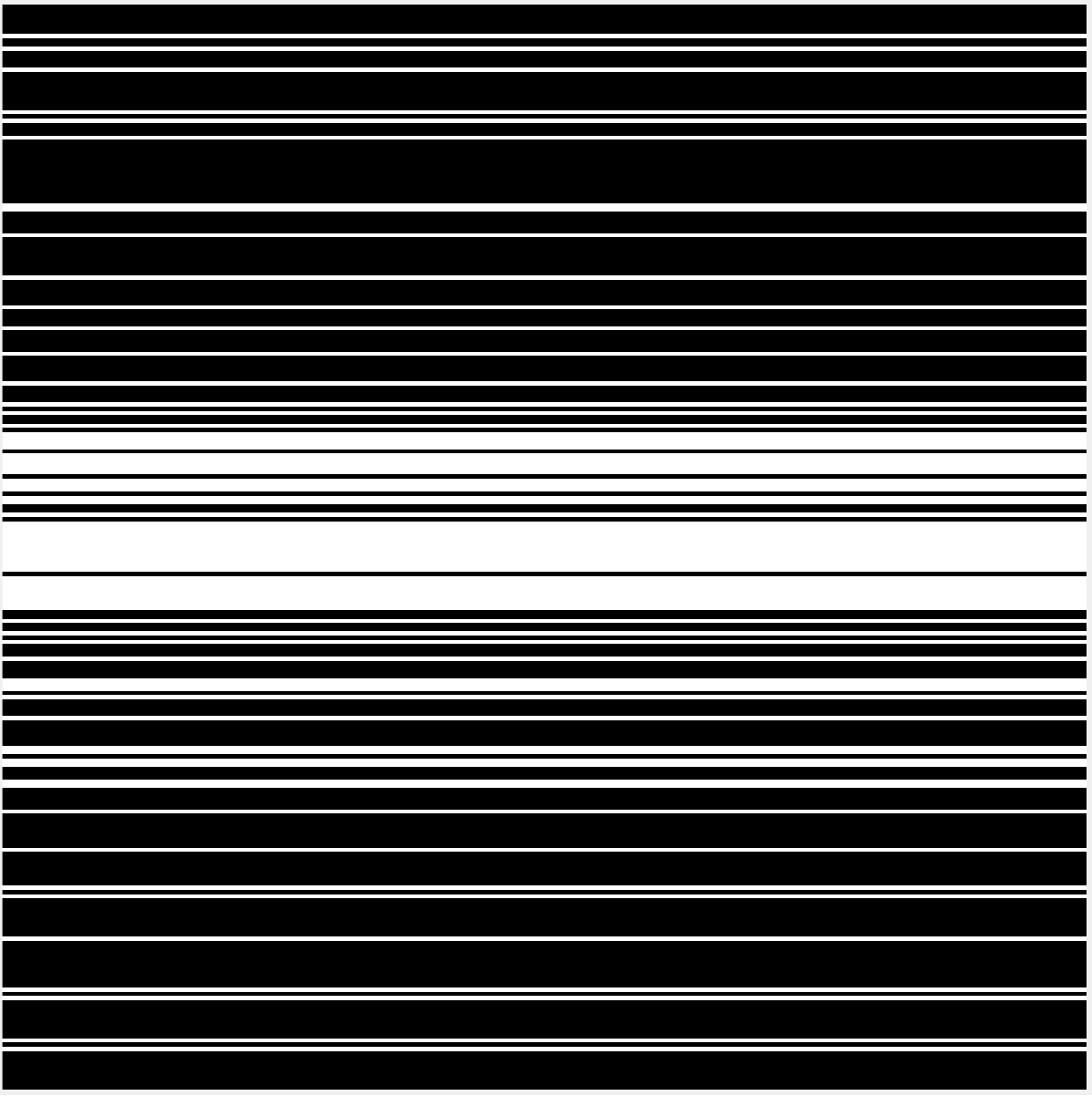} &
\hspace{-4mm}\includegraphics[width=.16\textwidth]{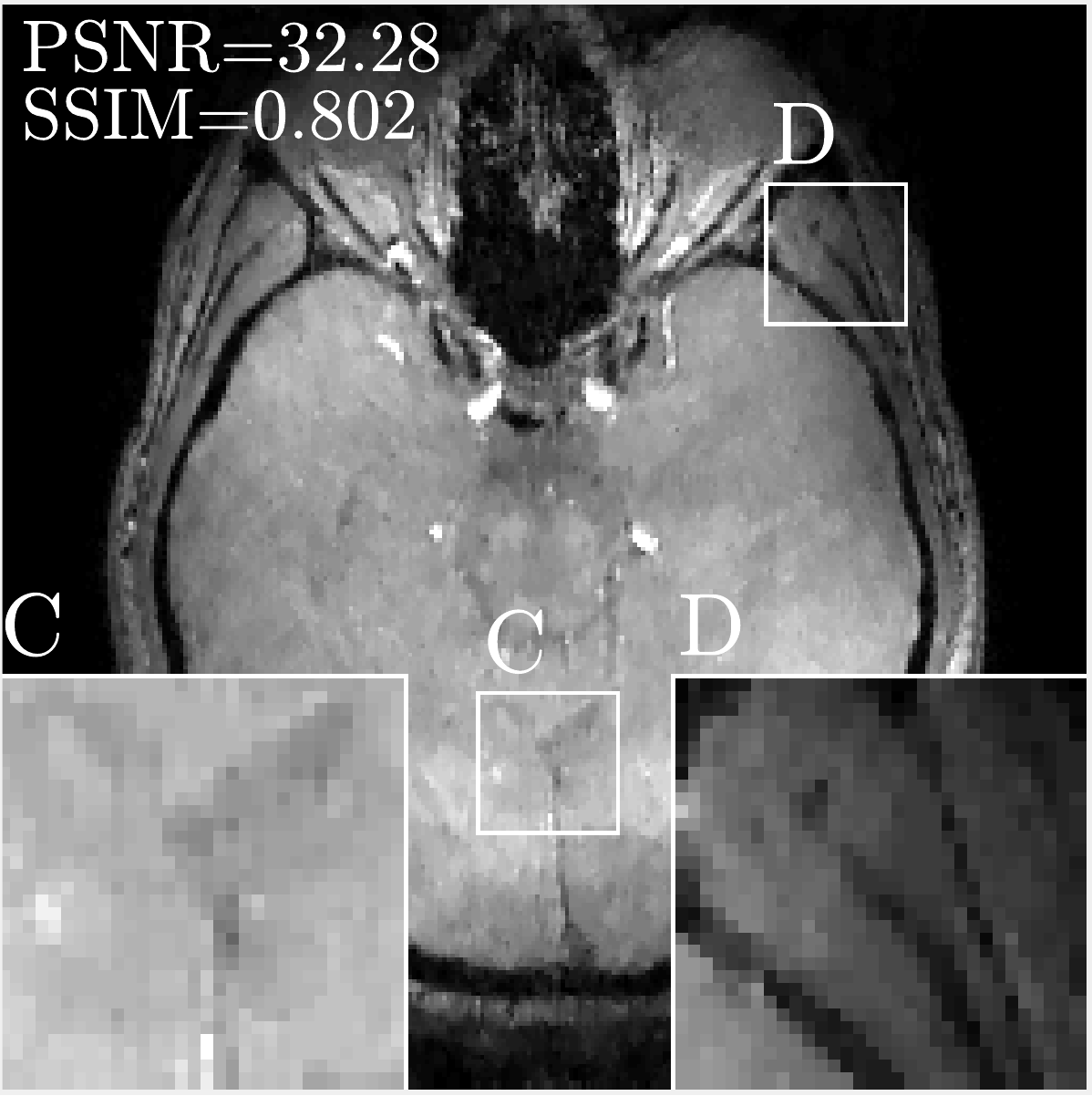}  & 
\hspace{-4mm}\includegraphics[width=.16\textwidth]{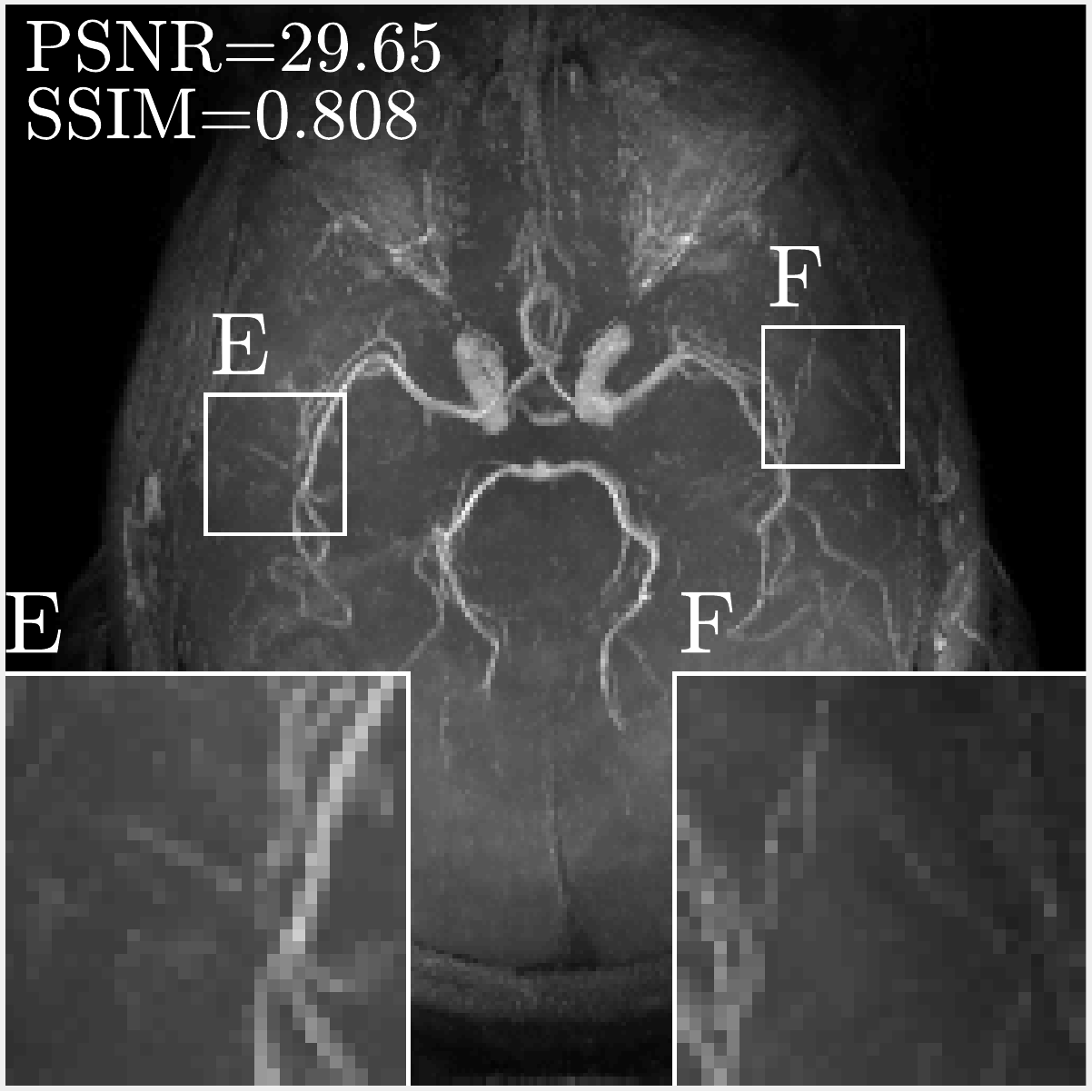} 
 \\ [-1mm]

\hspace{-6mm} \rotatebox{90}{\hspace{4mm} \textbf{SSIM-greedy}} &
\hspace{-5mm}\includegraphics[width=.16\textwidth]{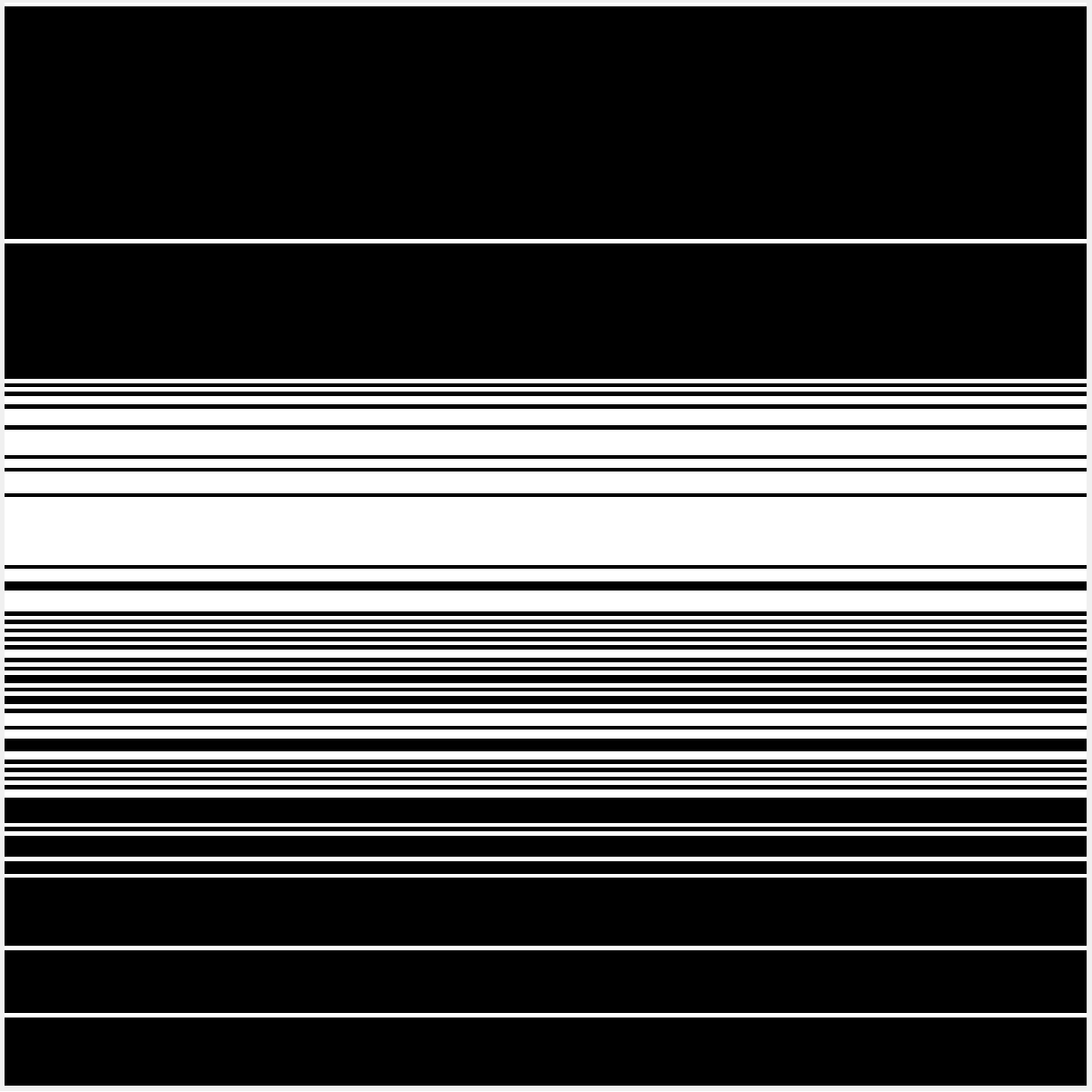} &
\hspace{-4mm}\includegraphics[width=.16\textwidth]{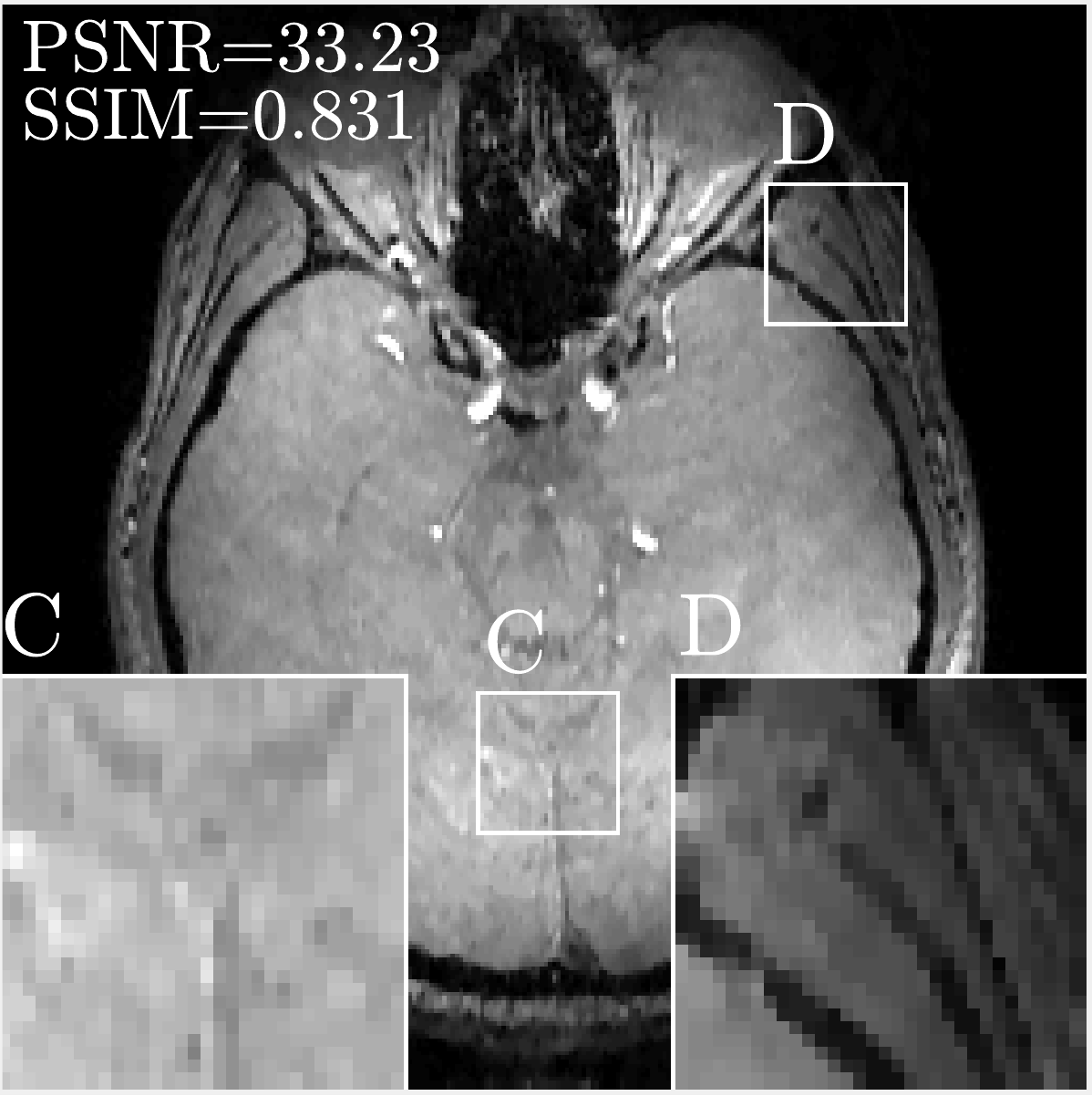} &
\hspace{-4mm}\includegraphics[width=.16\textwidth]{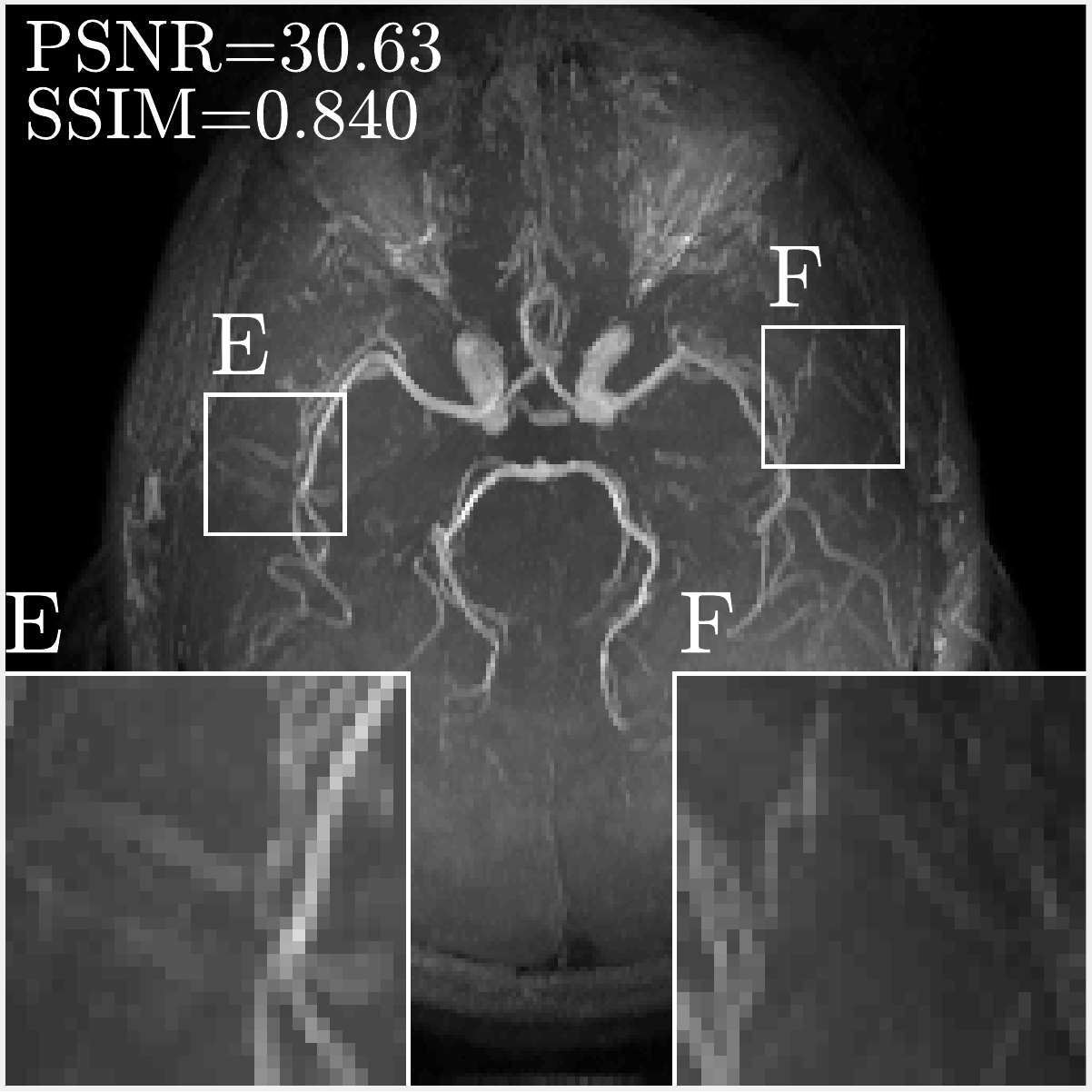}   \\ [-1mm]


\hspace{-5mm}\rotatebox{90}{\hspace{3mm} \textbf{PSNR-greedy}} &                         
\hspace{-5mm}\includegraphics[width=.16\textwidth]{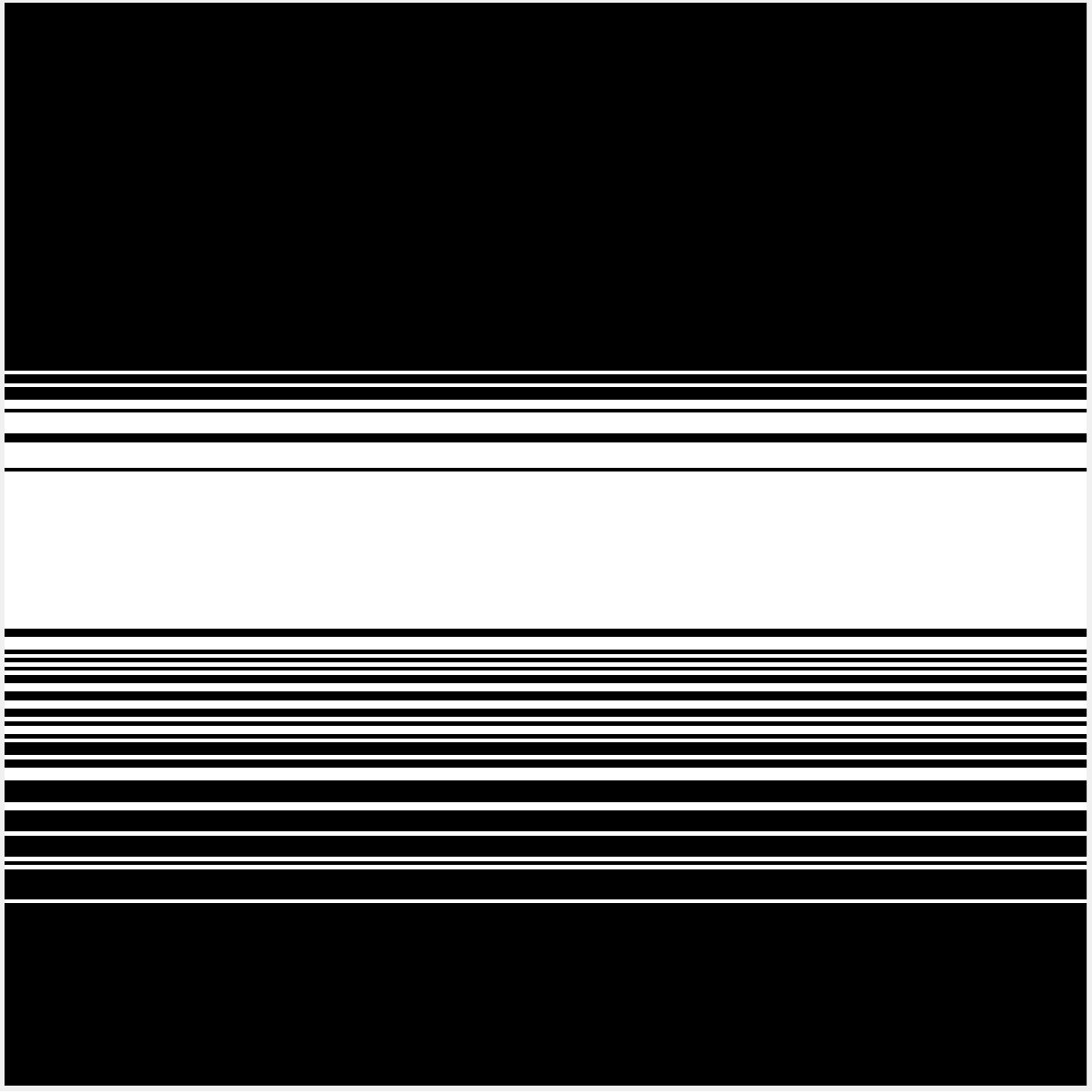} &
\hspace{-4mm}\includegraphics[width=.16\textwidth]{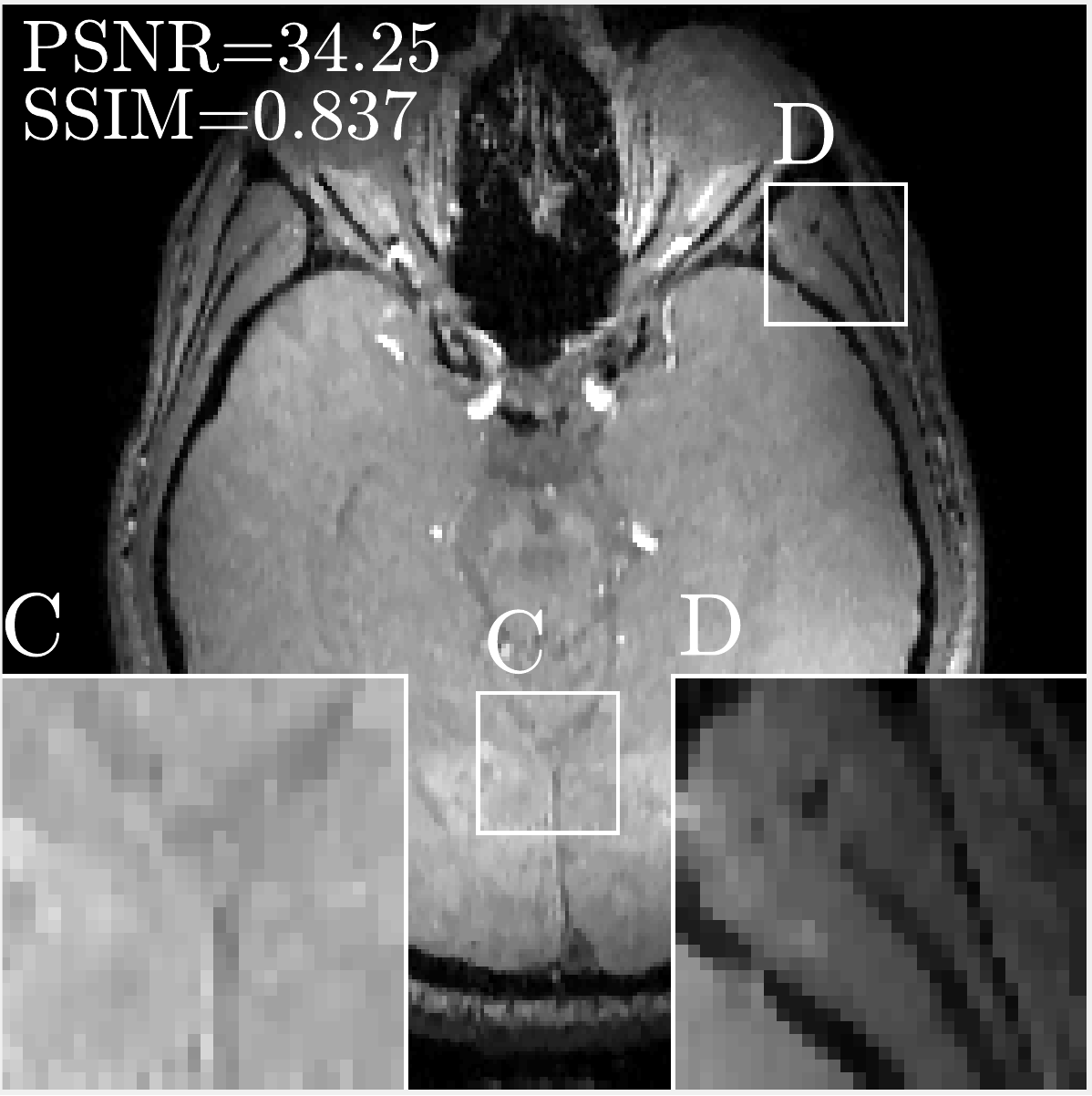}  & 
\hspace{-4mm}\includegraphics[width=.16\textwidth]{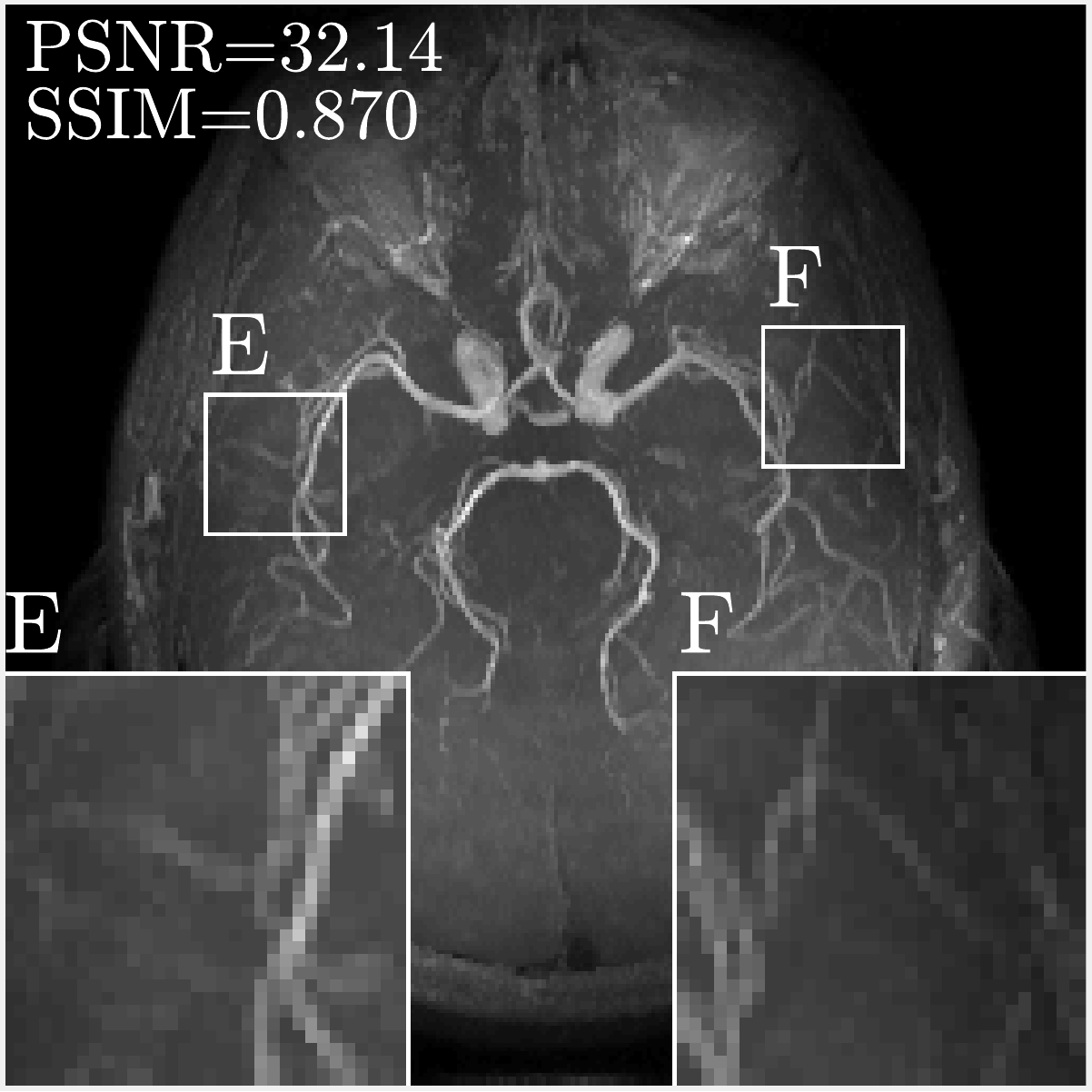}  \\ [-1mm]

\hspace{-5mm}\rotatebox{90}{\hspace{6mm} \textbf{\rev{Low pass}}} &                         
\hspace{-5mm}\includegraphics[width=.16\textwidth]{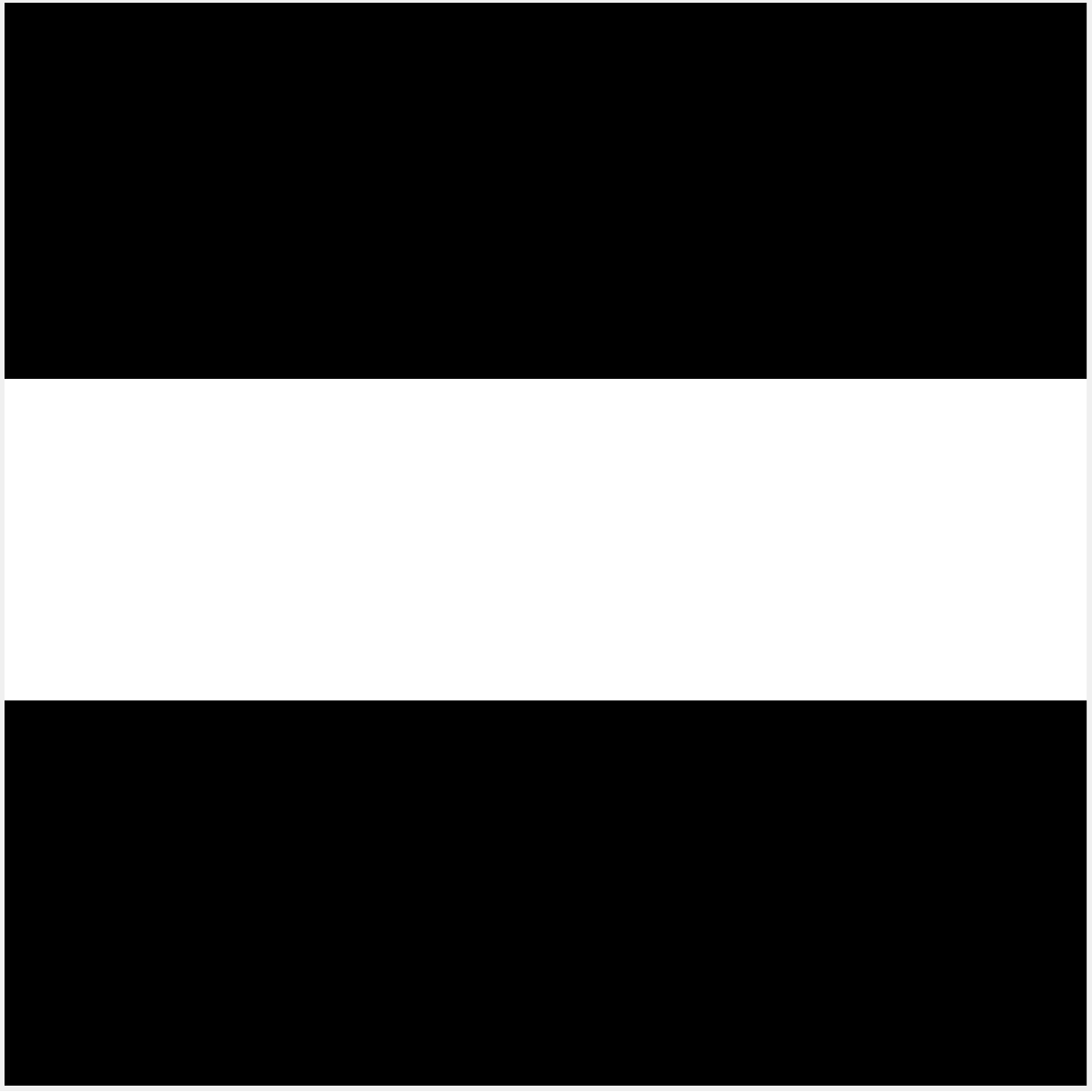} &
\hspace{-4mm}\includegraphics[width=.16\textwidth]{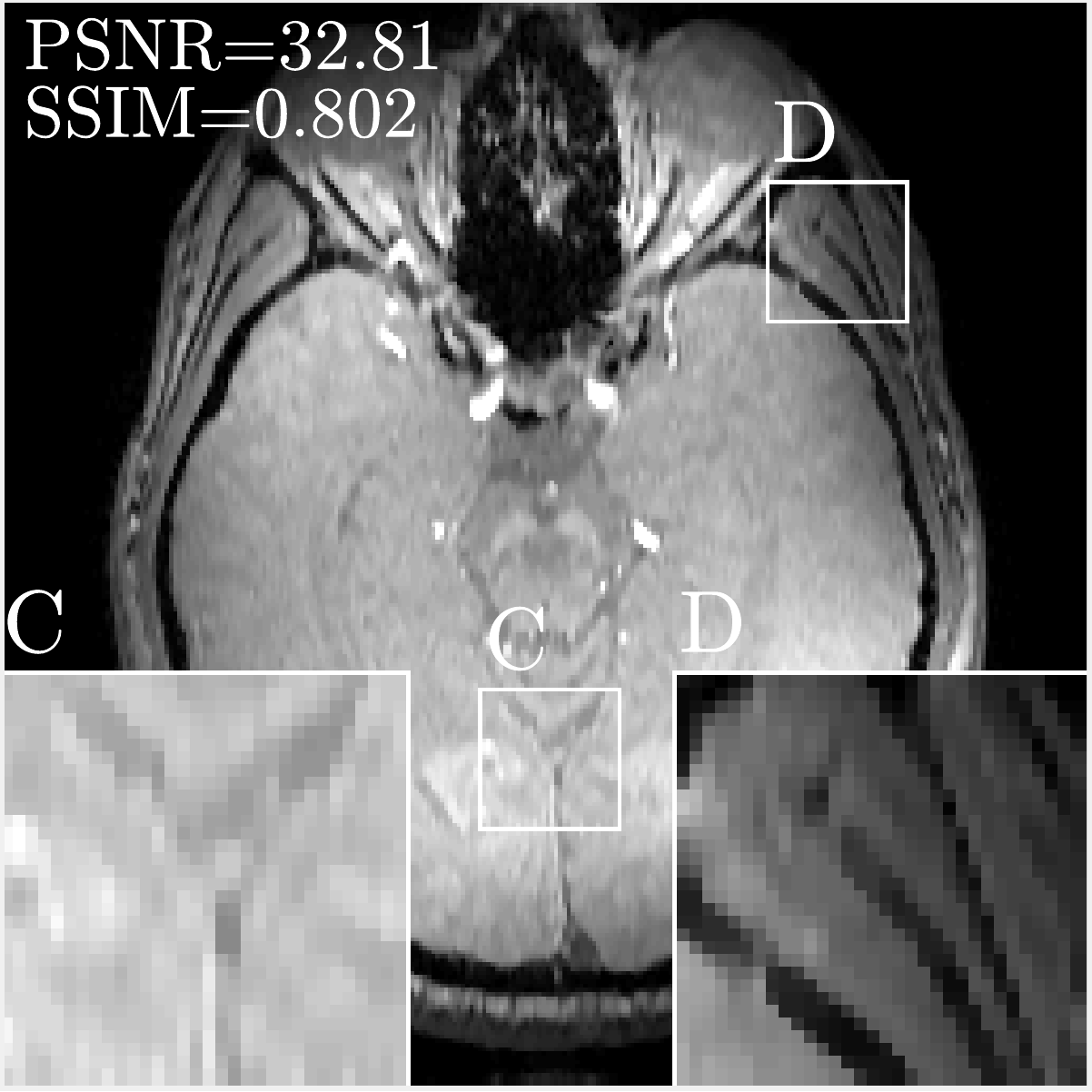}  & 
\hspace{-4mm}\includegraphics[width=.16\textwidth]{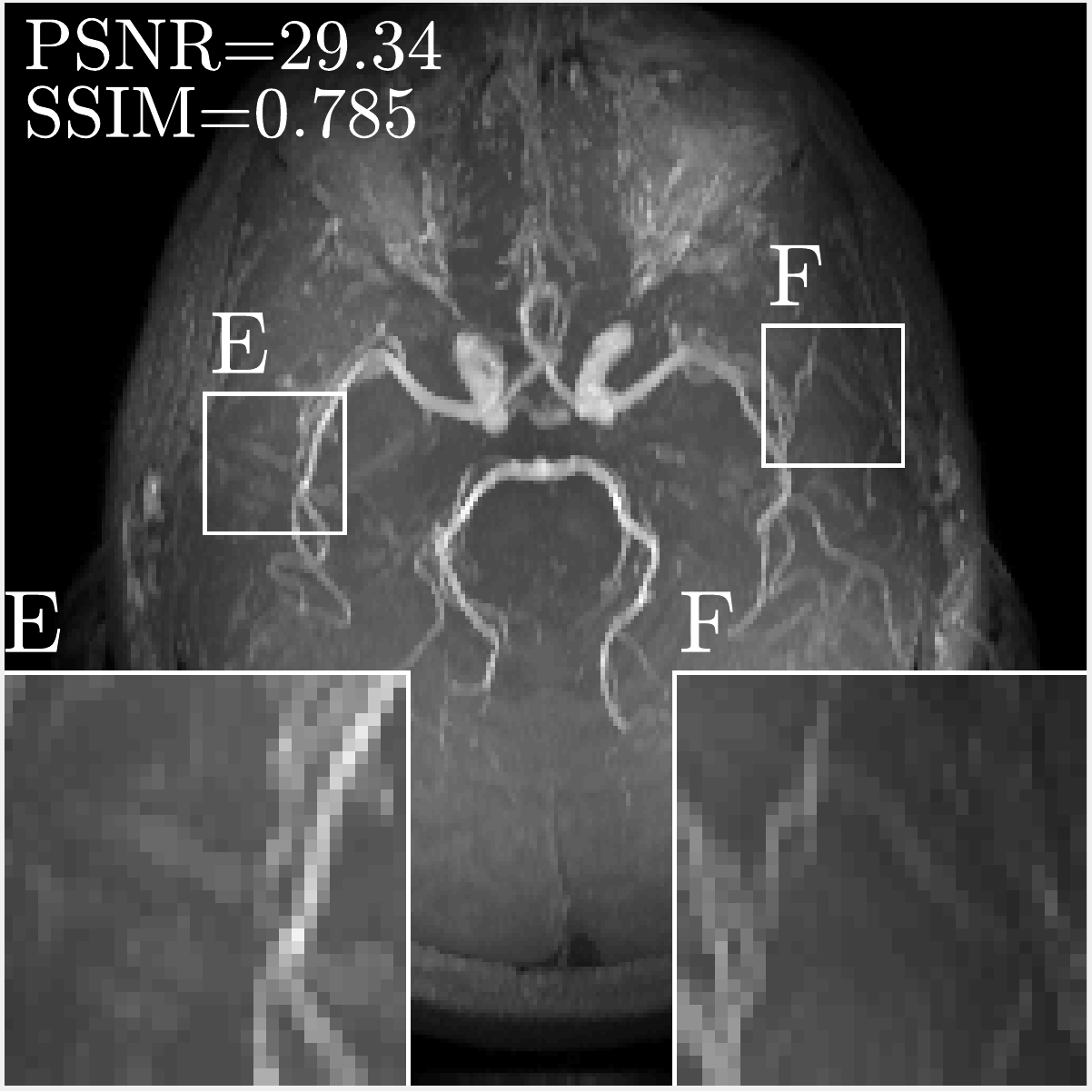}  \\ [-1mm]


\end{tabular}
\caption{\rev{Masks obtained, example reconstructions, and MIP views of a volume under TV decoding at 30\% sampling rate. The mask in the fourth row is obtained using the SSIM as the performance measure in Algorithm \ref{alg:1}, and the following mask is obtained using the PSNR. We also present the performances of the coherence-based \cite{lustig2007sparse} and single-image based \cite{vellagoundar15robust} masks. The last row shows the low-pass mask performance. The reconstruction shown is for subject 1, slice 15 in the middle column, and for the MIP of the whole brain in the last column. In the first row, we present the ground truth as a single slice and as MIP; these are used as references when computing the errors. \label{fig:PSNR_SSIM}}} 
\end{figure}

These observations are further illustrated in Figure \ref{fig:PSNR_SSIM} (only for TV decoder and its masks due to space constraints), where we show the optimized masks, the reconstructions on a single slice \rev{and as the maximum intensity projection (MIP) of the volume this slice belongs to \cite{wallis1989three}.}  We see in particular that the two masks are somewhat different, with that for the PSNR containing more gaps at higher frequencies and fewer gaps at lower frequencies. \rev{We also observe  that compared to the data used in the previous subsections, the angiographic data used in this experiment is more concentrated at the center of the $k-$space. The greedy algorithm is able to adapt to this change and obtain masks that have more lower frequencies. }

\begin{table} 
\centering
\caption{\label{tab:cross_recon} Reconstruction performances at 30\% subsampling rate averaged over 60 angio test slices.  The cases that the training is matched to the performance measure and decoder are highlighted in bold.}
\begin{tabular}{|l|||*{6}{c|}}\hline
{\hspace{4mm} Decoder}  & \multicolumn{2}{c|}{TV}  & \multicolumn{2}{c|}{\rev{BP}}  \\ \hline
\backslashbox{Mask}{Metric }
&\makebox[3em]{\rev{SSIM}}&\makebox[3em]{PSNR} &\makebox[3em]{{SSIM}}&\makebox[3em]{PSNR} \\\hline \hline  
\rev{Coherence-based} &0.738&29.82&0.698&27.85\\\hline
\rev{Single-Image} &0.766&30.72&0.744&28.85\\\hline
SSIM-greedy-TV & {\bf0.795}&31.90&0.779&30.50\\\hline
PSNR-greedy-TV &0.799&{\bf32.74}&0.806&32.97\\\hline
\rev{SSIM-greedy-BP} &0.797&32.72&{\bf0.806}&33.08\\\hline
\rev{PSNR-greedy-BP}&0.795&32.54&0.804&{\bf32.85}\\\hline
\rev{Low Pass} & 0.771&31.70&0.777&31.93\\\hline
\end{tabular}
\end{table}

\rev{

\subsection{Experiments with additive noise}

The data we used in the previous subsections has very low levels of noise. In order to test the validity of our claims in the noisy setting,  we add bivariate circularly symmetric complex random Gaussian noise to our normalized complex images, with a noise standard deviation of $\sigma = 3 \times 10^{-4}$ for both the real and imaginary components. Since the ground truth images are normalized, this noise level gives an average signal-to-noise ratio (SNR) of $25.68$ dB. We set the denoising parameter of NESTA  to \secondrev{$\epsilon$} $= 10$ for TV minimization  and to \secondrev{$\epsilon$} $ = 1.1$ for BP with shearlets which work well with the various masks and images used in this section. In Algorithm \ref{alg:1}, we measure the error at each iteration with respect to denoised image that is obtained using BM3D denoising algorithm \cite{dabov2009bm3d}.  Note that the ground truth should not be used in the learning algorithm, since it is unknown in practice.
On the other hand, in the testing part, we compute the errors with respect to the ground truth images.

As can be seen from Figure \ref{fig:NOISY} and Table \ref{tab:table_noisy}, the greedy algorithm is still capable of finding a better mask compared to the other baseline masks. Therefore, in this example, our approach is robust with respect to noise. Note that we train with respect to the PSNR, but also report the SSIM values. Note also that compared to the case where the noise levels were very low, the mask obtained in noisy setting is slightly closer to a low-pass mask.  \secondrev{The reason for this is that the noise hides the relatively weaker signal present at high frequencies, while only having a minimal effect on the stronger signal present at lower frequencies.}

\begin{figure} 
\centering
\begin{tabular}{cccc}
& \hspace{-5mm} \textbf{\rev{Mask}} & \hspace{-4mm} \textbf{\rev{TV decoder}} & \hspace{-4mm} \textbf{\rev{BP decoder}}    \\
\hspace{-7mm} \rotatebox{90}{\hspace{3mm} \textbf{\rev{Adding noise}}} &                         
\hspace{-5mm}\includegraphics[width=.16\textwidth]{FIGURES_REV/ABc_original_magnified_3.pdf} &
\hspace{-4mm}\includegraphics[width=.16\textwidth]{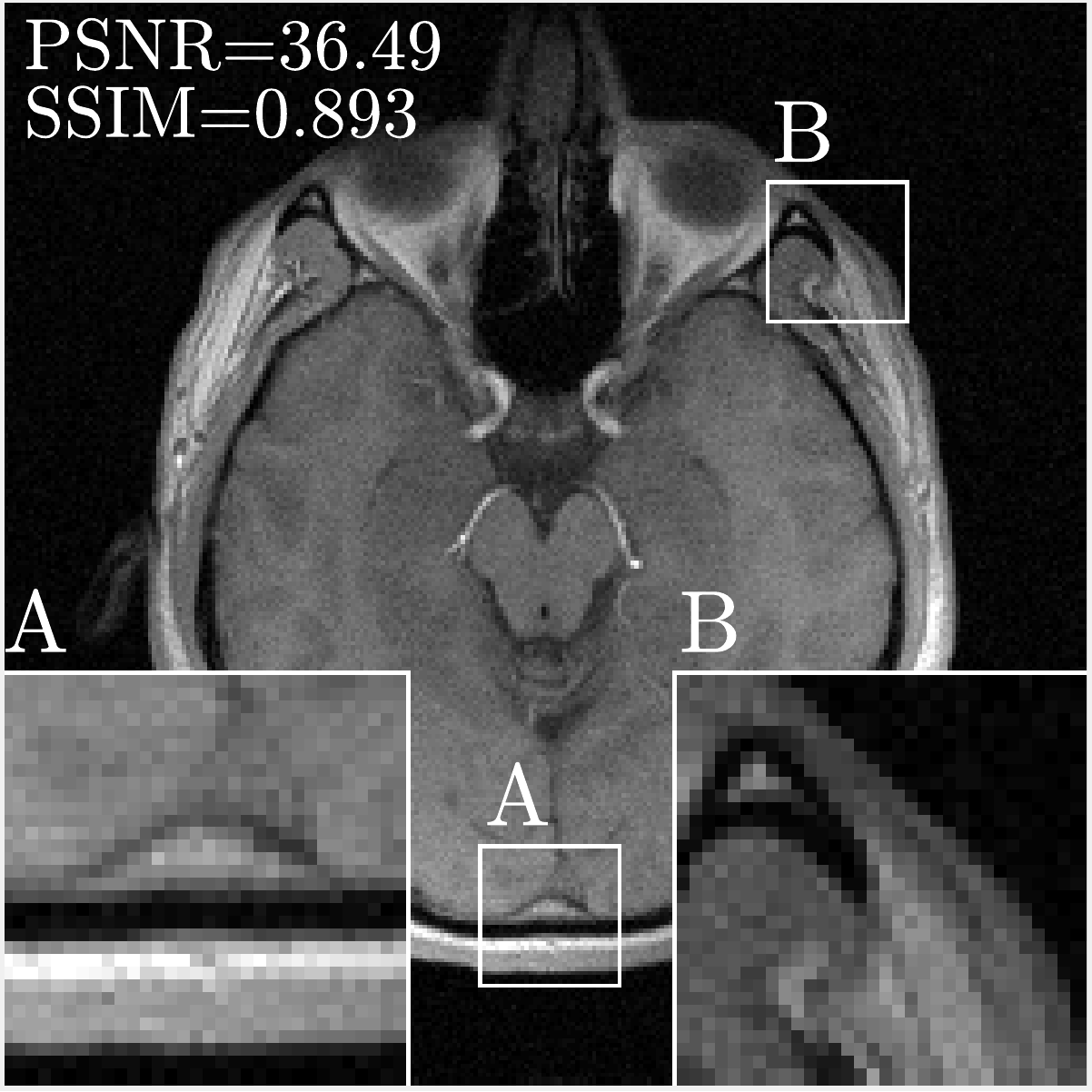}  & 
\hspace{-4mm}\includegraphics[width=.16\textwidth]{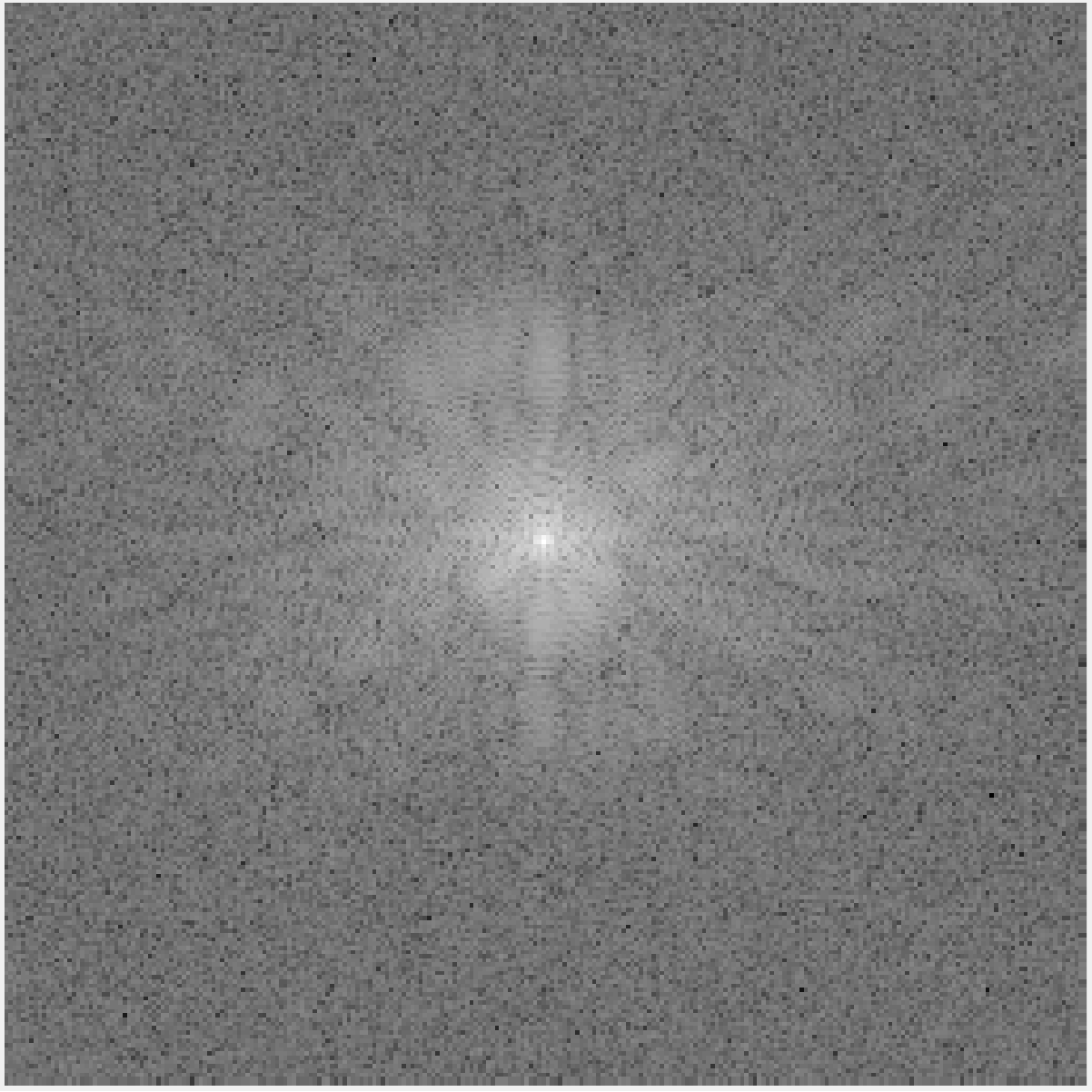}  \\ [-1mm]
 
\hspace{-7mm} \rotatebox{90}{\hspace{3mm} \textbf{\rev{Coher. based}}} &                         
\hspace{-5mm}\includegraphics[width=.16\textwidth]{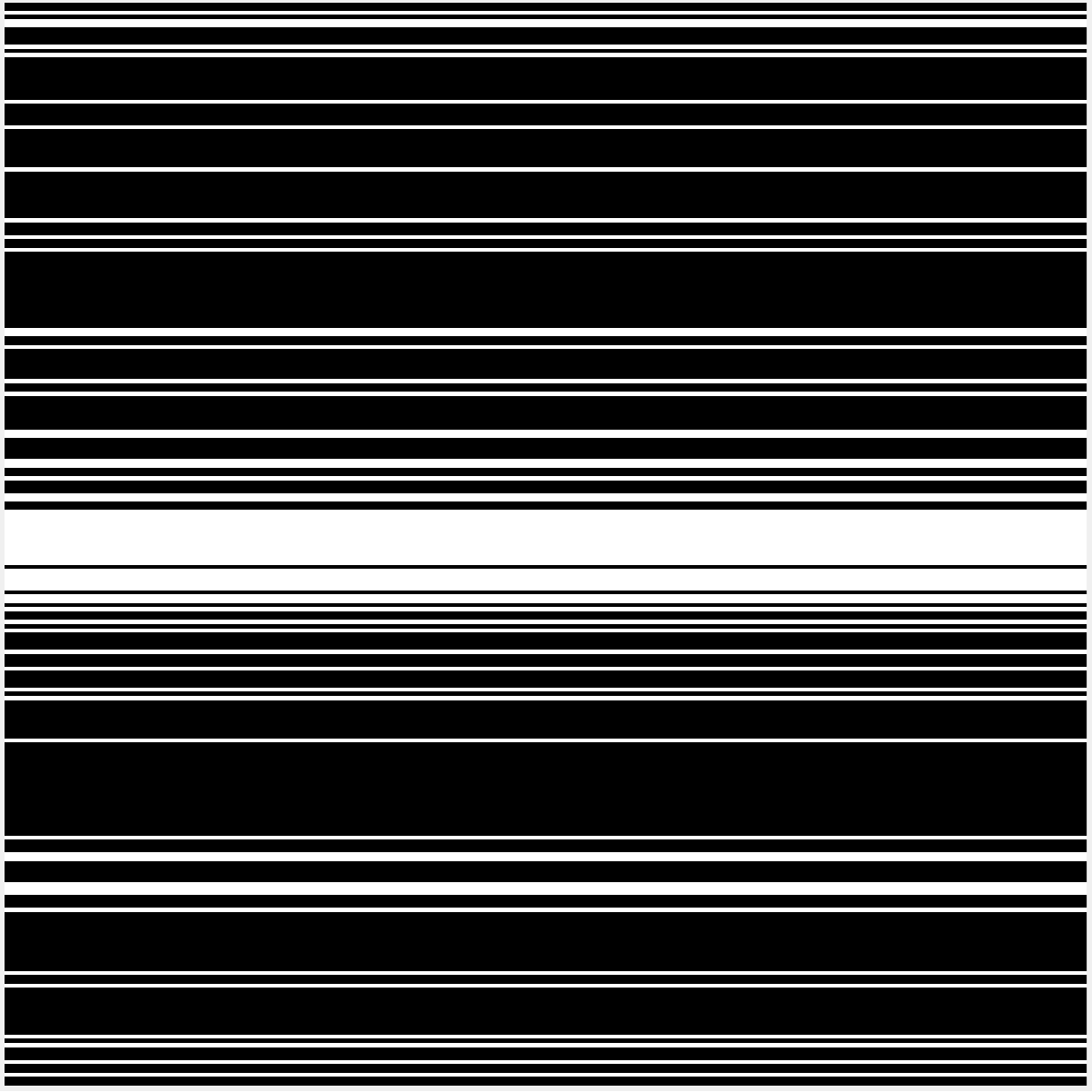} &
\hspace{-4mm}\includegraphics[width=.16\textwidth]{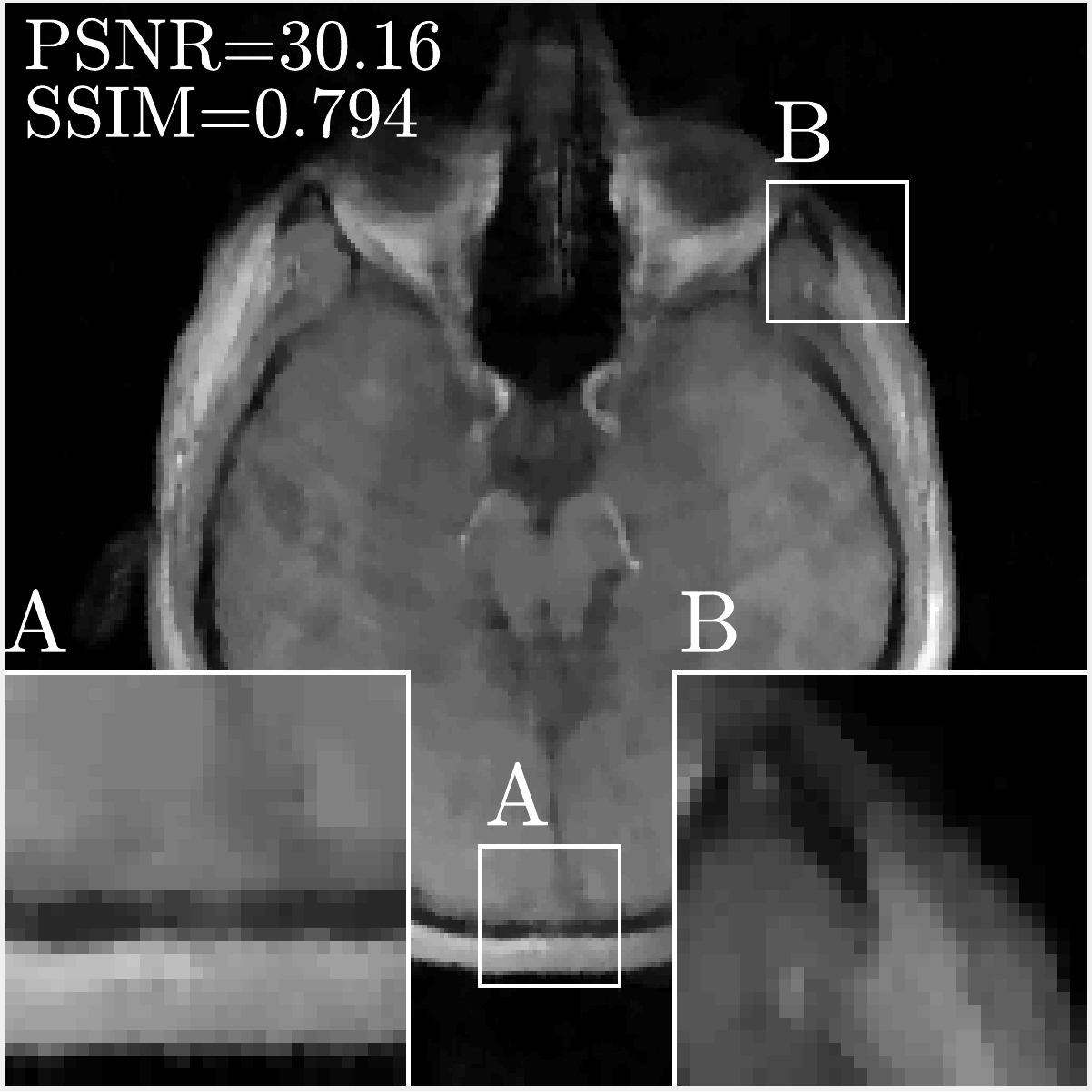}  & 
\hspace{-4mm}\includegraphics[width=.16\textwidth]{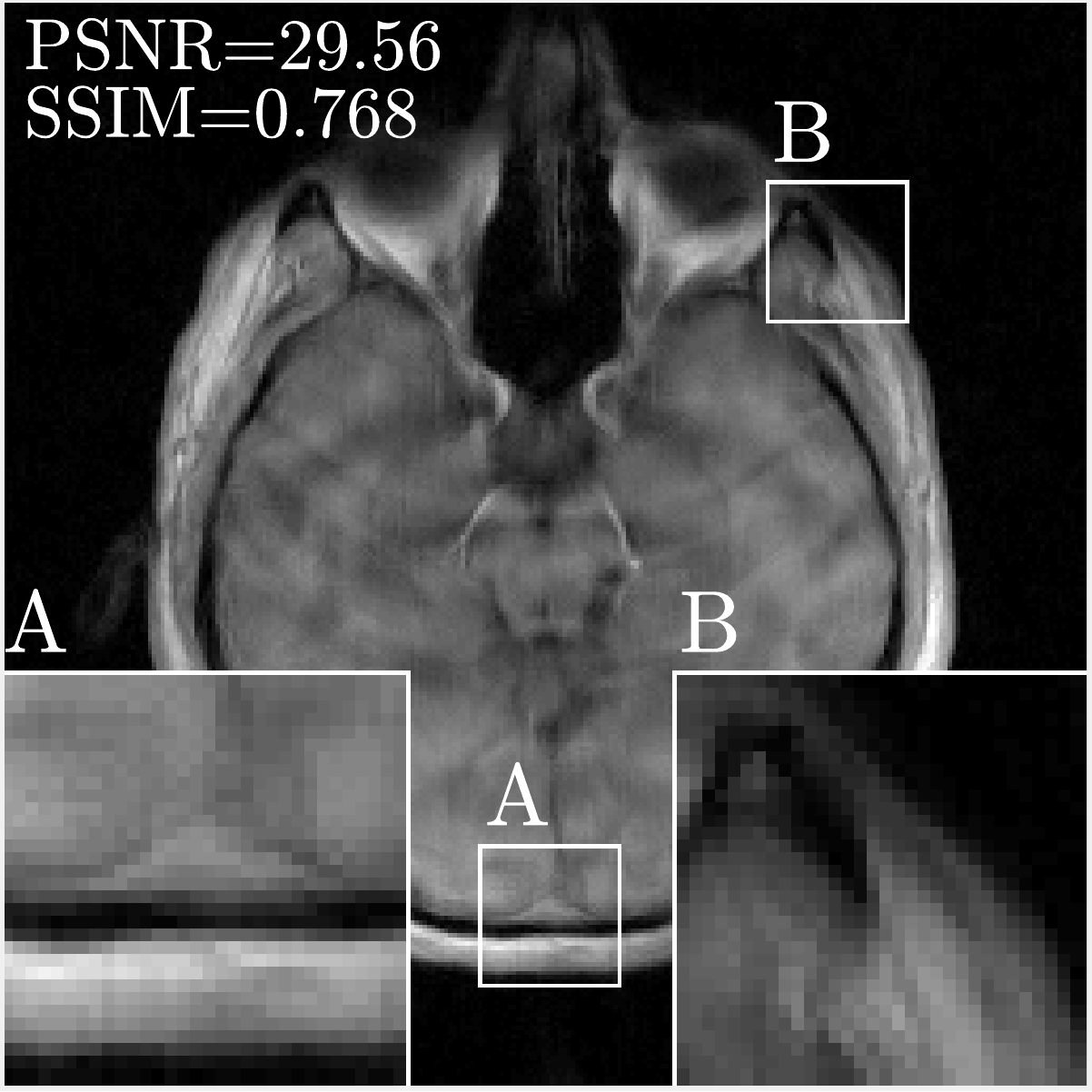}  \\ [-1mm]
\hspace{-7mm} \rotatebox{90}{\hspace{4mm} \textbf{\rev{Single image}}} &                         
\hspace{-5mm}\includegraphics[width=.16\textwidth]{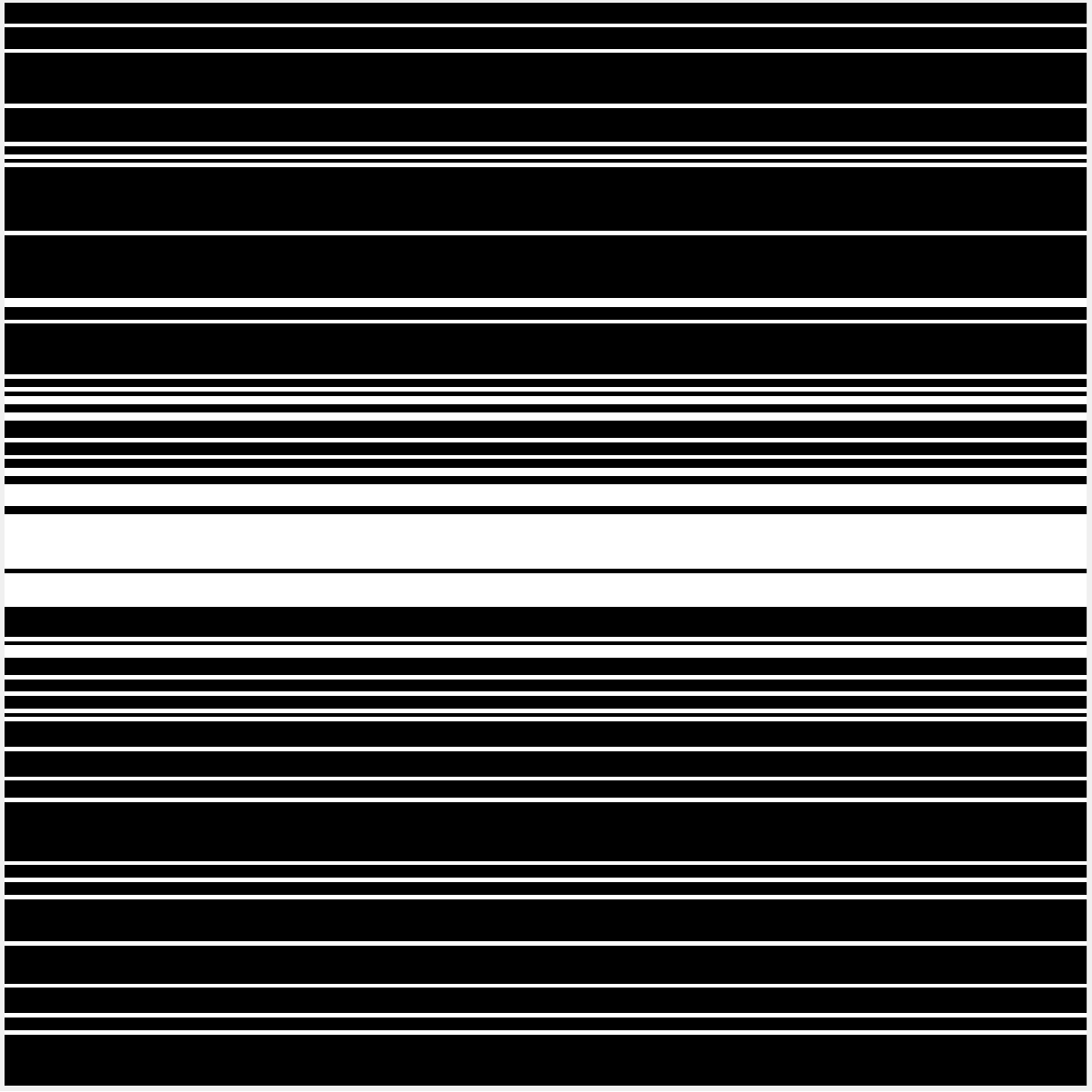} &
\hspace{-4mm}\includegraphics[width=.16\textwidth]{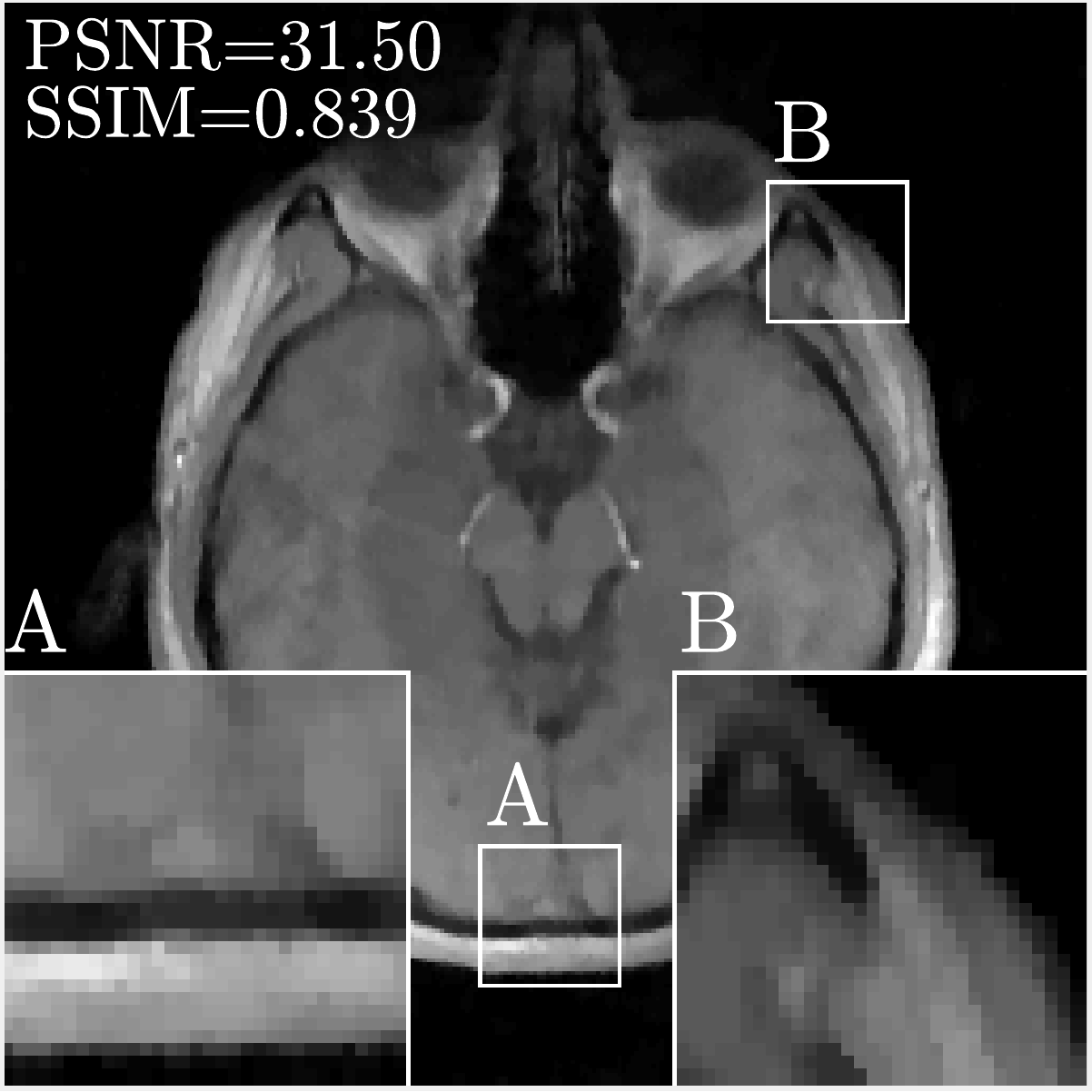}  & 
\hspace{-4mm}\includegraphics[width=.16\textwidth]{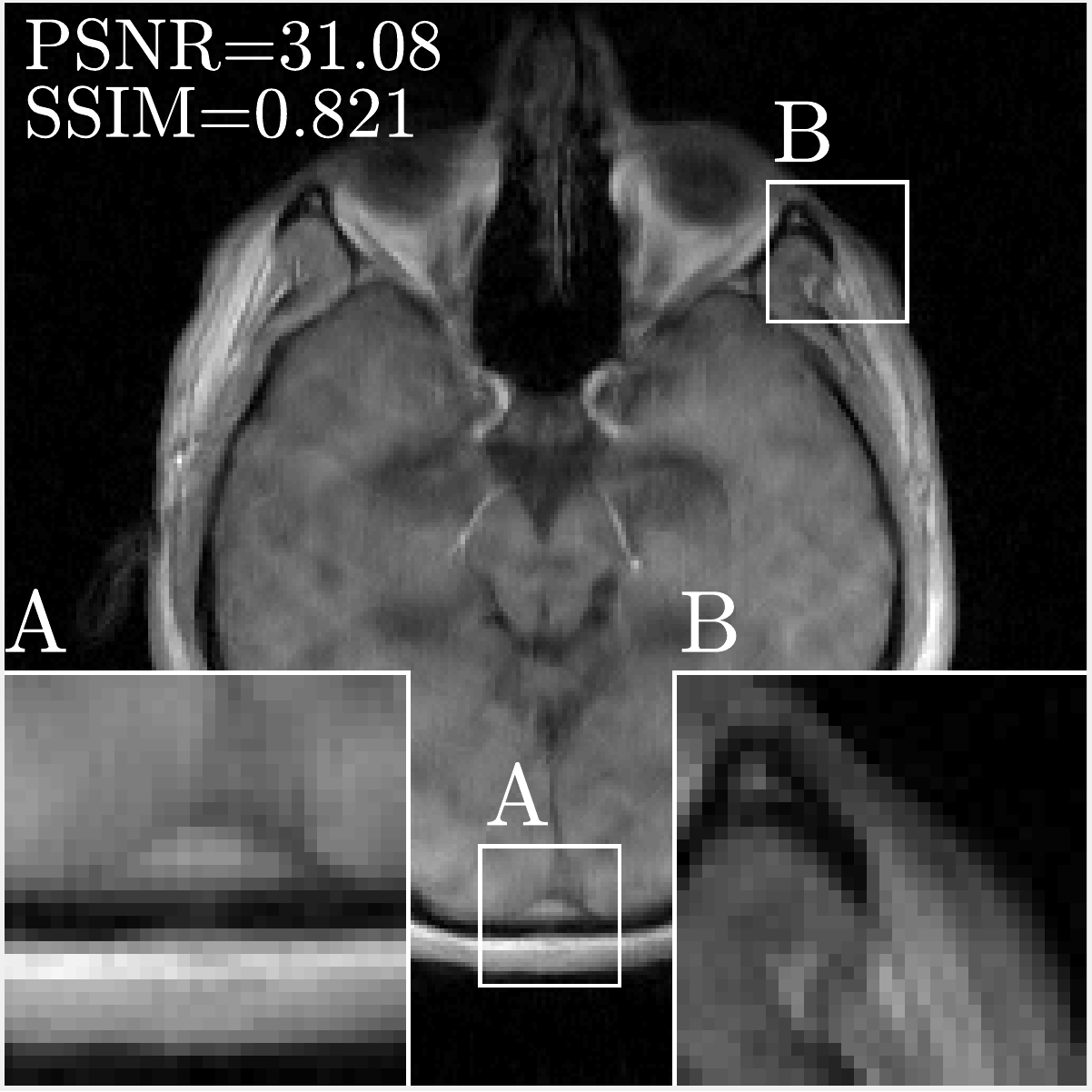} 
 \\ [-1mm]

\hspace{-6mm} \rotatebox{90}{\hspace{5mm} \textbf{\rev{TV-greedy}}} &
\hspace{-5mm}\includegraphics[width=.16\textwidth]{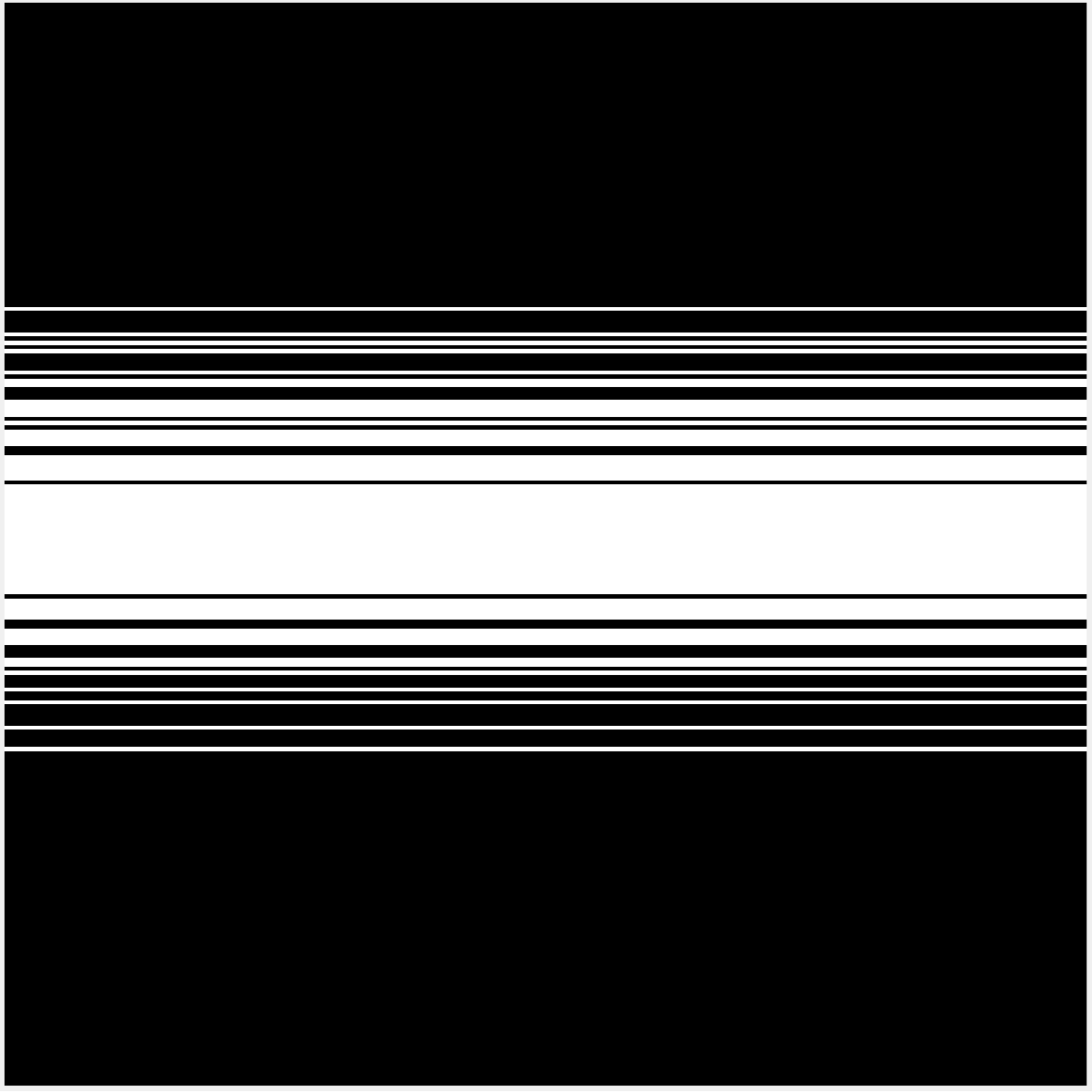} &
\hspace{-4mm}\includegraphics[width=.16\textwidth]{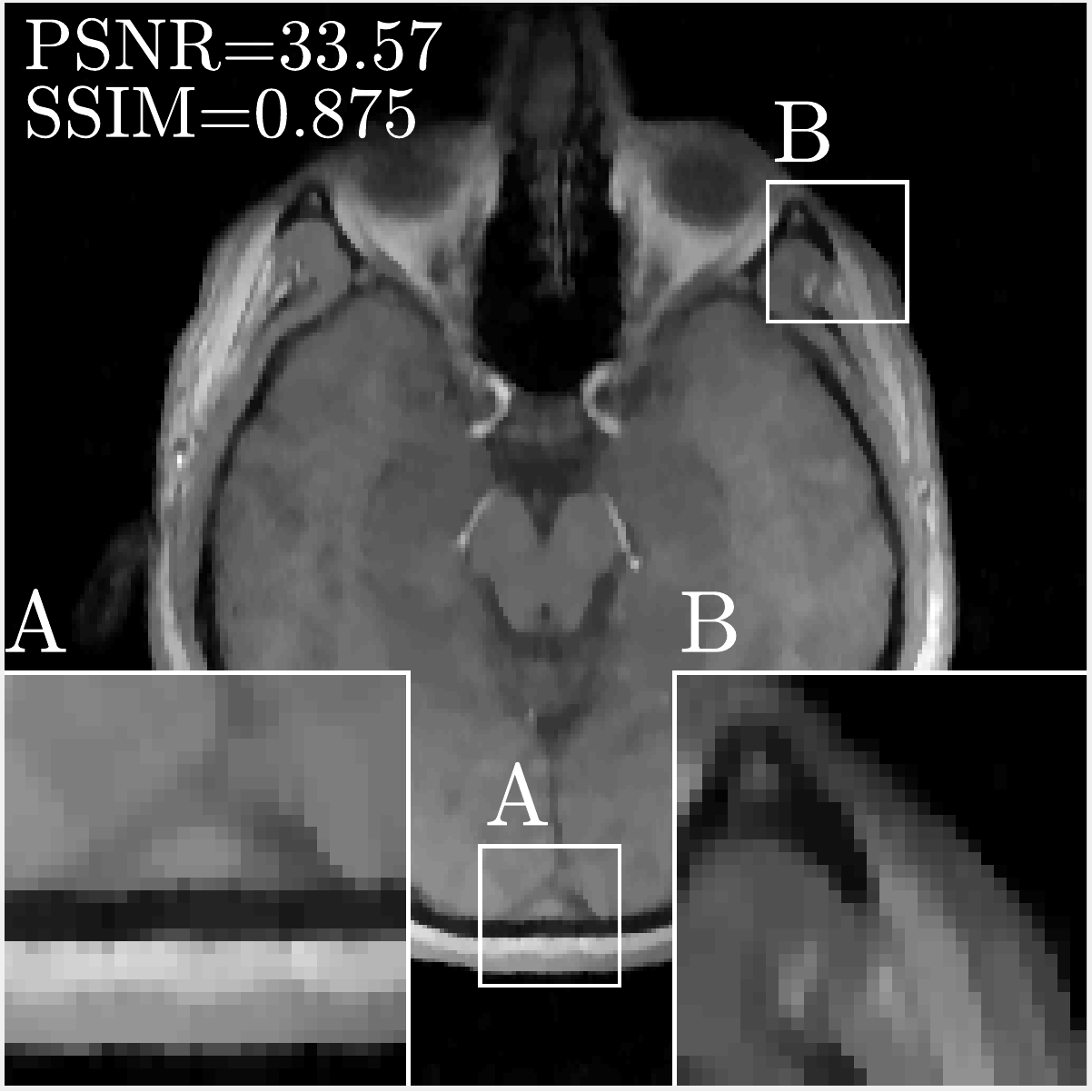} &
\hspace{-4mm}\includegraphics[width=.16\textwidth]{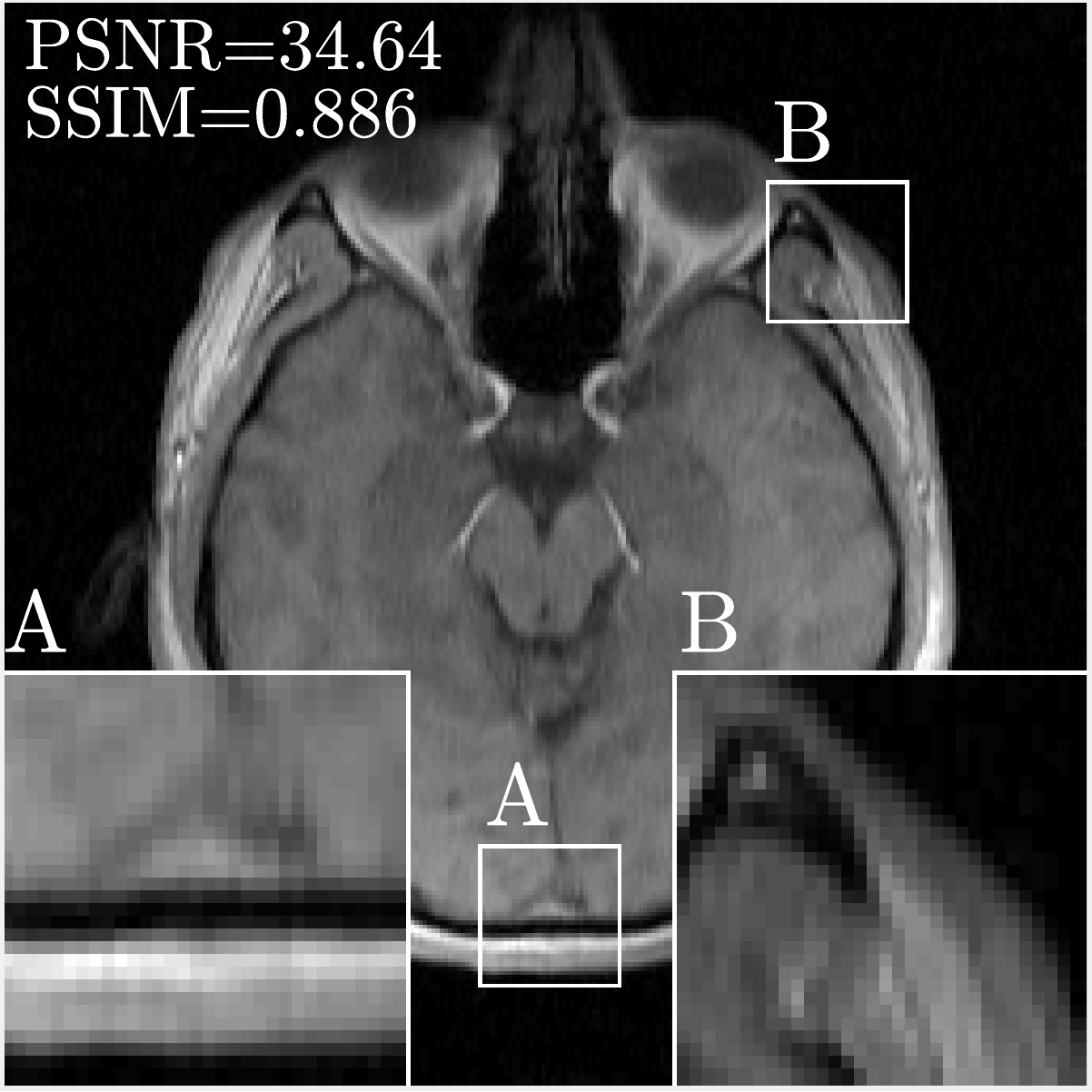}   \\ [-1mm]


\hspace{-5mm}\rotatebox{90}{\hspace{5mm} \textbf{\rev{BP-greedy}}} &                         
\hspace{-5mm}\includegraphics[width=.16\textwidth]{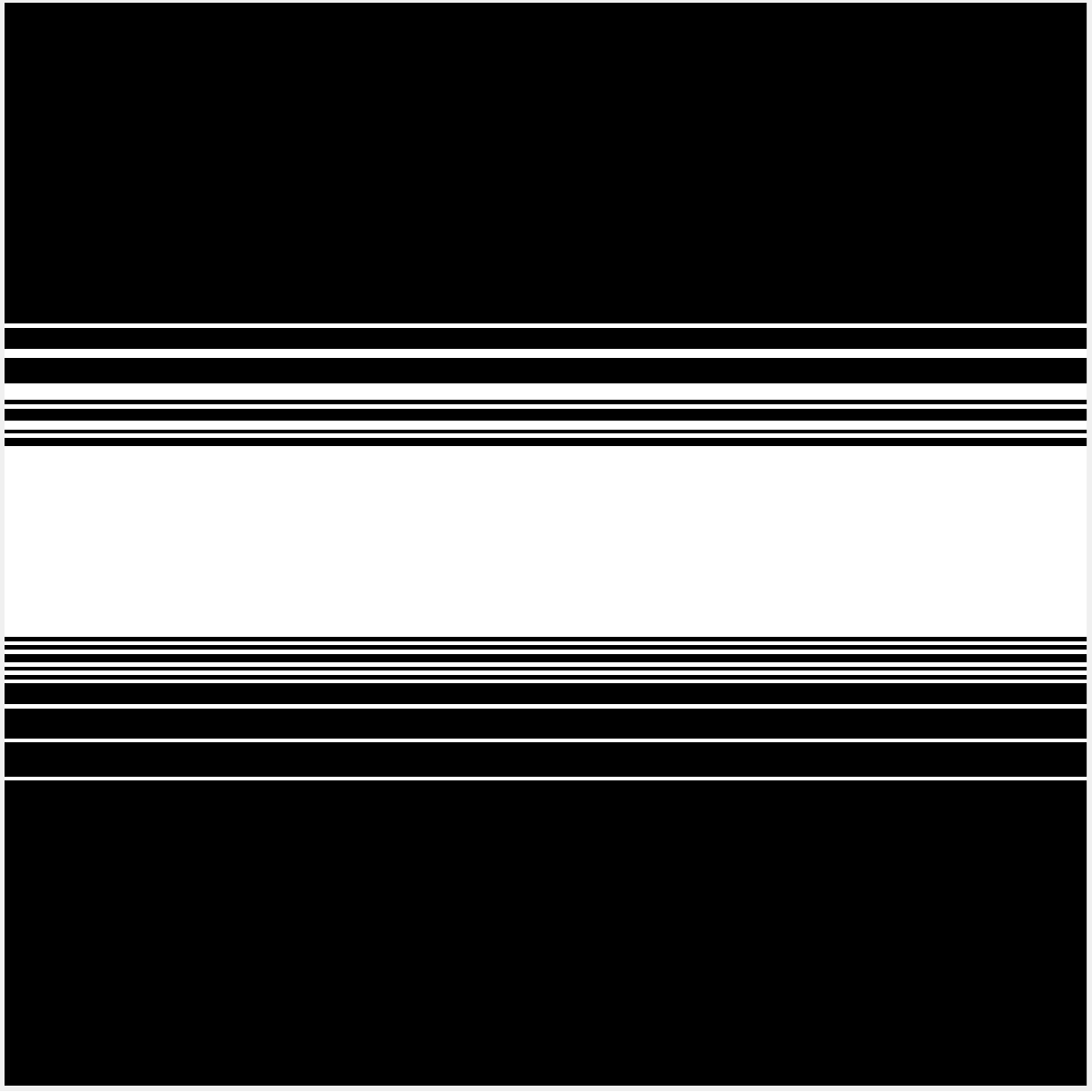} &
\hspace{-4mm}\includegraphics[width=.16\textwidth]{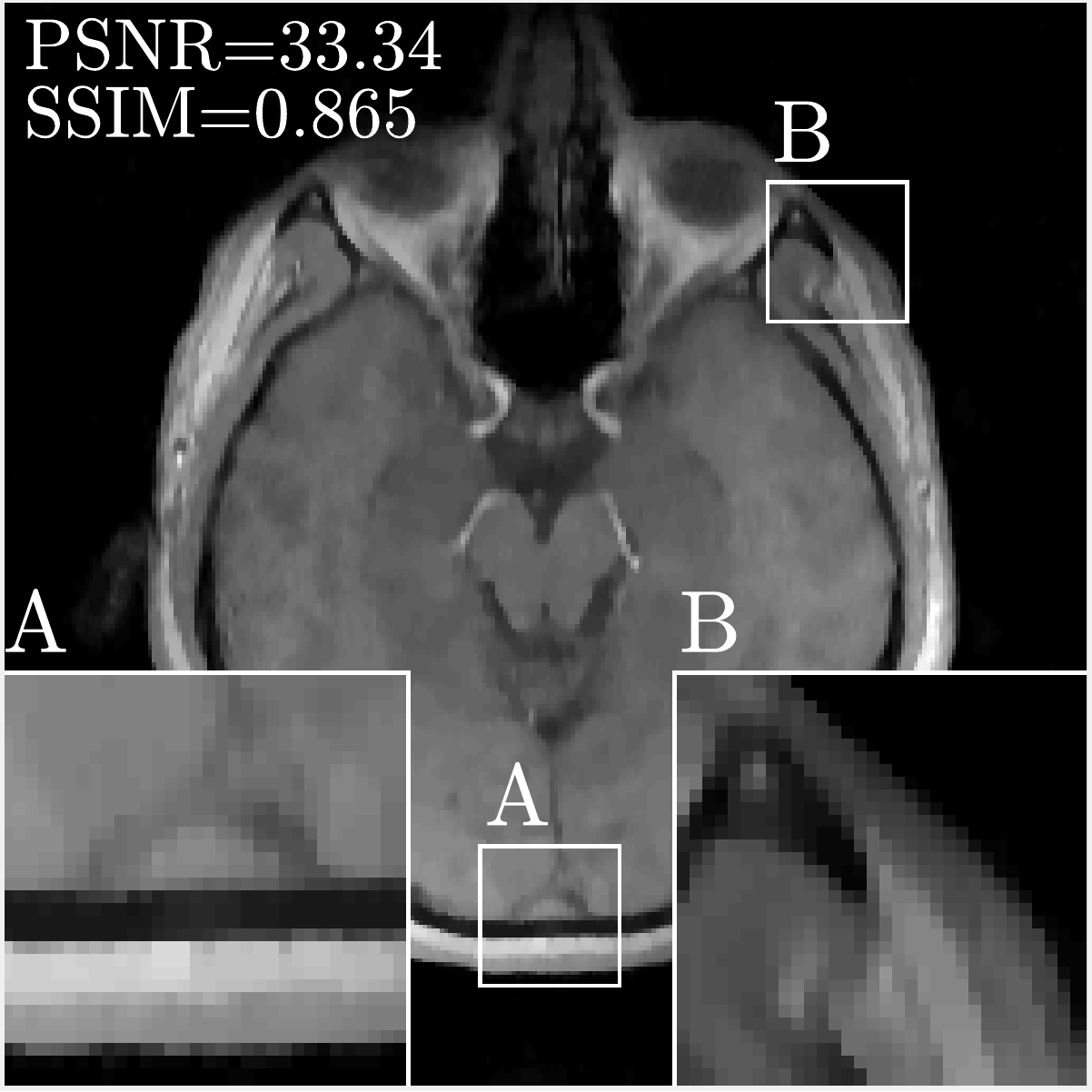}  & 
\hspace{-4mm}\includegraphics[width=.16\textwidth]{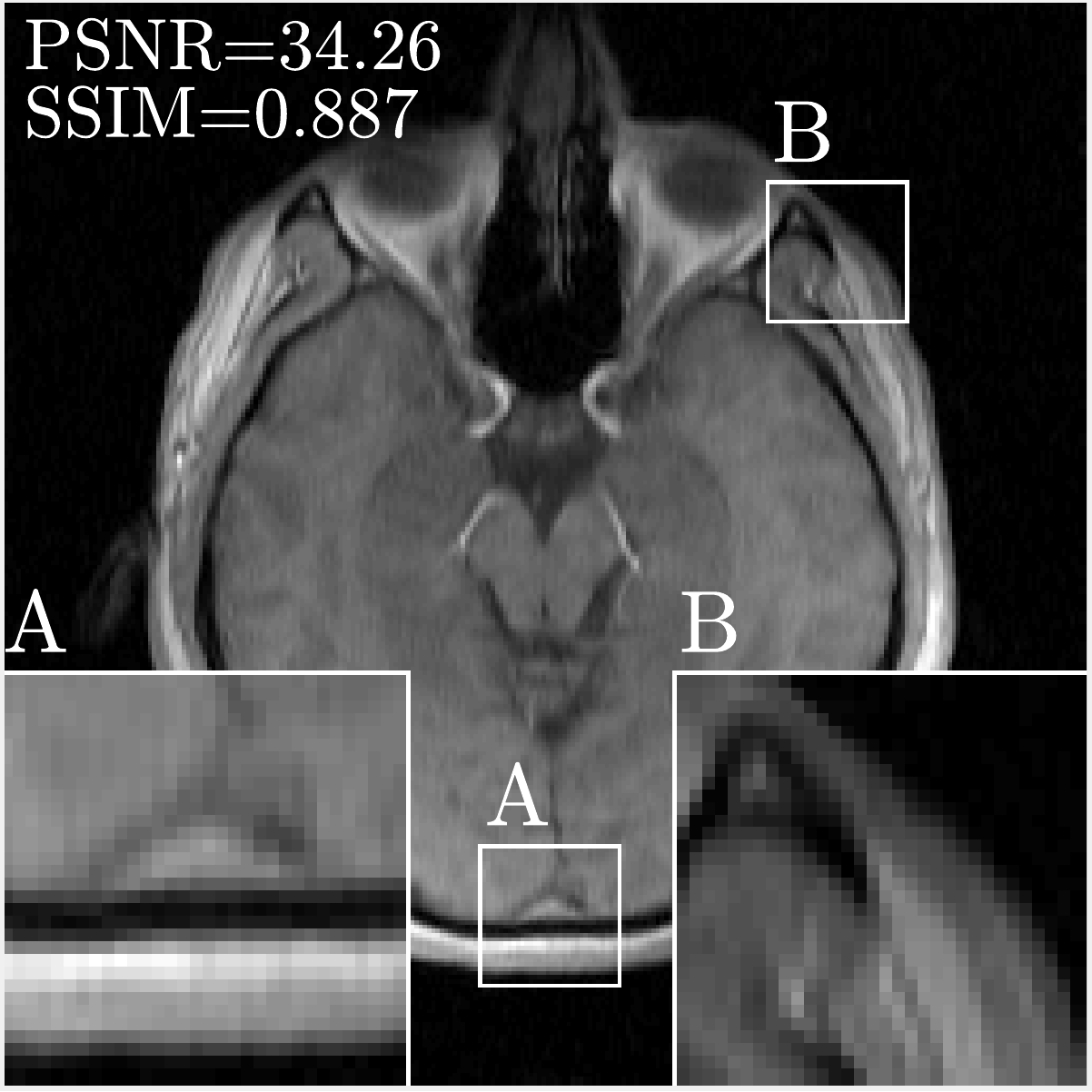}  \\ [-1mm]

\hspace{-5mm}\rotatebox{90}{\hspace{6mm} \textbf{\rev{Low pass}}} &                         
\hspace{-5mm}\includegraphics[width=.16\textwidth]{FIGURES_REV/mask_h=1_p=1_rate_25_LP.pdf} &
\hspace{-4mm}\includegraphics[width=.16\textwidth]{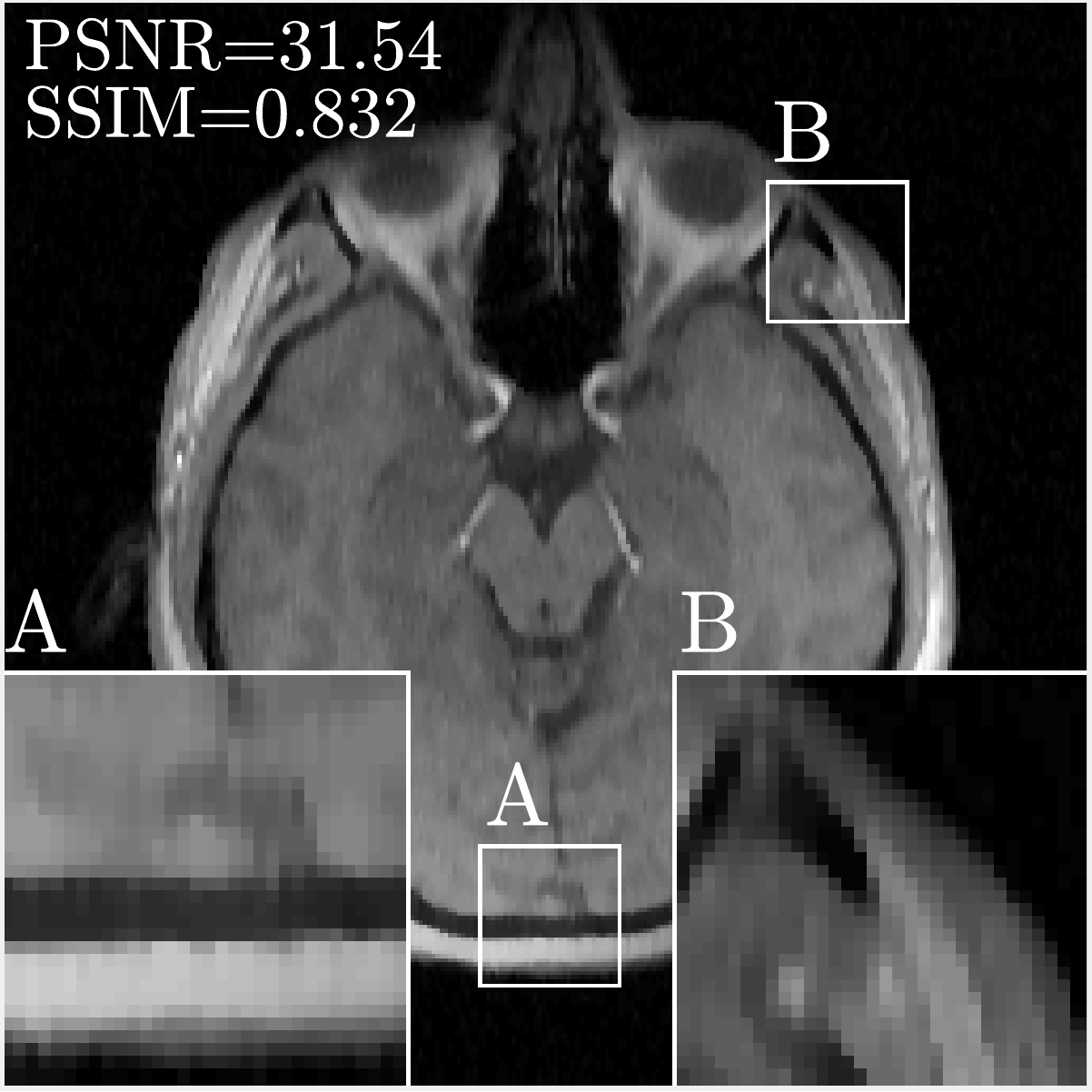}  & 
\hspace{-4mm}\includegraphics[width=.16\textwidth]{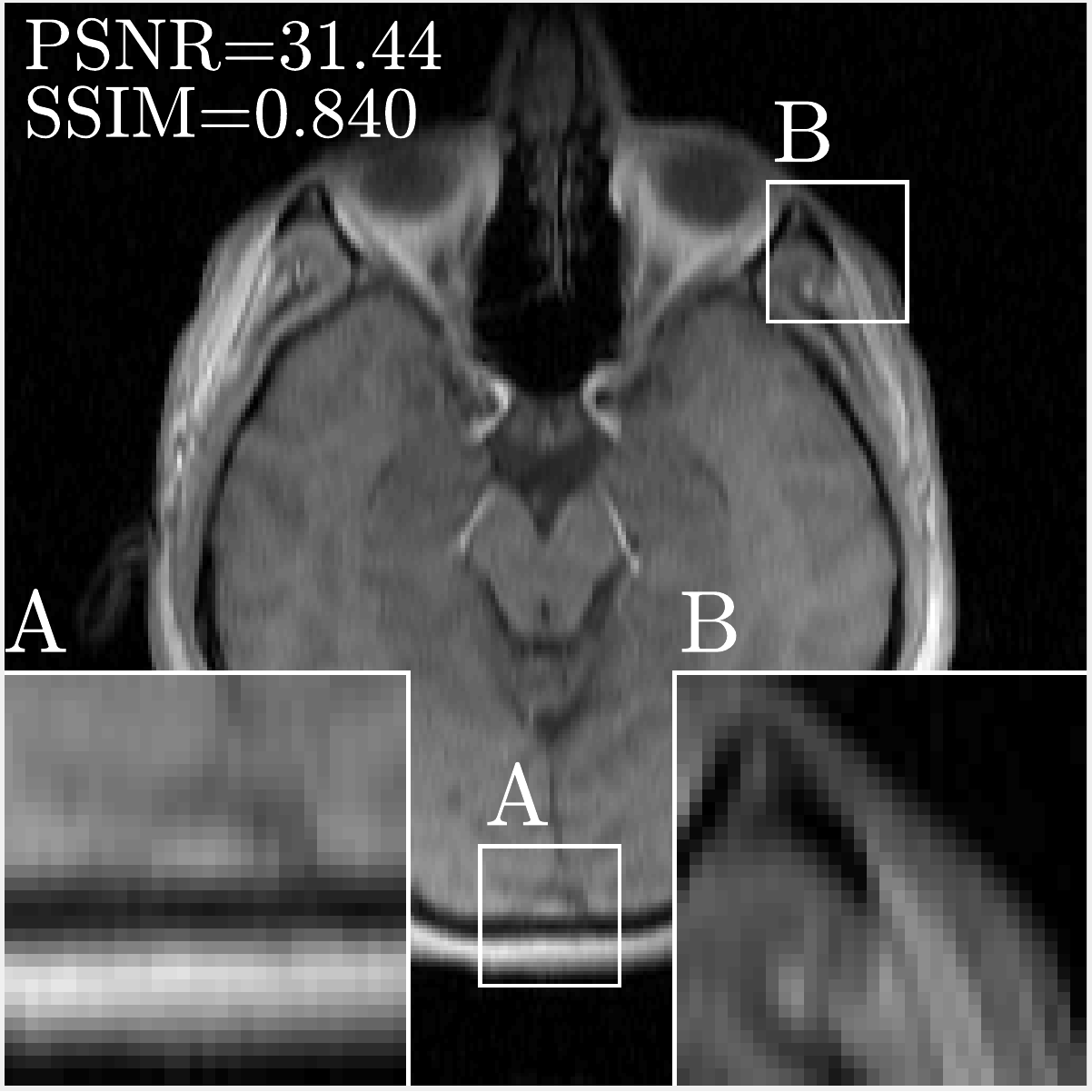}  \\ [-1mm]


\end{tabular}
\caption{\rev{Masks obtained and example reconstructions under TV and BP decoding at 25\% sampling rate. PSNR is used as the performance measure in the greedy method given in Algorithm \ref{alg:1}.  The reconstruction shown is for subject 2, slice 4. We also present the performances of coherence-based \cite{lustig2007sparse} and single-image based \cite{vellagoundar15robust} masks. In the first row, we present the ground truth, noisy ground truth, and its k-space.  The last row shows the low-pass mask performance.  \label{fig:NOISY}}} 
\end{figure}

\begin{table} 
\centering
\caption{\label{tab:table_noisy} \rev{ PSNR and SSIM performances at 25\% subsampling rate with additive noise, averaged over 60 test slices and 10 random noise draws. The cases that the training is matched to the performance measure and decoder are highlighted in bold.
 }}
\begin{tabular}{|l|||*{6}{c|}}\hline
{\hspace{4mm} Decoder}  & \multicolumn{2}{c|}{\rev{TV}}  & \multicolumn{2}{c|}{\rev{BP}}  \\ \hline
\backslashbox{Mask}{Metric }
&\makebox[3em]{SSIM}&\makebox[3em]{PSNR} &\makebox[3em]{SSIM}&\makebox[3em]{PSNR} \\\hline \hline  
\rev{Coherence-based} & 0.799&29.73&0.808&30.29\\\hline
\rev{Single-Image} & 0.828&30.86&0.843&31.58\\\hline
\rev{TV-greedy }& 0.878& {\bf33.07}&0.896&34.21\\\hline
\rev{BP-greedy}& 0.875&32.89&0.893&{\bf34.07}\\\hline
\rev{Low Pass} & 0.859&31.37&0.872&31.80\\\hline
\end{tabular}
\end{table}
}


\section{Conclusion} \label{sec:CONCLUSION}

We have presented a versatile learning-based framework for selecting masks for compressive MRI, using training signals to optimize for a given decoder and anatomy.  As well as having a rigorous \secondrev{justification via statistical learning theory, our approach is seen to provide improved performance} on real-world data sets for a variety of reconstruction methods.  Since our framework is suited to general decoders, it can potentially be used to optimize the indices for new reconstruction methods that are yet to be discovered. \rev{In this work, we focused on 1D subsampling for 2D MRI, 2D subsampling (via horizontal and vertical lines) for 2D MRI, and 1D subsampling for 3D MRI, but our greedy approach can potentially provide an automatic way to optimize the sampling in the settings of 2D subsampling for 3D MRI and non-Cartesian sampling, as opposed to constructing a randomized pattern on a case-by-case basis. \secondrev{For the setting of 3D MRI, there is an additional computational challenge to our greedy algorithm, since the candidate set is large.}}

In future studies, we will also seek to validate the performance under the important practical variation of {\em multi-coil} measurements, as well as applications beyond MRI such as computer tomography, phase retrieval, and ultrasound.  We finally note that in this paper, the number of subjects and training images used was relatively small, and we anticipate that larger data sets would be of additional benefit in realizing the full power of our theory.

\appendix



\subsection{Proof of Proposition \ref{prop:guarantee}} \label{sec:MOTIVATION}

    Using the fact that $\eta$ lies in $[0,1]$ and applying Hoeffding's inequality \cite{Massart2007}, we obtain for any $\Omega \in \mathcal{A}$ and $t > 0$ that
    \begin{equation}
        \left\vert \frac{1}{m} \sum_{j = 1}^m \eta_{\Omega}(\x_j) - \mathbb{E}_P \left[ \eta_{\Omega}(\x) \right] \right\vert \le t, \notag
    \end{equation}
    with probability at least $1 - 2 \exp ( - 2m t^2  )$. \rev{Since the probability of a union of events is upper bounded by the sum of the individual probabilities (i.e., the union bound)}, we find that the same inequality holds for \emph{all} $\Omega \in \mathcal{A}$ with probability at least $1 - 2 \abs{ \mathcal{A} } \exp ( - 2 m t^2 )$.
    The proposition follows by setting $\delta = 2 \abs{ \mathcal{A} } \exp ( - 2 m t^2 )$ and solving for $t$.

\subsection{Proof of Proposition \ref{prop:guarantee_noisy}} \label{sec:PF_NOISY}

By the fact that the Fourier transform is a unitary operation and i.i.d.~Gaussian vectors are invariant under unitary transforms, we have
\begin{align}
\etanoisy(\Omega) 
    &= \mathbb{E} \left[ \eta ( \x , \hat{\x} ( \P_\Omega \bPsi \x + \w ) ) \right]  \notag \\
    &= \mathbb{E} \left[ \eta ( \x , \hat{\x} ( \P_\Omega \bPsi ( \x + \v ) ) ) \right], \label{eq:w_to_v}
\end{align}
where $\hat{\x}(\b)$ denotes the estimator applied to the noisy output $\b$, and $\v$ has the same distribution as any given $\v_j$.  

Let $\vtil = \xi(\v)$ be the denoised version of $\v$.  Using the triangle inequality, we write
\begin{align}
& \left\vert \frac{1}{m} \sum_{j = 1}^m \eta ( \x_j + \vtil_j, \hat{\x} ( \P_\Omega \bPsi ( \x_j + \v_j ) ) ) -  \etanoisy(\Omega) \right\vert \notag \\
& = \Bigg\vert \frac{1}{m} \sum_{j = 1}^m \eta ( \x_j + \vtil_j, \hat{\x} ( \P_\Omega \bPsi ( \x_j + \v_j ) ) ) \notag \\
    &\qquad  -  \mathbb{E} \left[ \eta ( \x , \hat{\x} ( \P_\Omega \bPsi ( \x + \v ) ) ) \right] \Bigg\vert \notag \\
& \rev{\leq} \Bigg\vert \frac{1}{m} \sum_{j = 1}^m \eta ( \x_j + \vtil_j, \hat{\x} ( \P_\Omega \bPsi ( \x_j + \v_j ) ) ) \notag \\
    &\qquad  -  \mathbb{E} \left[ \eta ( \x + \vtil , \hat{\x} ( \P_\Omega \bPsi ( \x + \v ) ) ) \right] \Bigg\vert \notag \\
    & ~~+ \Big\vert \mathbb{E} \left[ \eta ( \x + \vtil , \hat{\x} ( \P_\Omega \bPsi \x + \w ) ) \right] -  \mathbb{E} \left[ \eta ( \x , \hat{\x} ( \P_\Omega \bPsi \x + \w ) ) \right] \Big\vert . \notag
\end{align}
Using \eqref{eq:w_to_v} and following the proof of Proposition \ref{prop:guarantee}, the first term is upper bounded by $\sqrt{ \frac{1}{2m} \log \big( \frac{2 \abs{ \mathcal{A} }}{\delta} \big) }$ with probability at least $1 - \delta$.  Moreover, by the continuity condition assumed in the proposition statement, the second term above is upper bounded by $L \, \mathbb{E} \left[ \norm{\vtil}_2 \right]$, thus completing the proof.

\section*{Acknowledgment}

This work has received funding from the European Research Council (ERC) under the European Union's Horizon 2020 research and innovation program (grant agreement n$^{\circ}$ 725594 - time-data), and from Hasler Foundation Program: Cyber Human Systems (project number 16066). It was also sponsored by the Department of the Navy, Office of Naval Research (ONR) under a grant number N62909-17-1-2111. 



\ifCLASSOPTIONcaptionsoff
  \newpage
\fi

\bibliographystyle{IEEEtran}
\bibliography{references_adaptive_MRI,yhli}


\end{document}